\documentclass{article}

\usepackage[T1]{fontenc}
\usepackage{palatino}
\usepackage{fullpage}
\usepackage{graphicx}
\usepackage{amsmath}
\usepackage{amsfonts}
\usepackage{amssymb}
\usepackage[dvipsnames,svgnames,x11names,hyperref]{xcolor}
\definecolor{lkcol}{RGB}{130,70,200}
\usepackage[hidelinks,colorlinks=true,linkcolor=lkcol,citecolor=blue,linktocpage]{hyperref}
\usepackage{cite}
\usepackage{csquotes}
\usepackage{caption}
\usepackage{subcaption}
\usepackage{nicematrix}
\usepackage{mathtools}
\usepackage{tabularray}
\usepackage{float}
\usepackage{amsthm}
\usepackage{dsfont}
\usepackage{calc}
\usepackage{fancyhdr}
\usepackage{titling}

\numberwithin{equation}{section}

\newcommand{\tSigma}{\widetilde{\Sigma}}

\newcommand{\IC}{\mathbb{C}}

\newcommand{\IN}{\mathbb{N}}
\newcommand{\IQ}{\mathbb{Q}}
\newcommand{\IR}{\mathbb{R}}
\newcommand{\IZ}{\mathbb{Z}}

\def\be{\begin{equation}}
\def\ee{\end{equation}}

\def\IR{{\mathbb{R}}}

\def\IZ{{\mathbb{Z}}}
\def\IP{{\mathbb{P}}}
\def\IC{{\mathbb{C}}}
\def\IN{{\mathbb{N}}}

\def\CH{\mathcal{H}}
\def\CO{\mathcal{O}}
\def\CQ{\mathcal{Q}}
\def\CM{{\mathcal{M}}}
\def\CN{{\mathcal{N}}}

\def\CW{\mathcal{W}}
\def\tCW{\widetilde{\mathcal{W}}}

\def\eff{\mathrm{eff}}

\def\top{{\mathrm{top}}}
\def\eff{{\mathrm{eff}}}

\def\open{{\mathrm{open}}}
\def\wr{{\mathfrak{wr}}}

\def\Tr{{\rm{Tr}}}

\def\eff{{\rm{eff}}}

\def\vortex{{\rm{vortex}}}
\def\kinkyvortex{{\rm{kinky\ vortex}}}
\def\top{{\rm{top}}}
\def\open{{\rm{open}}}
\def\disk{{\rm{disk}}}
\def\rel{{\rm{rel}}}

\def\theory{\mathrm{theory}}

\def\theory{\text{theory}}

\def\MS{{\mathrm{MS}}}

\newtheorem{remark}{Remark}
\newtheorem{conjecture}{Conjecture}

\usepackage{authblk}

\title{
Linking disks, spinning vortices and \\
exponential networks of augmentation curves\\[30pt]
}
\author[1]{Kunal Gupta}
\author[1,2,3]{Pietro Longhi}
\affil[1]{Department of Physics and Astronomy, Uppsala University, Box 516, 751 20 Uppsala, Sweden}
\affil[2]{Department of Mathematics, Uppsala University, Box 480, 751 06 Uppsala, Sweden}
\affil[3]{Centre for Geometry and Physics, Uppsala University, Box 516, 751 20 Uppsala, Sweden}
\date{}                     
\fancypagestyle{firstpage}
{
    \fancyhead[L]{}    
    \fancyhead[R]{UUITP-32/24}
}

\begin{document}

\maketitle
\thispagestyle{firstpage}

\begin{abstract}
We propose a mirror derivation of the quiver description of open topological strings known as the knots-quivers correspondence, 
based on enumerative invariants of augmentation curves encoded by exponential networks.
Quivers are obtained by studying M2 branes wrapping holomorphic disks with Lagrangian boundary conditions on an M5 brane, 
through their identification with a distinguished sector of BPS kinky vortices in the 3d-3d dual QFT. 
Our proposal suggests that holomorphic disks with Lagrangian boundary conditions
are mirror to calibrated 1-chains on the associated augmentation curve, whose intersections encode the linking of boundaries.
\end{abstract}

\newpage

\tableofcontents

\section{Introduction}

Topological string theory on local Calabi-Yau threefolds has long been known to conceal an underlying 
integral structure, predicted by its embedding into M-theory \cite{Gopakumar:1998ii, Gopakumar:1998jq}.
In the open string sector, the topological $A$-model computes contributions from holomorphic maps $u:(C,\partial C)\to (X,L)$ from the worldsheet to a Calabi-Yau threefold with special Lagrangian boundary conditions.
The free energy admits a resummation organized by contributions of M2 branes wrapping holomorphic curves with boundary on M5 branes wrapping $L$, and weighted by integer-valued LMOV invariants $N_{\lambda,\beta, s}$ \cite{Ooguri:1999bv, Labastida:2000zp, Labastida:2000yw} 
\be
	\sum_{g,h,\beta} g_s^{-2+2g+h} \, F^{\beta, r}_{g,h} \, Q^\beta\, \Tr V^{r} = 
	\sum_{\lambda,\beta,s} N_{\lambda,\beta, s}
	\sum_{d\geq 1} 
	\frac{1}{d} \frac{q^{2ds}Q^{d\beta}}{q^{d}-q^{-d}}\, \Tr_{\lambda} V^d.
\ee
Here $ F^{\beta, r}_{g,h}\in \IQ$ denotes the open topological string amplitude with genus $g$ and $h$ holes, mapping to a curve in relative homology class $(\beta,r)$ and $Q,V$ are K\"ahler moduli of $X$ and Chan-Paton factors for $L$. On the right hand side the sum runs over all representations $\lambda$ of the gauge group of branes on $L$, and $q=e^{g_s}$.

\begin{figure}[h!]
\begin{center}
\includegraphics[width=0.35\textwidth]{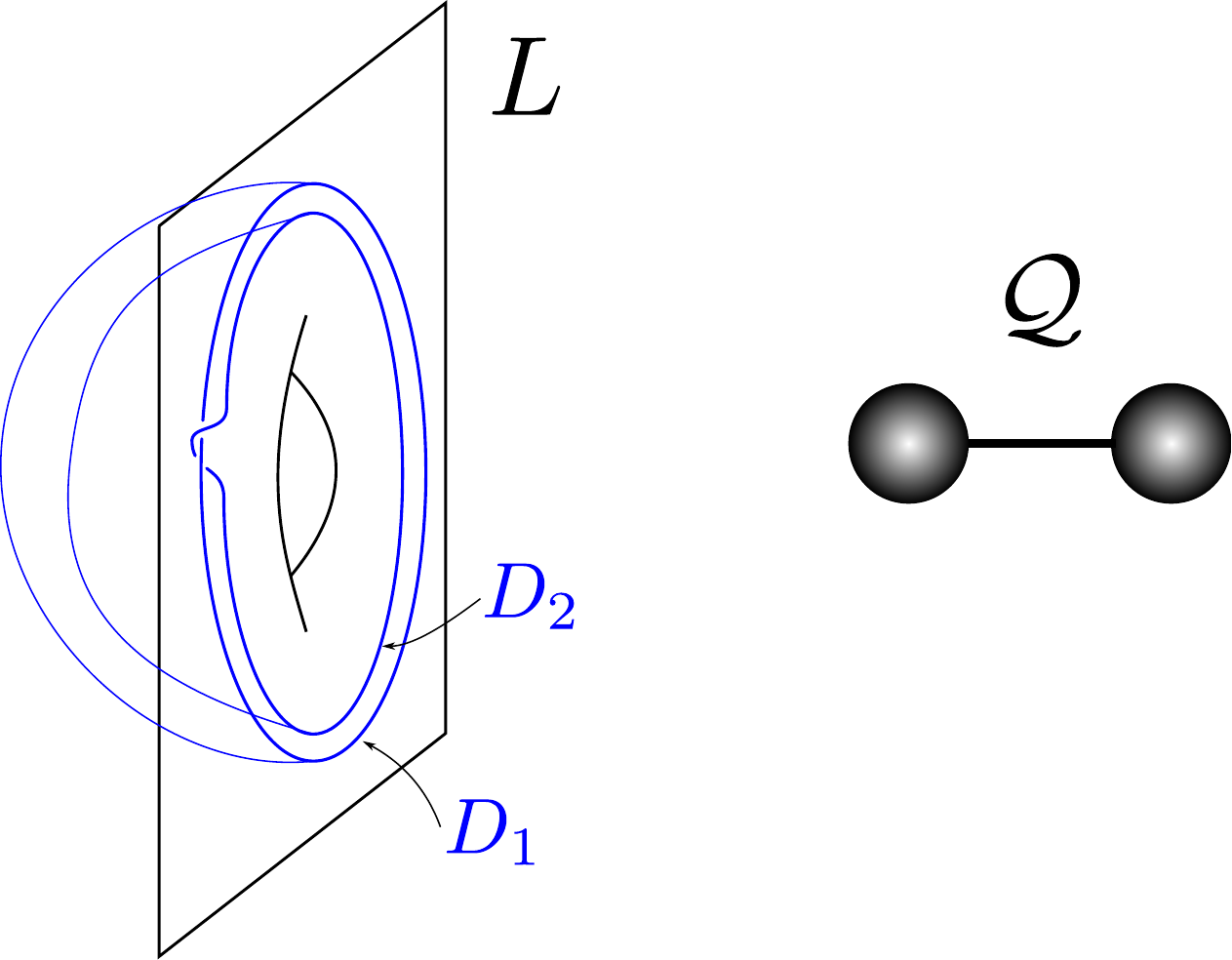}
\caption{Holomorphic disks $D_i$ with linking boundaries, and corresponding quiver.}
\label{fig:quiver-intro}
\end{center}
\end{figure}

A new perspective on LMOV integrality has emerged in recent years, starting with the observation of a novel and stronger underlying structure \cite{Kucharski:2017ogk}. 
For a certain class of Calabi-Yau threefolds and special Lagrangians, the LMOV spectrum is generated by a \emph{finite} set of M2 branes wrapping holomorphic disks $\{D_1,\dots, D_m\}$ \cite{Ekholm:2018eee}. 
Boundaries of M2 branes generate 
Wilson lines $\hat X_i = \exp \int_{\partial D_i} \hat {\mathcal A} $ in the worldvolume QFT on $L$,
whose algebra is determined by linking
\be\label{eq:Cij-intro}
	C_{ij} = \mathrm{lk}(\partial D_i, \partial D_j) \,.
\ee
The algebra of quantum holonomies encoded by $C_{ij}$ generates the LMOV spectrum as follows 
\be
	Z_{\open\ \top} \ =\  :  (\hat X_1;q^2)_\infty^{-1} \dots (\hat X_m;q^2)_\infty^{-1} : \,,
\ee
where $: \underline{\ \ \ } :$ denotes a normal ordering operation which relates the operator and the partition function, see \cite{Ekholm:2019lmb}. The linking matrix of basic disks can be encoded by a quiver with adjacency matrix $C_{ij}$, see Figure~\ref{fig:quiver-intro}.

In this paper we address the long standing question of how the quiver structure underlying $Z_{\open\ \top}$ can be deduced from $L$ itself. 
We propose a constructive algorithm to compute the adjacency matrix and the relative homology classes of basic disks $D_i$ from the augmentation curve $\Sigma$ of the special Lagrangian $L$.
The latter admits a presentation as an algebraic curve
\be
	\Sigma : \  \{(x,y)\in (\IC^*)^2, \ A(x,y,Q)=0\}\,,
\ee
and describes the moduli space of an $A$-brane on $L$ corrected by disk worldsheet instantons \cite{Aganagic:2013jpa}.
Our main result is Conjecture \ref{conj:QuiverSolitonCorrespondence}, which provides an explicit identification between the quiver and certain combinatorial data associated to $\Sigma$ by its exponential network \cite{Eager:2016yxd, Banerjee:2018syt}.

More precisely, we identify basic holomorphic disks $D_i$ with 1-chains on $\Sigma$ calibrated by the holomorphic symplectic Liouville 1-form $\log y\, d\log x$, and the linking number \eqref{eq:Cij-intro} with intersections of the 1-chains, suitably defined.
We summarize this in Table \ref{table:recap}, where $(ii,1)$ soliton paths denotes a distinguished sector of calibrated 1-chains, whose detailed definition can be found in Section \ref{sec:quivers}.
We test our conjecture in several examples, including framed toric branes in $\IC^3$ and the resolved conifold, and framed knot conormal Lagrangians of the trefoil and figure-eight knots. In all cases, we find a perfect match with the quiver description of the open topological string.

\begin{table}[h!]
\caption{}
\begin{center}
\begin{tabular}{|c|c|c|c|}
	\hline
	M-theory & Quiver & Augmentation curve & 3d $\CN=2$ QFT\\
	\hline
	M2 brane on $D_i$ & Vertices & $(ii,1)$ soliton paths $a_i$ & BPS vortices \\
	linking boundaries
	& $C_{ij}$ & intersections $\langle a_i,\ a_j\rangle$ 
	& $\kappa_{ji}^\eff :$ orbital spin
	\\
	\hline  
\end{tabular}
\end{center}
\label{table:recap}
\end{table}%

Our proposal is motivated by the physics of the 3d-3d correspondence \cite{Terashima:2011qi, Dimofte:2011ju, Dimofte:2011py}.
The dual QFT description of an M5 brane on $L\times \IR^2\times S^1$ is a 3d $\CN=2$ Chern-Simons matter QFT on $\IR^2\times S^1$, whose explicit description was given in \cite{Ekholm:2018eee} for all cases under study.
In particular, it was shown that the QFT is a $U(1)^{\times m} $ gauge theory with mixed Chern-Simons couplings $\kappa^\eff_{ij} \equiv C_{ij}$ given by the quiver adjacency matrix.
From the viewpoint of QFT, worldsheet instantons correspond to BPS vortices \cite{Dimofte:2010tz}. 
We argued in \cite{Gupta:2024ics} that the spectrum of BPS vortices is captured by a larger BPS sector computed by exponential networks and known as \emph{kinky vortices} \cite{Banerjee:2018syt}. The $(ii,1)$ sector of kinky vortices encodes, in a suitable corner of the parameter space of couplings, BPS states with unit vorticity which correspond to the $m$ basic disks
\be
	(a_1,\dots, a_m) \ \sim \ (D_1,\dots, D_m)\,.
\ee

Interactions among basic vortices are governed by the mixed Chern-Simons couplings $\kappa_{ij}^\eff$. In particular these determine the \emph{orbital} spin of a two-vortex boundstate, see Appendix \ref{app:interactionspin}. It is well-known that the orbital spin of BPS states in 4d $\CN=2$ gauge theories is captured by the Dirac-Schwinger-Zwanziger pairing, which in turn is captured by the intersection pairing of cycles on the corresponding Seiberg-Witten curve~\cite{Seiberg:1994rs}. 
We argue that the spin of BPS kinky vortices has a similar geometric counterpart\footnote{
In fact, this generalizes the relation between the spin of 2d framed-4d BPS states and spectral networks \cite{Galakhov:2014xba},
since the reduction of 3d $\CN=2$ QFT on $S^1$ to 2d $(2,2)$ QFT maps kinky vortices to BPS soliton kinks.} 
in the intersection pairing of calibrated 1-chains on $\Sigma$
\be
	\text{intersections }\langle a_i, a_j\rangle  \ \sim \   \text{2-vortex boundstate orbital spin}\,.
\ee

Finally, from a mirror symmetry viewpoint, our proposal suggests that holomorphic disk instantons of an $A$-brane on $L$ are dual to calibrated 1-chains on its augmentation variety, and that the linking structure of the basic disks is captured by the intersection pairing of the dual 1-chains. Additional data about holomorphic disks, in particular their relative homology classes, is also captured by the dual 1-chains, details can be found in Section \ref{sec:quivers}.

\paragraph{Organization of the paper}
Section \ref{sec:open-strings} contains an overview of useful background. Our main conjecture and the algorithm to compute quivers from augmentation varieties are presented in Section \ref{sec:quivers}. The conjecture is tested in several examples in Section \ref{sec:examples}.

\paragraph{Acknowledgements}
The work of KG is supported by the Knut and Alice Wallenberg foundation grant KAW 2021.0170, and by the Olle Engkvists Stiftelse Grant 2180108.
The work of PL is supported by the Knut and Alice Wallenberg Foundation grant KAW 2020.0307 and by the Swedish Research Council, VR 2022-06593, Centre of Excellence in Geometry and Physics at Uppsala University.

\section{BPS vortices:
from open strings to exponential networks
}\label{sec:open-strings}

In this section we review connections that emerged recently among topological strings, symmetric quivers, and exponential networks.

\subsection{Enumerative geometry and 3d $\CN=2$ BPS sectors}\label{eq:BPS-sectors}

A broad class of 3d $\CN=2$ QFTs can be engineered by wrapping M5 branes on a 3-manifold $L$~\cite{Terashima:2011qi, Dimofte:2011ju, Dimofte:2011py}.
With a suitable topological twist the resulting theory, henceforth denoted $T[L]$ and considered on $\IR^2\times S^1$, preserves four supercharges (one-quarter of the 6d $(2,0)$ parent QFT), and this allows to establish a relation between BPS sectors of the 3d QFT and certain enumerative invariants of $L$
\be\nonumber
\begin{array}{c|c|c}
	\text{Geometry} & \text{M-theory} & \text{3d $\CN=2$ QFT} \\
	\hline
	(X,L)& \text{M5 branes on $L$} & \text{$T[L]$} \\[4pt]
	\text{Enumerative Invariants} & \text{M2-M5 boundstates} & \text{BPS States} \\
\end{array}
\ee

An example of this correspondence is provided by M-theory on a Calabi-Yau threefold $X$ with $b_4=0$, where $L$ is a Lagrangian submanifold supporting a stack of M5 branes.\footnote{This M-theory engineering of $T[L]$ does not just produce an isolated 3d $\CN=2$ QFT, but also a 5d $\CN=1$ QFT engineered by $X$ \cite{Seiberg:1996bd, Morrison:1996xf, Douglas:1996xp, Intriligator:1997pq}. In fact, generally speaking, $T[L]$ should not be regarded as an isolated theory, but rather as a codimension-two defect for a 3d-5d system. 
We will restrict to Calabi-Yaus $X$ with $b_4=0$, since the absence of compact divisors implies that the 5d dynamics is trivial.
In these cases the 3d QFT can be regarded as an isolated theory, and all information about the 5d theory is encoded by global symmetries of $T[L]$.}
In this case M-theory features a BPS sector described by topological string theory on $X$ \cite{Gopakumar:1998ii, Gopakumar:1998jq}. 
If $S^1$ is viewed as the M-circle, its BPS spectrum features open-string instantons, which are dual to vortices localized at a point in $\IR^2$ and that wrap~$S^1$ \cite{Ooguri:1999bv, Dimofte:2010tz}. 
The correspondence between enumerative invariants and BPS states can be formulated as an isomorphism
between Hilbert spaces of open topological strings states and BPS vortices%
\be\label{eq:open-string-vortex}
	\CH^{(X,L)}_{\open\ \top} \simeq \CH^{T[L]}_{\vortex}\,.
\ee
Both sides carry a grading induced by symmetries of the theory, which is preserved by the isomorphism.

While relation \eqref{eq:open-string-vortex} between BPS vortices and open strings has been studied extensively, there is a less known, but broader class of BPS states that arises when the time direction is instead chosen as $\IR\subset\IR^2$. 
The main novelties arise due to the nontrivial topology of spacelike directions
$S^1\times\IR$.
Since $S^1\times \IR$ has two disconnected boundary components (as opposed to $\partial\IR^2$ which has only one), BPS sectors are graded by a \emph{pair} of vacua instead of one.
Moreover, while the flux of vortices in the plane is quantized, BPS particles on a cylinder are characterized by a \emph{relative} quantization condition.

For the sake of illustration, consider a $U(1)$ gauge field on $S^1\times \IR$, and suppose that its gauge-invariant holonomy at the two ends of the cylinder is constrained to a discrete set of values 
$\phi_i := \oint_{S^1} A $.\footnote{Up to an overall simultaneous shift of  $\phi_k$ by $2\pi i$, due to large gauge transformations.} 
In the absence of particles on the cylinder, the overall flux matches the change in holonomy from one end to the other
\be
	\int_{S^1\times\IR} F = \oint_{\ell_{+}-\ell_{-}} A = \phi_j - \phi_i \,.
\ee
On the other hand, when a particle carrying quantized flux (such as a BPS vortex) is present, the overall flux acquires an additional contribution 
\be\label{eq:kinky-vortex-flux}
	\int_{S^1\times\IR} F = \phi_{j} - \phi_i + 2\pi i \, n\,. \qquad (n\in \IZ)
\ee

\begin{figure}[h!]
\begin{center}
\includegraphics[width=0.4\textwidth]{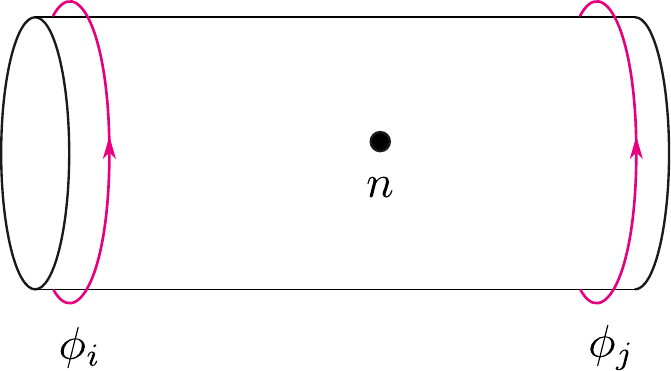}
\caption{}
\label{fig:vortex-on-cylinder}
\end{center}
\end{figure}

In 3d $\CN=2$ QFTs\footnote{For details about such theories, we refer to \cite{Intriligator:2013lca}.} on $S^1\times \IR^2$ the allowed values of $\phi_i$ are usually determined by a (twisted) effective superpotential\footnote{A holomorphic function of a complex-valued scalar $\sigma$, whose imaginary part is the gauge holonomy, and whose real part is a vectormultplet scalar.}
\be\label{eq:phi-i-crit-W}
	\text{crit} \left(\tCW_{\eff}\right) \simeq \{\phi_i\}_{i}\,.
\ee
Correspondingly, the spectrum of BPS vortex configurations is broadened to include not only vortices in a single vacuum $(i=j)$, but also vortices that interpolate between two distinct vacua $(i\neq j)$. 
This type of BPS sector arises when $T[L]$ is considered on a space-like cylinder, and is known as \emph{kinky vortices} \cite{Banerjee:2018syt}, since the corresponding particles resemble vortices near their core, 
but can interpolate between different vacua like a kink, see Figure 
\ref{fig:vortex-on-cylinder}.
The Hilbert space of kinky vortices is graded by $(ij,n)$ where $1\leq i,j\leq K$ label the vacua \eqref{eq:phi-i-crit-W} at the two ends of the cylinder, and $n$ labels the quantized flux of the particle as in \eqref{eq:kinky-vortex-flux}
\be
	\CH^{T[L]}_{\kinkyvortex}(z) \simeq \bigoplus_{i,j=1}^{K} \bigoplus_{n\in \IZ} \CH^{T[L]}_{(ij,n)}(z)\,.
\ee

When the circle shrinks, kinky vortices can either become infinitely heavy and leave the spectrum, or can descend to kinks of the resulting 2d $(2,2)$ QFT \cite{Cecotti:1992rm}. 
Like their 2d counterparts, the BPS spectrum of kinky vortices is characterized by wall-crossing phenomena. 
The spectrum varies piecewise continuously over the moduli space of couplings $z\in \CM$ of $T[L]$, and
jumps across real-codimension one loci 
\be\label{eq:H-wall-cross}
	\CH^{T[L]}_{\kinkyvortex}(z) \simeq \Psi\left[ \CH^{T[L]}_{\kinkyvortex}(z') \right]\,.
\ee
Here $\Psi$ is a generalization of Cecotti-Vafa wall-crossing in 2d $(2,2)$ QFT \cite{Cecotti:1992rm} to 3d $\CN=2$ QFT, whose decategorification was computed in \cite{Banerjee:2018syt}.\footnote{The computation is based on a connection between exponential networks \cite{Eager:2016yxd} and higher-dimensional $tt^*$ geometry \cite{Cecotti:1991me,Cecotti:2013mba}. See also \cite{Klemm:1996bj, Gaiotto:2011tf, Gaiotto:2012rg} for previous foundational work.
Note that this is different, but related, to jumps in the spectrum of BPS vortices which is controlled by Stokes phenomena in the space of Fayet-Iliopoulos couplings \cite{Garoufalidis:2020pax, Grassi:2022zuk, Alim:2022oll}.}

The BPS sectors of standard and kinky vortices are not entirely distinct, however. 
In \cite{Gupta:2024ics} we proposed that standard vortices arise as a sub-sector of kinky vortices in a specific region of their moduli space $\CM$.
To be more precise, let $T[L]$ be a $U(1)$ Chern-Simons matter theory with a Fayet-Iliopoulos coupling $\zeta$, and let 
$z = \zeta R$
be the dimensionless combination obtained by introducing the radius of $S^1$.
In \cite{Gupta:2024ics} we argued that a subsector of $(ii,n)$ kinky vortices, for a fixed but distinguished choice of $i$, agrees in the limit of large FI coupling with the spectrum of BPS vortices
\be\label{eq:ii-n-sector}
	\lim_{z\to\infty} \bigoplus_{n\in \IZ}\CH^{T[L]}_{(ii,n)}(z)\  \simeq\  \CH^{T[L]}_{\vortex} \simeq \CH^{(X,L)}_{\open\ \top}\,.
\ee
Note that taking the limit is important, since $\CH^{T[L]}_{(ii,n)}(z)$ changes due to wall-crossing \eqref{eq:H-wall-cross}.
It then follows from \eqref{eq:open-string-vortex} that the open topological string worldsheet instantons on $(X,L)$
are captured by a subsector of BPS kinky vortices.

\subsection{Integral structures and underlying quivers}

A well-known property of the open topological string partition function, is that it 
exhibits integrality properties predicted by its embedding into M-theory \cite{Ooguri:1999bv, Labastida:2000yw, Labastida:2000zp}.
These extend to an integral structure on the space of BPS vortices of $T[L]$ through the
3d-3d correspondence \eqref{eq:open-string-vortex}.
We shall restrict to considering a single brane on $L$ from now on.
The integral structure then arises from a factorization of the open string/vortex partition function of the following form
\be\label{eq:LMOV-factorization}
	Z^{T[L]}_{\vortex}(x,u,q) = \prod_{r,\beta,s} \, \exp\left[ N_{r,\beta,s} \sum_{d\geq 1} \frac{1}{d} \frac{(q^{s}u^{\beta}x^r)^d}{q^{d}-q^{-d}}\right]\,,
\ee
where $(r,\beta,s)\in \IN^{b_1(L)}\times \IZ^{b_2(X)}\times \IZ$ are the vorticity, the flavor charge, and the spin of BPS states.\footnote{The flavor symmetry group is $G_F\simeq U(1)^{b_2(X)}$ for CY3 with $b_4=0$, which is where our discussion takes place.} The numbers $N_{r,\beta,s}\in \IZ$ are known as LMOV invariants, and reorganize open strings, or vortices, into `counts' of M2 branes ending on the M5 brane wrapping $L$.
In particular $\beta$ and $r$ represent homology classes of a holomorphic curve $\upsilon$ and of its boundary
\be\label{eq:disk-charges}
	r = [\partial \upsilon] \in H_1(L)\,,
	\qquad
	\beta = [\upsilon] \in H_2^{\rel}(X,L) \,.
\ee

\begin{figure}[h!]
\begin{center}
\includegraphics[width=0.8\textwidth]{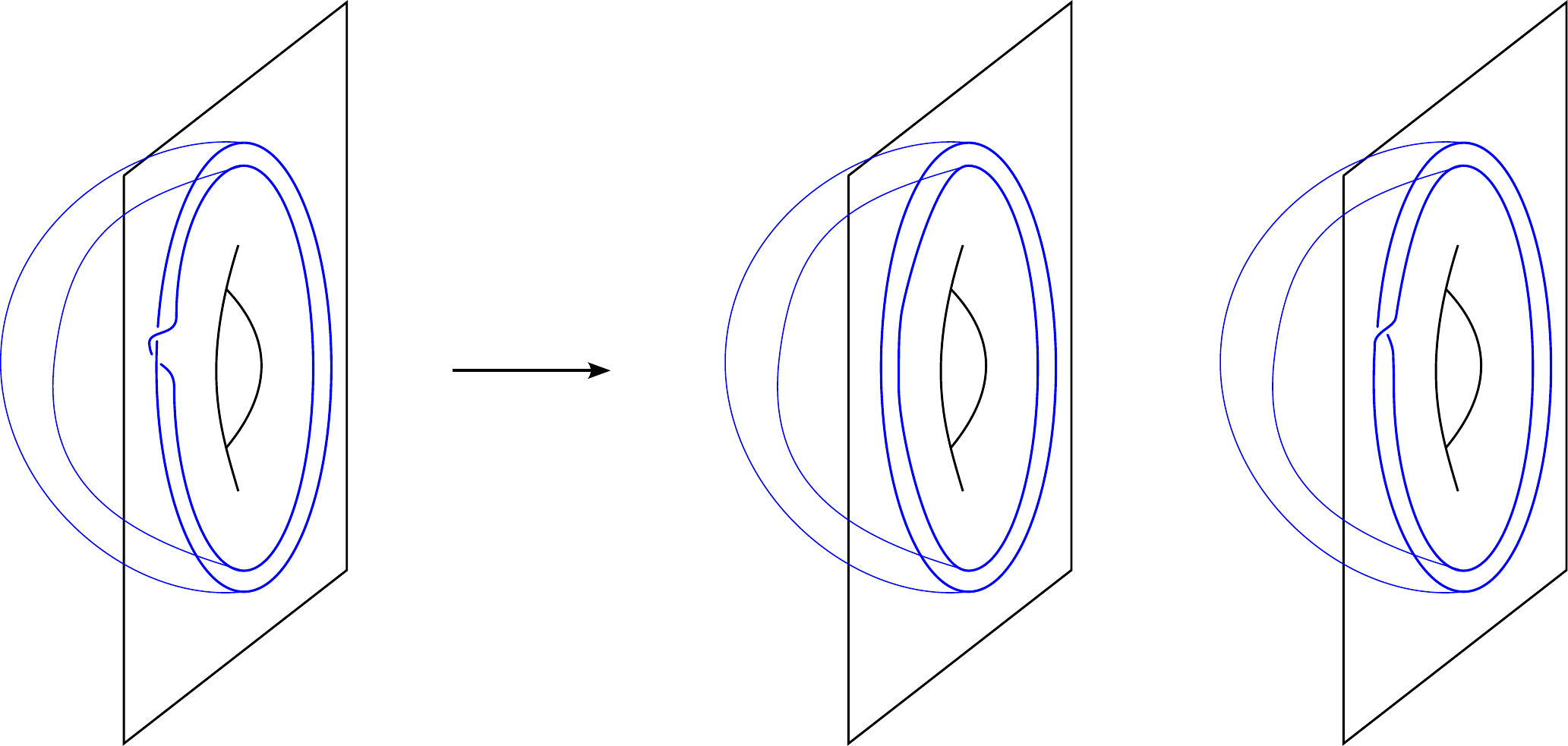}
\caption{The multi-cover skein relation.}
\label{fig:multi-cover-skein}
\end{center}
\end{figure}

Evidence for an even stronger structure on the space of BPS vortices has emerged more recently \cite{Kucharski:2017ogk}.
Based on this, it was argued in \cite{Ekholm:2018eee,Ekholm:2019lmb} that the entire spectrum of LMOV invariants is in fact generated by a \emph{finite} set of M2 branes wrapping holomorphic disks 
\be\label{eq:LMOV-generation}
	\{D_1,\dots, D_m\} 
	\quad\mathop{\rightsquigarrow}^{L}\quad
	N_{r,\beta,s}\,.
\ee
The $m$ basic disks correspond to BPS states with unit winding 
\be
	|N_{1,\beta_i,C_{ii}}| = 1\,, 
\ee
where $r\in \IZ$ should be understood to hold under the restriction $b_1(L)=1$, and the meaning of $C_{ii}\in \IZ$ will be clarified shortly. 
By equation \eqref{eq:LMOV-generation}, we mean that the LMOV spectrum is generated through interactions among the basic disks, which are mediated by the M5 brane on $L$.
In fact, boundaries $\partial D_i$ source Wilson lines in the worldvolume Chern-Simons theory on $L$ (the 3d-3d dual theory to $T[L]$). This algebra gives rise to \emph{multi-cover skein relations} among basic disks, which 
generate the entire LMOV spectrum of states with winding $r\geq 1$ \cite{Ekholm:2019lmb}, see Figure \ref{fig:multi-cover-skein}.\footnote{This is related to the creation and decay of boundstates of M2 branes mediated by M5 branes in wall-crossing theory  \cite{Gaiotto:2009hg}.}

The reorganization of LMOV invariants to a finite set of generators \eqref{eq:LMOV-generation} is known to hold for specific types of Lagrangians and Calabi-Yau threefolds studied in earlier literature \cite{Aganagic:2000gs, Aganagic:2001nx, Iqbal:2004ne, Ooguri:1999bv}, 
such as for toric branes in Calabi Yau threefolds with $b_4=0$ \cite{Panfil:2018faz},
for certain knots and link conormal Lagrangian in the resolved conifold \cite{Kucharski:2017ogk}, 
and for certain knot complements viewed as zero-sections of $T^*S^3$ \cite{Ekholm:2021irc}.\footnote{For certain knots and links it is however known that generalizations of the quiver need to be considered \cite{Ekholm:2021gyu}.}
In all these cases the open string partition function coincides with
a specialization of the
the motivic Donaldson-Thomas partition function of a quiver $\CQ$ \cite{PhysRevD.96.121902}
\be\label{eq:KQ-corr}
	Z^{(X,L)}_{\open\ \top}(x,u,q) 
	=
	Z^{T[L]}_{\vortex}(x,u,q) 
	=
	\left[P^\CQ(x_1,\dots, x_m,q)\right]_{x_i = 
	u^{\beta_i} q^{q_i}x}\,,
\ee
which takes the form \cite{2011arXiv1103.2736E}
\be\label{eq:PQ-fn}
	P^\CQ(x_1,\dots, x_m,q)
	=
	\sum_{d_1\dots, d_m\geq 0} q^{\sum_{i,j=1}^{m}C_{ij} d_i d_j} \prod_{i=1}^{m} \frac{x_i^{d_i}}{(q^2;q^2)_{d_i}}\,.
\ee
A physical interpretation of the quiver was proposed in \cite{Ekholm:2018eee}, 
by identifying vertices with the generating set of M2 branes wrapping basic disks $\{D_i\}$, and the adjacency matrix with the linking matrix
\be\label{eq:basic-disks-linking}
	\{D_i\}  \simeq \text{vertices of }\CQ\,,\qquad {\rm{lk}}(\partial D_i,\partial D_j)=C_{ij}\,.
\ee
The specialization of the formal variables $x_i$ in \eqref{eq:KQ-corr} was moreover understood to encode topological charges of basic disks \eqref{eq:disk-charges}, and the self-linking of their boundaries.

\subsection{$\CQ$-deformations of $T[L]$}\label{sec:Q-deform}

\begin{figure}[h!]
\begin{center}
\includegraphics[width=0.5\textwidth]{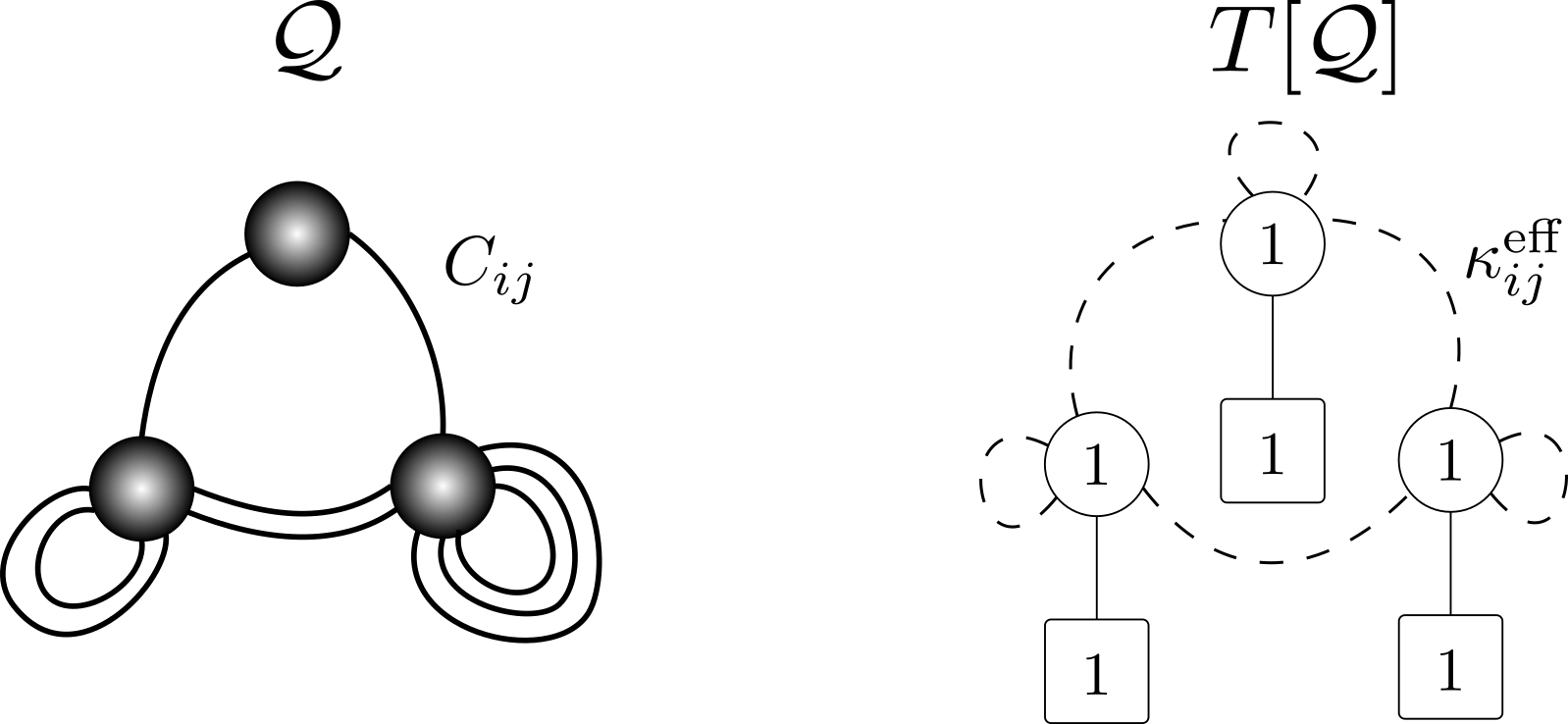}
\caption{The theory $T[\CQ]$ corresponding to a quiver.}
\label{fig:TQ-QFT}
\end{center}
\end{figure}

In addition to providing a structural reorganization of LMOV invariants, quivers also provide a Lagrangian description for theories of class $T[L]$.
Indeed quiver partition functions like \eqref{eq:PQ-fn} coincide with vortex partition functions of a certain class of Abelian Chern-Simons matter QFTs \cite{Ekholm:2018eee}.
If $\CQ$ is a symmetric quiver with $m$ vertices, the corresponding theory $T[\CQ]$ consists of $m$ copies of a chiral multiplet with unit charge under a $U(1)$ gauge field.
These $m$ sectors interact through mixed Chern-Simons couplings whose (effective) levels are encoded by the adjacency matrix $C_{ij}$, see Figure \ref{fig:TQ-QFT}.\footnote{Unless otherwise stated, when drawing $\CQ$, $k$ edges connecting vertices $i$ and $j$ corresponds to $C_{ij}=k >0$.}

Each sector carries a global `topological' symmetry $U(1)_{\top}$, with associated flavour fugacities denoted by $(x_1,\dots, x_m)$ corresponding to exponentiated Fayet-Ilioupoulos couplings.
A BPS vortex carries a quantized magnetic flux (i.e. $U(1)^{m}_{\top}$ charge) encoded by non-negative integers $(d_1,\dots, d_m)\in \IN^{m}$.
The corresponding grading of $\CH_{\vortex}$ is reflected by the quiver partition function $P^{\CQ}$, see \eqref{eq:PQ-fn}, which indeed coincides with the $\IR^2\times S^1$ index of the theory
\be
\begin{split}
Z^{T[\CQ]}_{\vortex}(x_1,\dots, x_m,q) 
& = P^\CQ(x_1,\dots, x_m,q) \,.
\end{split}
\ee
The correspondence \eqref{eq:KQ-corr} can be understood as an equivalence between $T[\CQ]$ and $T[L]$, when $U(1)_{\top}^m$ is broken to a $U(1)$ subgroup, e.g. by dynamical effects such as a superpotential which enforces a relation among flavour fugacities
\be
	T[L]\simeq T[\CQ]_{x_i=u^{\beta_i} q^{q_i}x} \,.
\ee

The broader class of theories $T[\CQ]$ has interesting features in its own right,
even without imposing relations among $x_i$. 
In particular, multi-cover skein relations such as the one in Figure {fig:multi-cover-skein} have a dual QFT intepretation in terms of dualities \cite{Ekholm:2019lmb}.
In terms of quivers, the example considered above corresponds to an equivalence between quivers $\CQ, \CQ'$ with adjacency matrices
\be
	C= 
	\begin{pmatrix}
           0 & 1 \\
           1 & 0
        \end{pmatrix}
        \,,
        \qquad
        C= 
	\begin{pmatrix}
           0 & 0 & 0 \\
           0 & 0 & 0 \\
           0 & 0 & 1
        \end{pmatrix}
        \,.
\ee
From the viewpoint of 3d $\CN=2$ QFT this expresses a duality between $U(1)^2$ and $U(1)^3$ QFTs with effective mixed Chern-Simons couplings
\be\label{eq:pentagon-TQ}
	T[\raisebox{0pt}{\includegraphics[width=0.04\textwidth]{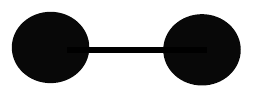}}] \simeq T[\raisebox{-3pt}{\includegraphics[width=0.05\textwidth]{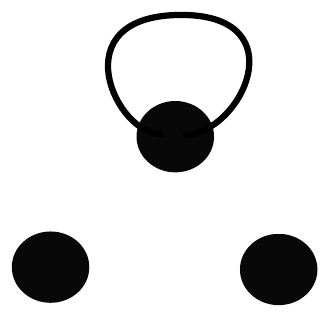}}]\,.
\ee 
In particular, vortex partition functions coincide upon a suitable identification of fugacities
\be
	P^{(\raisebox{0pt}{\includegraphics[width=0.02\textwidth]{two-node-Q}})}(x_1, x_2,q) = P^{(\raisebox{-3pt}{\includegraphics[width=0.025\textwidth]{three-node-Q}})}(x_1, x_2, q^{-1}x_1 x_2,q) \,.
\ee
This relation can be viewed from several angles
\begin{itemize}
\item
Mathematically this can be understood as a consequence of the pentagon relation for quantum dilogarithms \cite{Faddeev:1993rs}, by promoting $P^{\CQ}$ to an operator valued in a quantum torus algebra \cite{Ekholm:2019lmb}.
\item
The M-theoretic origin of \eqref{eq:pentagon-TQ} lies in the multi-cover skein relation involving M2 boundaries on M5 branes, from Figure \ref{fig:multi-cover-skein}. 
\item
From a QFT perspective, it can be understood as a consequence of known 3d $\CN=2$ dualities, such as combinations of particle-vortex duality and SQED/XYZ duality.
\end{itemize}
Multi-cover skein dualities vastly generalize \eqref{eq:pentagon-TQ} to a web of relations among quivers with different number of vertices and linking structure, corresponding to QFTs with different gauge groups. See \cite{Ekholm:2019lmb} for an exhaustive description.

\begin{figure}
\begin{center}
\includegraphics[width=0.4\textwidth]{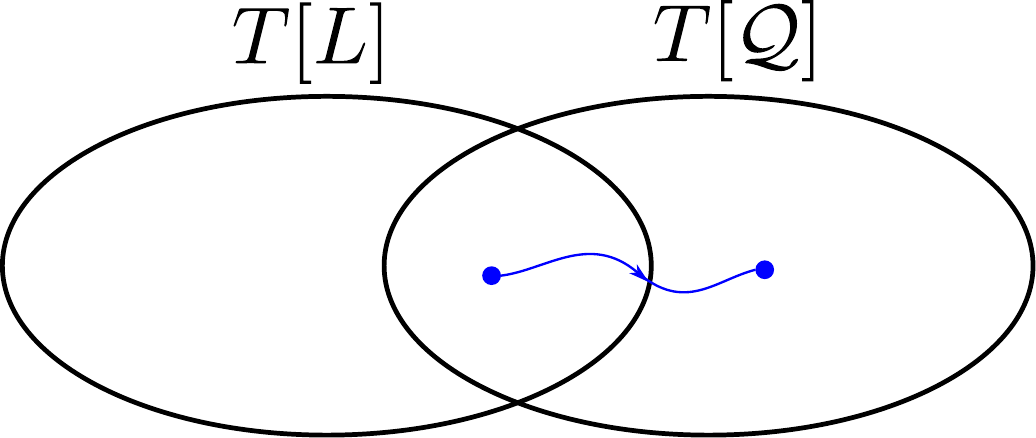}
\caption{A $\CQ$-deformation of $T[L]$.}
\label{fig:TL-TQ-relation}
\end{center}
\end{figure}

The main role of $T[\CQ]$ theories in the present work will be to provide a certain deformation of $T[L]$. 
Our main interest will be the overlap region in Figure \ref{fig:TL-TQ-relation}, namely certain theories $T[L]$ engineered by M5 branes on Lagrangian subamanifolds which admit a dual quiver description. 
However, in some cases it will be useful to consider a continuous deformation of $T[L]$ into the complement $T[\CQ]\setminus (T[\CQ]\cap T[L])$.
Suppose $T[L]$ admits a dual Lagrangian description $T[\CQ]$, and let us choose this (among the many possible ones related by multiple-cover skein dualities) so that $x_i \sim x$ are linear in $x$.\footnote{The choice may or may not be unique, see \cite{Ekholm:2018eee,Ekholm:2019lmb}.}
We define the $\CQ$-deformation of $T[L]$ as the theory $T[\CQ]$ with \emph{generic} flavour fugacities $x_i$, which is defined by the following data
\be\label{eq:TQ-data}
\begin{split}
	\text{Data of $T[\CQ]$:}
\end{split}
\ \ \left\{
\begin{split}
	&\text{$m$ vertices of $\CQ$} \\
	&\text{adjacency matrix }C_{ij} \in \IZ^{m^2}\\
	&\text{fugacities }c_k \in \IC^*\\
	&q_k \in \IZ\\
\end{split}
\right.
\ee

\subsection{Augmentation curves and exponential networks}

\begin{figure}[h!]
\begin{center}
	\includegraphics[width=0.45\textwidth]{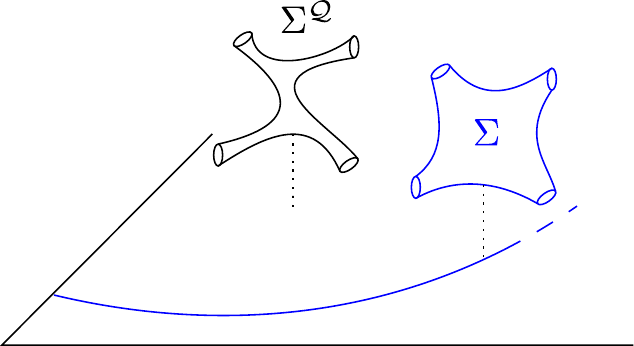}
\caption{The family of $\CQ$-deformations of the augmentation variety.}
\label{fig:Sigma-Q}
\end{center}
\end{figure}

The open string free energy admits a genus expansion in powers of $g_s = \log q^2$, whose leading term at $g_s\to 0$ is given by disk instantons
\be
	\log Z^{(X,L)}_{\open\ \top} = \frac{1}{g_s} W_{\disk}(x, u) + \dots
\ee
The open string modulus $x$ corresponds to a flavour fugacity for the QFT $T[L]$ akin to an FI coupling through the correspondence \eqref{eq:KQ-corr}. 
In fact, the disk potential is Legendre-dual to $\tCW_{\eff}$, and its critical points determine the genus-zero corrected moduli space of an $A$-brane on $L$, 
which is 
the \emph{augmentation curve} of the brane \cite{Aganagic:2013jpa}
\be
	\Sigma:=\left\{ (x,y)\in \IC^*\times \IC^* ; A(x,y,u)=0\right\}\,.
\ee
When $T[L]$ admits a $\CQ$-deformation $T[\CQ]$ one may consider the critical points of
\be
	\log P^{\CQ} = \frac{1}{g_s} W^{\CQ}(x_1,\dots, x_m) + \dots
\ee
defined by 
\be\label{eq:A-i-poly}
	y_j = \exp\left(\frac{\partial W^{\CQ }}{\log x_j}\right) = 1 - x_j (-y_j)^{C_{jj}}  \prod_{j \neq k} y_k^{C_{jk}}\,,
\ee
which gives an $m$-dimensional variety $Y_m$ (the twisted chiral ring of $T[\CQ]$).
We define the $\CQ$-deformed augmentation variety $\Sigma^\CQ$ to be a 1-dimensional subvariety $\Sigma^\CQ\subset Y_m$ obtained by setting $x_i = c_i x$ and $y = \prod_{i=1}^m y_i$, see \cite{Ekholm:2018eee}
\be\label{eq:Q-deformed-curve}
	\Sigma^{\CQ}:=\left\{ (x,y)\in \IC^*\times \IC^* ; A^{\CQ}(x,y,c_1,\dots c_m)=0\right\}\,.
\ee
The moduli space of $\Sigma^{\CQ}$ is $m$-complex dimensional, and is locally parameterized by $\{c_i\}$.
The original augmentation variety is recovered by restricting to the locus determined by the change of variables in \eqref{eq:KQ-corr} (specialized to $q=1$) as shown in Figure \ref{fig:Sigma-Q}
\be
	\lim_{c_i \to  u^{\beta_i}} A^{\CQ}(x,y,c_1,\dots, c_m) = A(x,y,u) \,.
\ee

We will view $\Sigma$ as a finite-degree ramified covering of $\IC^*$ with base coordinate $x$.
A choice of $(x,Q)$ fixes a theory $T[L]$, and sheets $y_i(x)$ correspond to a discrete set of massive vacua for the theory.\footnote{The discussion generalizes in a straightforward way to $\Sigma^{\CQ}$ and the theory $T[\CQ]$.}
For a fixed choice of moduli, the kinky vortices discussed in Section \ref{eq:BPS-sectors} arise as soliton field configurations interpolating between vacua $y_i(x)$ and $y_j(x)$. 
A kinky vortex has BPS central charge given by a 1-chain integral
\be\label{eq:Z-BPS}
	Z_a = \frac{1}{2\pi R} \int_a \log y\, d\log x\,.
\ee
where $R$ is the $S^1$ radius, and $a$ is a path connecting the two vacua, up to relative homology.
More precisely, the integrand in \eqref{eq:Z-BPS} is single-valued on a $\IZ$-covering of $\Sigma$ denoted $\tSigma$, obtained by passing to the universal covering for $\log y$.
Sheets of $\tSigma$ are labeled by $(i,N) \in \{1,\dots,K\} \times \IZ$ indexing critical points of $W_{\disk}$ as $\log y_i +2 \pi i N$. 
The BPS spectrum of kinky vortices between vacua $(i,N)$ and $(j,M)$ only depends on $M-N\equiv n$, and we 
characterize the charge of a soliton by four labels \cite{Gupta:2024ics}
\be
	a \sim (ij, n,k,\beta)\,,
\ee
where $k$ is the Kaluza-Klein momentum along $S^1$, and $\beta$ encodes possible contributions from additional flavour symmetries (twisted masses) in $T[L]$.
The spectrum is defined by the CFIV index \cite{Cecotti:1992qh}, which assigns a number to each soliton charge
\be\label{eq:CFIV}
	\mu_a \in \IQ\,.
\ee

\begin{figure}[h!]
\begin{center}
\includegraphics[width=0.4\textwidth]{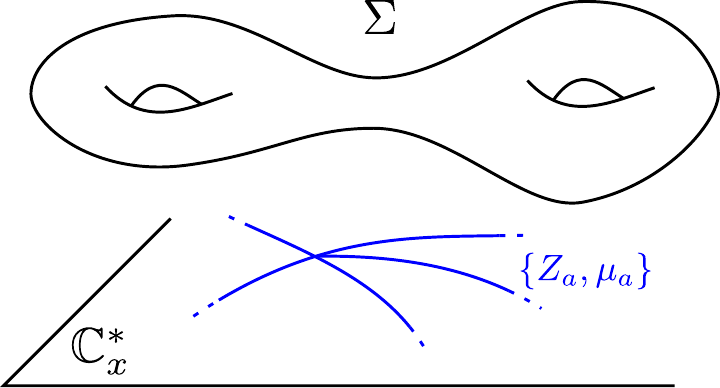}
\caption{A schematic picture of exponential networks.}
\label{fig:networks-cartoon}
\end{center}
\end{figure}

The geometry of the augmentation curve $A(x,y,Q)$ encodes the entire BPS spectrum of kinky vortices, i.e. the set of all central charges \eqref{eq:Z-BPS} and CFIV indices \eqref{eq:CFIV} for a given theory labeled by moduli $(x,Q)$
The computation of the spectrum from geometric data is achieved through exponential networks \cite{Eager:2016yxd}, in particular by means of the nonabelianization map established in \cite{Banerjee:2018syt}
\be
	A(x,y,u) \ \ \rightsquigarrow \ \ \{Z_a,\mu_a\} \,.
\ee
In a nutshell, the exponential network is a collection of trajectories on $\IC^*_x$ that originates at branch points of the covering induced by $\Sigma$, singularities of $\log y$, or at intersections of trajectories.
A schematic example is provided in Figure \ref{fig:networks-cartoon}, and much more detailed ones will be studied in later sections. Part of the data associated to each trajectory includes a collection of soliton paths $a$ and their CFIV index $\mu_a$. In particular, each trajectory can only carry solitons of a fixed type $(ij,n)$.
Exponential networks encode the spectrum of kinky vortices of $T[L]$ for all values of $(x,Q)$, including wall-crossing phenomena \eqref{eq:H-wall-cross}.

A specific feature of the augmentation curves that we will be studying is that they have a branch of solutions to $A(x,y,Q)=0$ with the property that $\lim_{x\to 0} y_i(x)=1$.\footnote{This follows from the fact that we restrict to theories that admit a quiver description, but may also be engineered more generally by a suitable reparametrization.}
The sectors of $(ii,n)$ kinky vortices are of particular interest, see \eqref{eq:ii-n-sector}, and we will often refer to this branch as the distinguished branch.
In particular, we showed in \cite{Gupta:2024ics} that the spectrum of $(ii,n)$ kinky vortices reproduces that of standard BPS vortices in the limit $x\to 0$, which can equivalently be referred to as the large radius limit. The corresponding charges $a\sim (ii,n,k,\beta)$ correspond to vortices with vorticity $n$, KK momentum $k$ along $S^1$, and flavour charge $\beta$.

\section{Quivers from augmentation varieties}\label{sec:quivers}

In this section we study the following question:
\be\nonumber
	\boxed{
	\begin{split}
	&\text{If $(X,L)$ admits a quiver description, can $\CQ$ can be recovered }\\
	& \text{from the geometry of the augmentation variety?}
	\end{split}
	}
\ee
We will approach this first by formulating the general idea, then formalizing it as a conjecture, and finally providing a constructive definition of the quiver $\CQ$ based on data of the augmentation polynomial $A(x,y,u)$.

\subsection{Heuristics behind our proposal}\label{sec:heuristics}

\begin{figure}[h!]
\begin{center}
\includegraphics[width=0.7\textwidth]{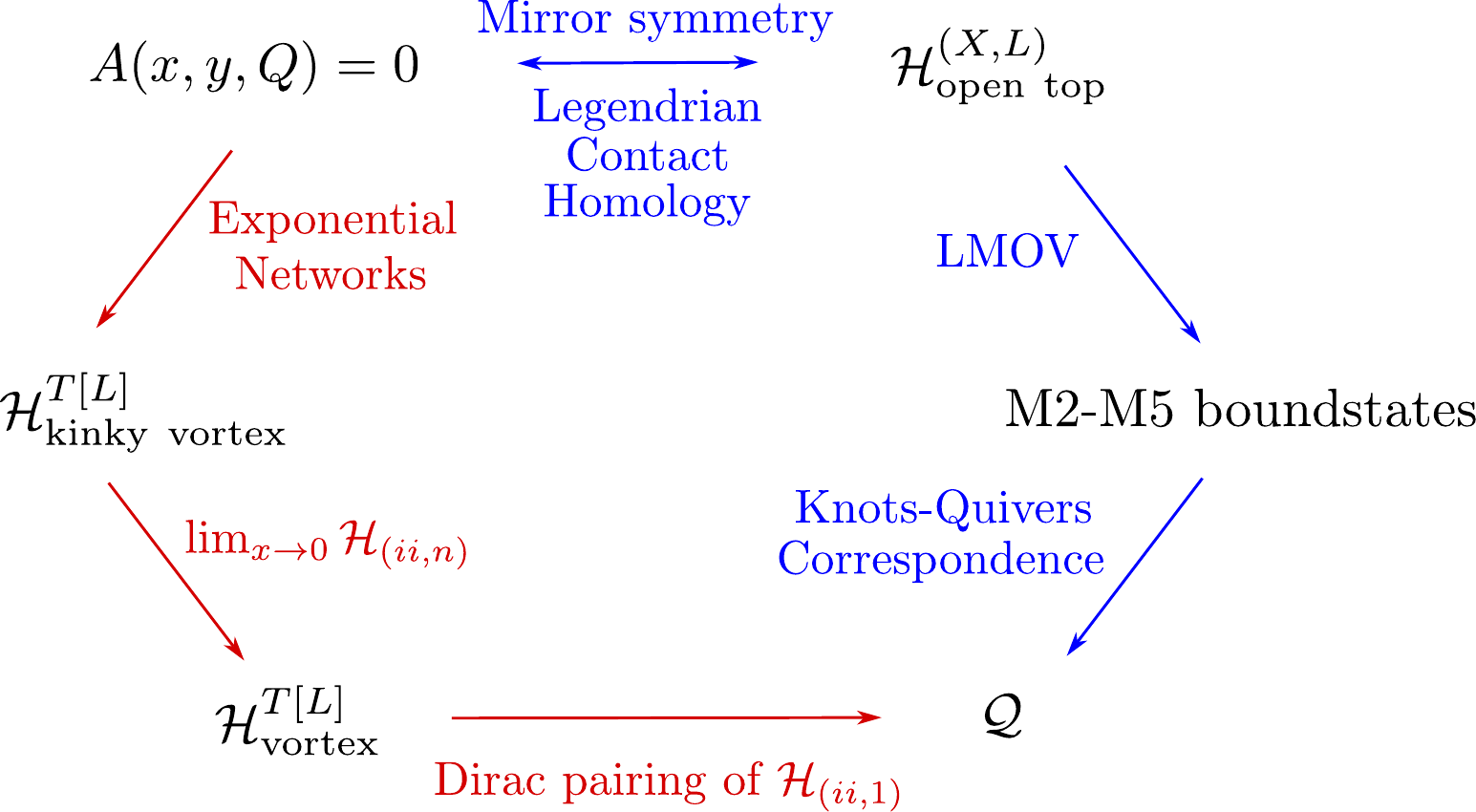}
\caption{Relations between augmentation varieties and quivers. In blue: the standard route \cite{Aganagic:2013jpa,Ooguri:1999bv,Labastida:2000yw,Labastida:2000zp,Kucharski:2017ogk}. In red, our proposal.}
\label{fig:logic-flow}
\end{center}
\end{figure}

BPS vortices of 3d $\CN=2$ QFTs have emerged in our discussion so far as the common thread connecting various topics related to the problem addressed in this paper: from open topological strings, to symmetric quivers, to exponential networks.
Following through the various connections, summarized in Figure \ref{fig:logic-flow}, naturally leads to the idea that, if topological strings on $(X,L)$ admit a quiver description, then $\CQ$ must be encoded by the exponential network.

More specifically, when the open topological string partition function admits a quiver description as in \eqref{eq:KQ-corr}, the whole BPS spectrum of M2 branes captured by LMOV invariants is generated by a finite set of basic BPS states.
The basic states correspond to M2 branes wrapping holomorphic disks $D_i$ in $X$ with boundary on $L$, and their interactions are mediated by the $A$-brane in a way that is controlled by the linking of their boundaries \eqref{eq:basic-disks-linking}. Together this data defines a quiver $\CQ$ with adjacency matrix 
\be
	C_{ij} = {\rm lk}(\partial D_i, \partial D_j)\,.
\ee

The 3d-3d correspondence provides a dual QFT description involving a 3d $\CN=2$ QFT $T[L]$, for which $\CQ$ encodes a Lagrangian description $T[\CQ]$ as a $U(1)^m$ Chern-Simons matter gauge theory.
The basic M2 branes map to BPS vortices with unit charge, and the linking matrix maps to effective Chern-Simons couplings, see Figure \ref{fig:TQ-QFT}.
Moreover Chern-Simons couplings govern the dynamics of BPS vortices: in particular the orbital spin of a 2-vortex boundstate is proportional to $\kappa_{ij}^\eff$, see Appendix \ref{app:interactionspin} for a derivation
\be
	C_{ij}  = \kappa^{\eff}_{ij} \quad \leftrightarrow\quad \text{orbital spin of 2-vortex boundstates}\,.
\ee

The above chain of logic leads to the conclusion that, if one can compute all BPS vortices with unit charge of $T[L]$, together with their spin and the spin of their boundstates, this information should encode the quiver~$\CQ$.
We propose that all this information can be extracted directly from the geometry of the augmentation variety $\Sigma$, which arises as the moduli space of the $A$-brane on $L$ in the framework of Legendrian Contact Homology \cite{Aganagic:2013jpa}, or equivalently as the moduli space of vacua of the 3d-3d dual theory $T[L]$.

More specifically, exponential networks encode the BPS spectrum of kinky vortices, whose $(ii,n)$-sector corresponds to standard BPS vortices with vorticity $n$ and zero Kaluza-Klein momentum.\footnote{Not only the zero modes, but the whole Kaluza-Klein tower is captured by exponential networks, see \cite{Gupta:2024ics}.}
Moreover, for a specific choice of vacuum $i$, in the limit $x\to 0$ the spectrum of $(ii,n)$ vortices agrees with the prediction from open topological string theory through \eqref{eq:ii-n-sector}. This settles the question of computing the BPS vortices with unit charge. Next we discuss how their spin can be computed.

The spin of BPS kinks in 2d $(2,2)$ QFTs is encoded by spectral networks \cite{Galakhov:2014xba}, as the \emph{writhe} (i.e. the algebraic self-intersection number) of the path $a$ encoding the soliton charge
\be
	\text{spin of BPS vortices}  = \wr(a) \,.
\ee
This definition has a natural lift to exponential networks, viewing them as spectral networks for the corresponding KK theory \cite{Banerjee:2018syt}.

\begin{figure}[h!]
\begin{center}
\includegraphics[width=0.8\textwidth]{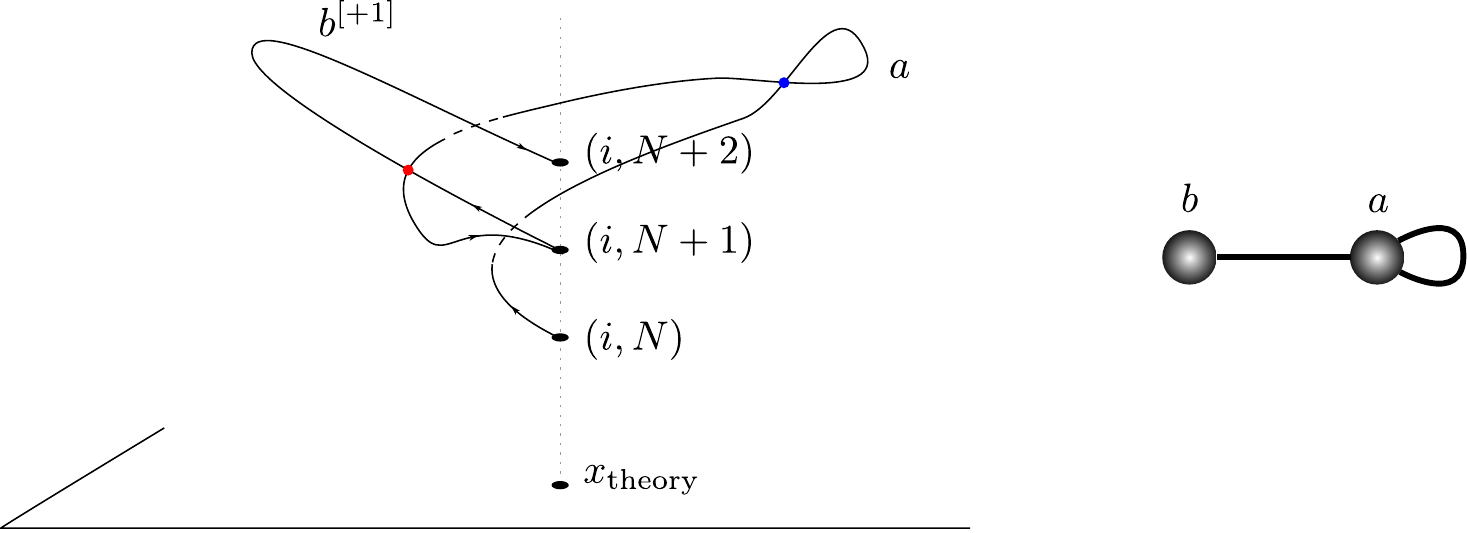}
\caption{Left: Two solitons $a,b$ of type $(ii,1)$. To compute mutual intersections (red dot) we consider the concatenation obtained by shifting $b$ by one unit in the logarithmic label of its endpoints, as explained in Sections \ref{sec:vertices-fugacities}-\ref{sec:adjacency-matrix}. The self-intersection of $a$ denoted by a blue dot does not contribute to the quiver matrix, since it appears both in $\wr(a)$ and in $\wr(ab)$ in \eqref{eq:int-pair}. 
Right: The corresponding quiver diagram encodes the mutual intersection between $\langle a,b^{[+1]}\rangle=+1$ as well as a self-intersection $\langle a,a^{[+1]}\rangle=+1$ computed by concatenation of $a$ with its own shift.}
\label{fig:conjecture-cartoon}
\end{center}
\end{figure}

Observe in fact that the writhe of two concatenated open paths computes their mutual intersection
\be\label{eq:int-pair}
	\langle a,b\rangle = \wr(ab) - \wr(a) - \wr(b)\,.
\ee
Therefore the intersection between (concatenatable) paths $a,b$ measures the difference between the spin of their boundstate $ab$ and the spin of the constituents, namely the (appropriately quantized) \emph{orbital} angular momentum of the 2-vortex system
\be
	\langle a,b\rangle  = \text{orbital spin}\,.
\ee
This parallels the well-known statement that the Dirac pairing, which measures the orbital angular momentum of dyons in 4d, is captured by intersections of paths on Seiberg-Witten curves \cite{Seiberg:1994rs}.
This leads us to the following conjecture, illustrated in Figure \ref{fig:conjecture-cartoon}

\begin{conjecture}\label{conj:QuiverSolitonCorrespondence}
Let $\Sigma$ be an algebraic curve
\be\label{eq:aug-curve-conj}
	\Sigma = \{(x,y)\in \IC^*\times \IC^*;\ A(x,y,u) = 0\}\,.
\ee
where $u = \{u_s\}$ is a generic set of complex moduli.
Suppose that $\Sigma = \Sigma^\CQ$ admits a presentation as a quiver augmentation curve as defined in \eqref{eq:Q-deformed-curve}.
The underlying quiver structure, including vertices, adjacency matrix and fugacities (see \eqref{eq:TQ-data}), can be obtained as follows.
\begin{enumerate}
	\item 
	Let $x_{\theory} \in \IC^*$ be a point sufficiently close to $0$, and let $i$ denote the distinguished vacuum.
	The spectrum of BPS kinky vortices with unit charge in vacuum $i$ is generated by 
	a distinguished basis of $m = \dim\CH_{(ii,1)}(x_\theory)$ states with CIFV index of unit norm $|\mu_{a_j}|=1$, 
	which are in 1-1 correspondence with the vertices of $\CQ$
\be\nonumber
	\text{vertices of }\CQ \ \ \leftrightarrow\ \ \text{basis of }\lim_{x_\theory\to 0}\CH_{(ii,n)}(x_\theory)\,.
\ee
	Each basis state corresponds to a calibrated path $a_j$ on $\tSigma$, with $j=1,\dots, m$, which is computed by exponential networks at the phase of its central charge $\vartheta_j = \arg Z_{a_j}$.
	
	\item
	Quiver fugacities of $T[\CQ]$ are encoded by central charges computed by 1-chain integrals \eqref{eq:Z-BPS} on the basis paths
	\be
		c_j = \frac{1}{x_\theory} \exp\left(\frac{Z_{a_j}}{2\pi i}\right)
		\,,\qquad j=1\dots m\,.
	\ee

        \item The intersection matrix of the $m$ basic paths $\langle a_i, a_j\rangle$ coincides with the adjacency matrix $C_{ij}$ of the quiver
\be\nonumber
	C_{ij} = \langle a_i, a_j\rangle\,.
\ee
    \end{enumerate}
\end{conjecture}

This conjecture applies in particular to the computation of quivers $\CQ$ associated with theories $T[L]$ arising from toric branes and knot conormals, among other examples.
By viewing $T[\CQ]$ as a deformation of $T[L]$ as explained in Section \ref{sec:Q-deform},
the quiver data for $T[L]$ is obtained by specializing to the limit $c_i \to Q^{\beta_i}$.
That is, applying our conjecture directly to the augmentation curve $A(x,y,Q)$, the quiver can be obtained from exponential networks.

\subsection{Algorithm}\label{sec:algorithm}

To complement and sharpen our main conjecture, we next explain how the data of the quiver
should be computed from an algebraic curve in $(\IC^*)^2$. Here we summarize the algorithm, and later we will develop each point in more detail.

\begin{enumerate}

\item
Fix a suitable choice for the \emph{theory point}, i.e. a point $x_\theory \in \IC^*_x$\footnote{This encodes the choice of modulus for $T[L]$, whose parameter space is  the $x$-plane with a puncture at the origin.},
in such a way that the spectrum of $(ii,1)$ BPS kinky vortices for the distinguished vacuum $i$ is \textit{stabilised} under wall-crossing~\eqref{eq:H-wall-cross}. 
Concretely, this implies choosing $x_\theory$ sufficiently close to $0$, as demonstrated in \cite{Gupta:2024ics}.

\item
Compute the spectrum of $(ii,1)$ kinky vortices using exponential networks, by keeping track of all network trajectories of type $(ii,1)$ that swipe across $x_\theory$ as $\vartheta$ varies in $[-\pi,0)$. 
Denoting the corresponding soliton paths (a precise definition is given below) by $a_k$ for $k=1,\dots, m$, each is associated to a vertex of $\CQ$.

\item
The central charge of each soliton admits a decomposition
\be\label{eq:Z-ak}
	Z_{a_k} = 2\pi i \left[\log x_\theory + \sum_s (\beta_k)_s \, \log u_s\right]\,,
\ee
in terms of integers $(\beta_k)_s$, which
determines the quiver fugacities $c_k$ in \eqref{eq:TQ-data}
\be\label{eq:ck-def}
	c_k = \prod_s  {u_s}^{(\beta_k)_s}\,.
\ee

\item
The intersection matrix of $(ii,1)$ soliton paths, defined by path concatenation as in \eqref{eq:int-pair}, determines the quiver adjacency matrix 
\be
	C_{kl} = \langle a_k , a_l \rangle \,.
\ee
\end{enumerate}

The implementation of each step involves certain subtleties, which we address next.

\subsection{Quiver vertices and fugacities}\label{sec:vertices-fugacities}

We begin by discussing the steps that give the vertices of $\CQ$ and the associated fugacities $c_i$.

\subsubsection{Identification of kinky vortex charges with paths on $\tSigma$}

We recall some details regarding the definition of charges of kinky vortices in terms of 1-chains on $\tSigma$, more details can be found in \cite{Banerjee:2018syt, Gupta:2024ics}.

Given $\Sigma$ as in  \eqref{eq:aug-curve-conj}, we view it as a (possibly ramified) covering of $\IC^*_x$, with sheets given locally by the roots $y_j(x)$ for $j=1,\dots, K$. 
We further introduce a logarithmic covering $\tSigma$ of $\Sigma$ branched at punctures corresponding to $y=0,\infty$, with sheets $\log y_j(x) +2\pi i \, N$ labeled by $N\in \IZ$. 
A choice of trivialization for these coverings, i.e. a system of branch cuts and a globally defined labeling $(j,N)$ of sheets of $\tSigma$ as a covering of $\IC^*_x$ will be understood from now on.

Charges of kinky vortices, or solitons, are classified by relative homology classes of paths on $\tSigma$ interpolating between vacua $(i,N)$ and $(j,N+n)$. 
We label soliton charges only by the flux $n$ that they carry, and not by the logarithmic label $N$ of the vacuum they start from, since only the difference is physically observable due to the overall shift symmetry $N\to N+1$ arising from large gauge transformations on $S^1\times \IR^2$.
The action of large gauge transformations on soliton paths is implemented by a shift map constructed in \cite{Banerjee:2018syt}.
Consider a class of paths $a^{[N]}$, with $N\in \IZ$, interpolating respectively between vacua 
\be
	\mathrm{beg}(a^{[N]}) = (i,N)\,,
	\qquad
	\mathrm{end}(a^{[N]}) = (j,N+n)\,,
\ee
where $(i,N)$ denotes the point $(x, \log y_j(x)+2\pi i N)\in \tSigma$.
They have the same CFIV indices and central charges if and only if they are related by the shift map,\footnote{We assume moduli of $\Sigma$ and $x$ are chosen generically.} and in this case we identify them as charges
\be\label{eq:shift-equiv}
	\mu_{a^{[N]}} = \mu_{a^{[M]}}\,,
	\quad
	Z_{a^{[N]}} = Z_{a^{[M]}}
	\qquad\Rightarrow\qquad
	a^{[N]}\sim a^{[M]}\,.
\ee
We denote the equivalence class of equivalent paths simply as $a$, omitting $\delta$. 
When it will be important to discuss the choice of a representative, we will sometimes denote by $a$ the representative $a^{[0]}$ and its shifted copies by $a^{[N]}$.

\subsubsection{Fixing the theory point by stabilization of the $(ii,1)$ sector}

The BPS spectrum of kinky vortices, i.e. the collection $\{\mu_a, Z_a\}$ of CFIV indices and central charges for all soliton charges $a$, depends on a choice of $x\in \IC^*$ due to wall-crossing phenomena in the parameter space of couplings of the 3d $\CN=2$ QFT, for which $x$ plays the role of a (complexified and exponentiated) Fayet-Ilioupoulos coupling. 

The spectrum jumps when $x$ crosses lines of marginal stability, whose definition is 
\be
	\MS(a,b) = \{x \in \IC^*; \mu(a) \mu(b) \neq 0, \ Z_a /  Z_b\in \IR_{>0},\   \mathrm{end(a)}=\mathrm{beg}(b)\}\,.
\ee
In other words, MS walls of two solitons $a,b$ are loci in $\IC^*$ where both states have nonzero CFIV index, their central charges have the same phase, and solitons can be concatenated so that the vacuum corresponding to the end of $a$ coincides with the vacuum corresponding to the beginning of $b$. 

In general, $\mathrm{MS}(a,b)\neq \mathrm{MS}(b,a)$ due to concatenation ordering. 
A soliton of type $(ij,n)$ i.e. interpolating between vacuum $(i,N)$ and $(j,N+n)$ can only be concatenated with a soliton of type $(jk,m)$, for arbitrary $k\in 1,\dots, K$ and $m\in \IZ$.
An important exception occurs for solitons of types $(ii,n)$: if both $a,b$ are of this type, then concatenation is possible with both orderings and $\MS(a,b)=\MS(b,a)$.

The structure of MS walls for kinky vortices of type $(ii,n)$ has been studied extensively in \cite{Gupta:2024ics}.
As $x$ approaches $0$, the central charge of $(ii,n)$ solitons is dominated by
\be
	\lim_{x\to 0} Z_a  = \frac{1}{2\pi i}\int_a\log y \, d\log x \approx n \log x +\dots
\ee
On the other hand the pattern of CFIV indices is typically rather intricate, 
due to an infinite sequence of MS walls that are located near $0$. 
Denoting by $\MS_{(ii,n)}$ all marginal stability walls where $(ii,n)$ solitons can be generated
\be
	\MS_{(ii,n)} = \bigcup_{ab \sim (ii,n)} \MS(a,b)\,,
\ee
we observed that these are all contained in an annular region of finite size around $x=0$
\be
	r^-_{ii,n} < \{|x|; x\in \MS_{(ii,n)}\} < r^+_{ii,n}\,.
\ee

Therefore the CFIV index of a soliton $a$ of type $(ii,n)$ will undergo a sequence of transitions, but will eventually stabilize once $|x_\theory| < r_{(ii,n)}^-$.
We denote the stabilized spectrum by $\mu_a$ 
\be
	\mu_a := \mu_a(x) \quad \forall |x|<r_{(ii,n)}^-\,.
\ee

\begin{figure}
    \centering
    \includegraphics[width=0.5\linewidth]{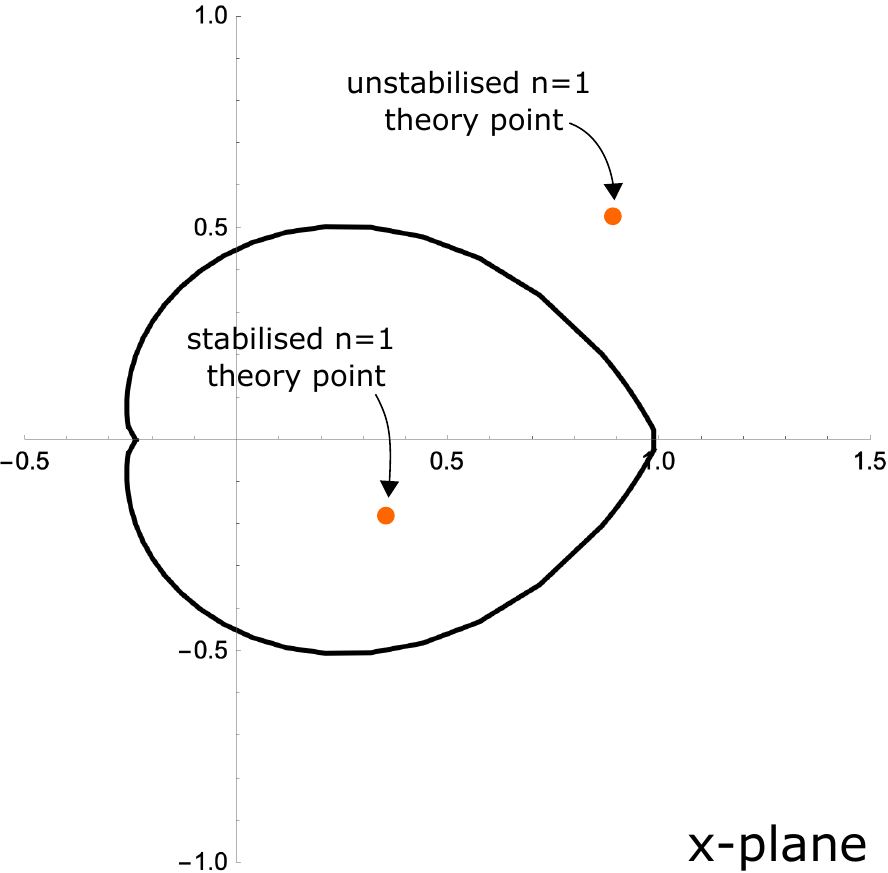}
    \caption{Walls of marginal stability, and stabilised and unstabilised theory points for $(ii,1)$ solitons on $\IC^*_x$ in the theory described by the mirror of $\mathbb{C}^3$, see \cite{Bullimore:2018yyb,Gupta:2024ics} for more details.}
    \label{fig:stabilisation}
\end{figure}

The spectrum of stabilized kinky vortices is expected to agree with the spectrum of standard BPS vortices, with a suitable identification of their partition functions \cite{Gupta:2024ics}.
An example ins shown in Figure \ref{fig:stabilisation}.
For the purpose of computing quivers from augmentation curves, we will only be interested in the stabilized spectrum of $(ii,1)$ solitons.

\subsubsection{Computation of the stabilised spectrum}

Having fixed a suitable theory point, the next task is to compute the stabilized spectrum of $(ii,1)$ kinky vortices.
This can be achieved by a standard application of exponential networks, see in particular \cite{Banerjee:2018syt} for details on the computation of CFIV indices. 
Given $\Sigma$, the exponential network $\CW(\vartheta)$ is a web of trajectories on $\IC^*_x$ corresponding to solutions of the BPS soliton equations, whose geometric counterpart is the ODE
\be\label{eq:E-walls}
	(\log y_j - \log y_i +2\pi i \, n) \frac{d\log x}{dt} = e^{i\vartheta}\,.
\ee
The equation is integrated starting from branch points $y_i(x)=y_j(x)$, or from punctures where $|\log y(x)| \to \infty$, or from intersections of trajectories.
The shape of the network depends on $\vartheta$.

Kinky vortices for the QFT whose FI coupling is $x_\theory$ are encoded by trajectories of $\CW$ that pass through $x_\theory$ for any value of $\vartheta$. In fact, if the BPS spectrum contains a kinky vortex with charge $a$, then there will be a trajectory of $\CW$ that passes through $x_\theory$ exactly when the phase is that of its central charge
\be\label{eq:soliton-phase}
	\vartheta_a := \arg Z_a(x_\theory) \,.
\ee
Recall that $Z_a$ is a 1-chain integral \eqref{eq:Z-BPS} on $\tSigma$.
The class of models considered in this paper, which includes toric branes of toric CY3 with $b_4=0$ and knot conormal branes in the resolved conifold, has the special property that any closed periods of $\tSigma$ are linear combinations of complex moduli
\be
	Z_{\gamma} = 2\pi i  \sum_s k_{\gamma,s} \, \log u_s\,,\qquad k_{\gamma,s}\in \IZ\,.
\ee
Focussing on $(ii,n)$ BPS states, and given their physical interpretation as vortices with unit flux on the cylinder \cite{Gupta:2024ics}, it follows that their central charge is determined by a collection of integers $k_{a,s}$
\be\label{eq:soliton-Z-decomposition}
	a \text{ is of type }(ii,n) \quad\Rightarrow\quad
	Z_{a} = 2\pi i \left[n\log x_\theory + \sum_s k_{a,s} \, \log u_s\right]\,.
\ee

\begin{figure}
\centering
\begin{subfigure}{.4\textwidth}
  \centering
  \includegraphics[width=1\linewidth]{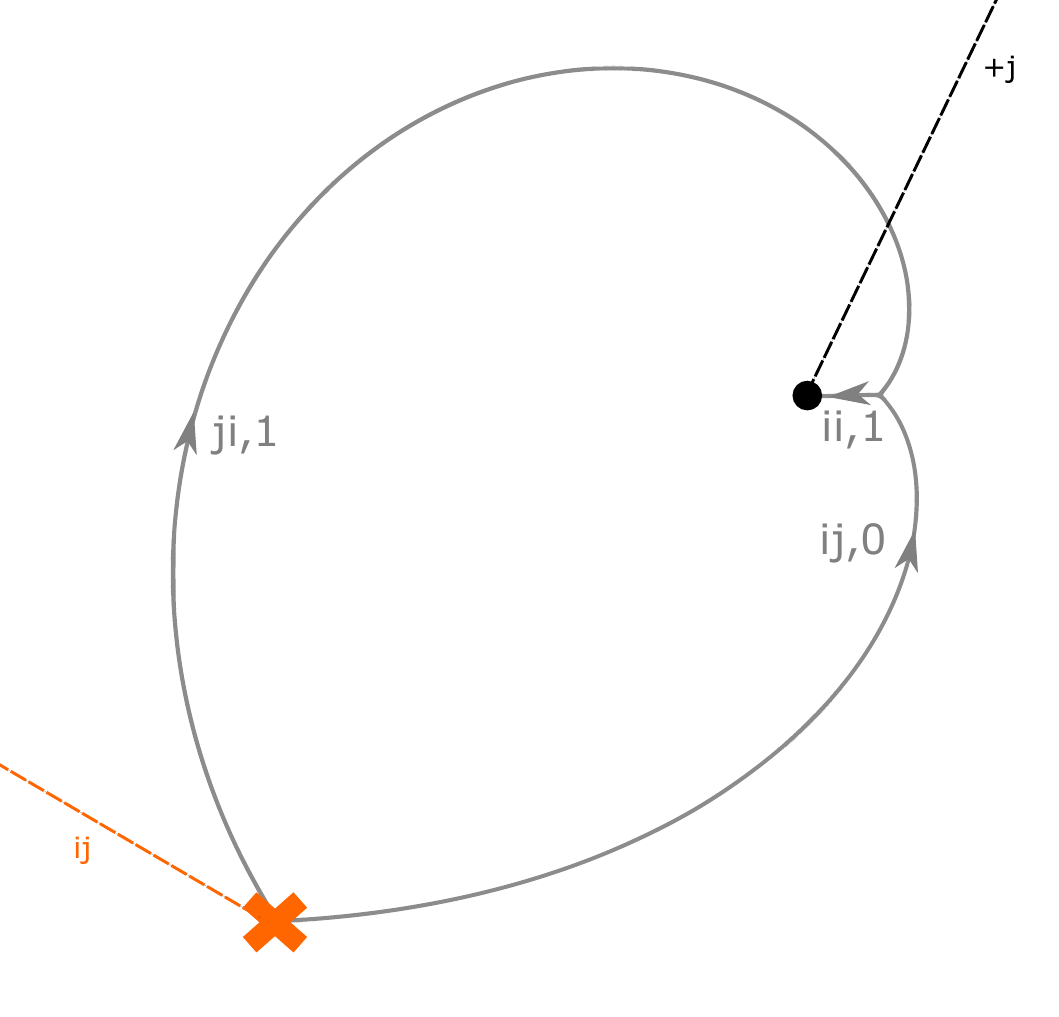}
  \caption{Soliton web}
  \label{fig:exampleweb1}
\end{subfigure}%
\begin{subfigure}{.4\textwidth}
  \centering
  \includegraphics[width=1\linewidth]{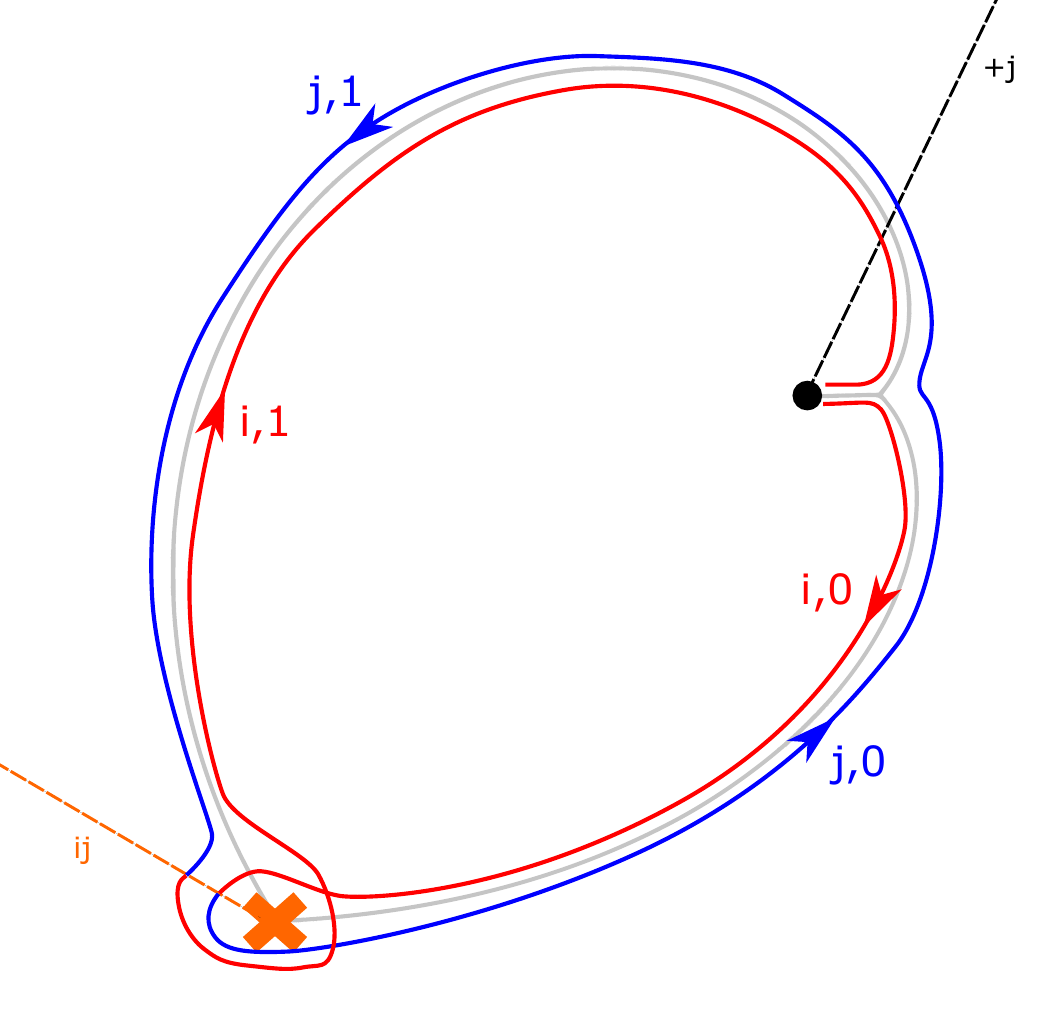}
  \caption{Resolved projected soliton paths.}
  \label{fig:examplestring1}
\end{subfigure}
\caption{An $(ii,1)$ soliton web from the exponential network for the mirror curve of the resolved conifold, see Section \ref{sec:unknotconormal}. 
The orange cross is a branch point, the dashed orange line is a polynomial branch cut, the black dot is the puncture, the black dashed line is a logarithmic branch cut, the orange point is the theory point and the solid lines are the relevant trajectories.
Although not shown here, we will use a black cross for logarithmic branch points away from punctures.}
\label{fig:examplestep2}
\end{figure}

Solitons of type $(ii,n)$ are carried by trajectories of the same type, which can be generated either at punctures with $|\log y|\to\infty$, or at intersections of other trajectories (but not at branch points). In either case, a calibrated path on $\tSigma$ of class $a$ is obtained by lifting to appropriate sheets of $\tSigma$ the trajectories that underlie the soliton: this includes the $(ii,n)$ trajectory, but also any of its parents, see Figure \ref{fig:exampleweb1} for an example.\footnote{Sometimes, trajectories are counted with multiplicities. Details are discussed in \cite{Banerjee:2018syt}.} The lift involves taking two copies of an $(ij,n)$ trajectory to sheets $(i,N)$ and $(j,N+n)$ with opposite orientations. The choice of $N$ is irrelevant as discussed earlier.

Since an exponential network $\CW(\vartheta)$ includes the CFIV index of each path $\mu_a$ as part of its data,
by scanning over all phases $\vartheta$, and keeping track of trajectories that sweep across $x_\theory$, the entire spectrum of kinky vortices at $x_\theory$ can be obtained. 
Computing the spectrum for a value of $x_\theory$ for which $\CH_{(ii,1)}$ is stabilized ensures that we detect all states corresponding to kinky vortices with unit vorticity $n=1$.
Let $a_{k}$ with $k=1,\dots, m$ denote the (equivalence classes) of calibrated paths of type $(ii,1)$ computed by the exponential network, we define a set of vertices of the quiver $\CQ$ in 1-1 correspondence with $\{a_k\}_{k=1}^m$.
The fugacities $c_i$ associated to vertices of $\CQ$ are defined by the decomposition of the central charges $Z_{a_k}$ given in \eqref{eq:Z-ak}-\eqref{eq:ck-def}.

\begin{remark}[Uniqueness of $a_k$]
The exponential network $\CW$ computes the spectrum $\{Z_{a_k},\mu_{a_k}\}_{k=1}^{m}$, and additionally determines calibrated paths on $\tSigma$ in class $a_k$ for each of the $(ii,1)$ basis states.
However, there may be more than one calibrated path for a given charge $a_k$.
This is because the CFIV index is integer-valued and soliton-anti-soliton pairs can give rise to cancellations.
Since our algorithm for building $\CQ$ requires us to determine a unique path for each $a_k$, it is essential to address how to deal with the possibility of cancellations. We will argue how to do this by a deformation argument.

First we observe that $|\mu_{a_k}|=1$ is necessarily true, because of the following reason.
Suppose that $\Sigma$ admits a quiver description $\Sigma^\CQ$. Then it is possible to write it as an algebraic curve $A(x,y,c_1,\dots, c_m)=0$ and by taking $c_j\to 0$ except for a distinguished value of $j$, turns the curve into $1-y-c_j (-y)^{C_{jj}} x=0$, as it follows from \eqref{eq:A-i-poly} that all other $y_{k\neq j}\to 1$ in this limit.
Then, the resulting curve corresponds to the mirror of $\IC^3$, which was analyzed in \cite{Gupta:2024ics}, where it was shown that there is a unique $(ii,1)$ kinky vortex close to $x=0$, with $\mu_a=\pm 1$.

Next, we can turn back on the other $c_{j\neq k}\neq 0$ up to the original values that determine $\Sigma$ as a specialization of $\Sigma^\CQ$. In this process, it may happen that new stable solitons arise in the charge sector $a_j$, but by invariance of the CFIV index these must come in particle-antiparticle pair. Keeping track of these, we can distinguish the original soliton from the canceling pairs and select a unique calibrated path in class $a_k$. We will see an example of this in Section \ref{sec:CFIVcancellations}.

\end{remark}

\subsection{The adjacency matrix}\label{sec:adjacency-matrix}

The computation of $C_{ij}$ from intersections of open paths as defined in \eqref{eq:int-pair}  involves several delicate steps:
\begin{itemize}
\item
A soliton charge $a$ corresponds to an equivalence class of (relative homology classes) paths on $\tSigma$ established by the shift map \eqref{eq:shift-equiv}.
To define the class of the concatenation of two paths $a_j$, $a_k$ it is necessary to choose compatible representatives $a_j^{[N]}, a_k^{[M]}$, which admit a concatenation. If follows from definitions that the equivalence class (under shift map) of the latter is independent of the choice of representatives.  
\item
To define a smooth concatenation it is necessary to choose a capping path that connects smoothly the endpoint of the first path to the starting point of the second one. This choice is a priori non-canonical, and must be considered as part of the definition of the intersection pairing.
\item
An additional subtlety that arises with $(ii,1)$ paths is that both $a_j a_k$ and $a_k a_j$ are valid concatenations. 
This raises the question of whether the intersection pairing is symmetric under exchange $\langle a_j,a_j\rangle = \langle a_k, a_j\rangle$, which is needed for consistency with the symmetry of the quiver's adjacency matrix.
\end{itemize}

\subsubsection{Resolution of projected soliton paths}

While paths are defined on $\tSigma$, having chosen a trivialization with global labels $(j,N)$ for its sheets over $\IC^*_x$, it will be convenient to work with projections of the soliton paths to the base. 
The projection is however degenerate, and to compute intersections it will be useful to adopt a choice of resolution.

As shown in Figure \ref{fig:exampleweb1}, a soliton path in class $a_k$ is associated to a collection of trajectories in $\CW$, which we call the \emph{soliton web} of $a_k$ and which we denote by $a_k^{\CW}$.
The soliton web coincides, in fact, with the projection of $a_k$ down to $\IC^*_x$ (possibly with multiplicities attached to each of its edges).
Therefore to define a resolution of the projection of $a_k$ it is sufficient to define a resolution of the associated soliton web.

Let $a$ be the charge of a kinky soliton of type $(ii,1)$. 
Its projection to $\IC^*_x$ consists of a collection of arcs on $\IC^*_x$ obtained by taking, for each trajectory of type $(jk,n)$ in the soliton web $a^\CW$, a copy labeled by $(j,N)$ and a copy labeled by $(k,N+n)$ oriented respectively anti-parallel and parallel to the trajectory itself.\footnote{The two copies might be taken with additional multiplicities, depending on $a$.}
The label $N$ associated to each arc is determined by compatibility with gluing at junctions with other arcs, as determined by a choice of representative for $a$ itself.

A resolution of the soliton web $a^{\CW}$ is then given by a choice of capping paths to glue the above  collection of arcs on $\IC^*_x$, according to the following rules:

    \begin{enumerate}
        \item Away from branch points, the $(j,n+N)$-labeled arc is on the right side while the $(k,N)$-labeled arc is on the left side of the trajectory, viewed with respect to the direction of increasing $t$ in \eqref{eq:E-walls}.
        \item 
        Around polynomial branch points, the computation of intersection number involves choosing a resolution for the projection of soliton paths. This presents a potential source of ambiguity because the resolution of each path can turn either clockwise or counterclockwise around a quadratic branch point. In appendix \ref{app:intersections1} we show that the intersection number is however invariant. 
        For higher-order polynomial branch points the choice of resolution is dictated by the monodromy itself, and no such ambiguity is present.           
        \item At junctions of trajectories of types $(ij,n)$, $(jk,m)$ and $(ik,m+n)$ the resolution of soliton paths involves smoothing cusps. The choice of smoothing can introduce extra intersections (kinks) in the path, however these drop out of the mutual intersection number defined via \eqref{eq:int-pair}. Therefore, any choice of smoothing can be used. For simplicity we will adopt the choice that does not involve additional kinks.
        \item At a logarithmic branch point, the projected soliton paths can be smoothed in such a way that it turns either clockwise or counterclockwise around the branch point. The choice is uniquely fixed by the signature of the logarithmic shift around the branch point, see Appendix \ref{app:intersections1}.
    \end{enumerate}
    An example is shown in Figure \ref{fig:examplestring1}.

\subsubsection{Concatenation}

Next we define the concatenation of paths of type $(ii,1)$ by using their resolved projections.
Naturally, the concatenation takes place at $x_\text{theory}$ where both resolved paths have their endpoints. 
However, the issue that we need to solve is that the tangents at the endpoints of the paths to be concatenated typically do not match. Therefore additional arcs need to be introduced to define a concatenation.

\begin{figure}
\centering
\begin{subfigure}{.4\textwidth}
  \centering
  \includegraphics[width=0.7\linewidth]{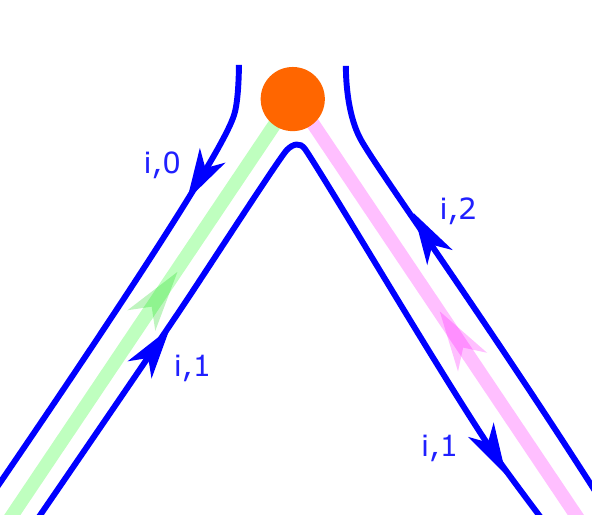}
  \caption{Clockwise concatenation}
  \label{fig:capping1}
\end{subfigure}%
\begin{subfigure}{.4\textwidth}
  \centering
  \includegraphics[width=0.7\linewidth]{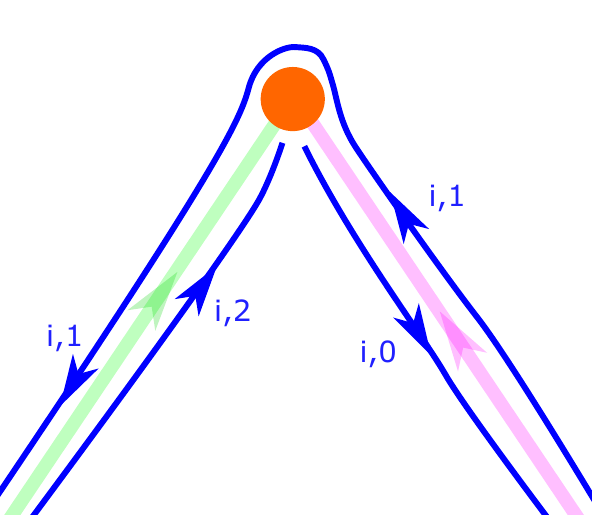}
  \caption{Counterclockwise concatenation}
  \label{fig:capping2}
\end{subfigure}
\caption{Zoomed in near $x_\text{theory}$, the translucent green and pink trajectories belong two different $(ii,1)$ soliton webs, with their base logarithmic indices chosen so that concatenation of soliton paths is possible. As mentioned in the text, the counterclockwise concatenation is the correct choice for obtaining the quiver.}
\label{fig:capping}
\end{figure}

The tangent of a projected soliton at its endpoint $x_\theory$ depends on the phase $\vartheta_a$ of its central charge~\eqref{eq:soliton-phase}.
In fact, solutions of \eqref{eq:E-walls} for $(ii,n)$ solitons are actually independent of $i$ and $n$ (up to rescaling of $t$), and are spirals given by
\begin{equation}\label{eq:spiral}
    x(t) = x_{\text{theory}} e^{\frac{e^{i \vartheta} (t-t_0)}{2\pi i}}\,.
\end{equation}
Note that requiring $\lim_{t\to+\infty}x(t) = 0$ implies that $\vartheta_{a_j} \in [-\pi, 0]$ for all $j = 1,2,\dots, m$. 
\footnote{In fact, when $x_\theory$ is in the stabilization regime, i.e. near $0$, the central charge of all $(ii,1)$ solitons is dominated by $\log x_\theory$, see \eqref{eq:soliton-Z-decomposition}.
Therefore in this limit
$$
    \lim_{x_\text{theory} \to 0} \vartheta_{a_j} = -\frac{\pi}{2}, \hspace{0.5cm} j = 1,2,\dots,m.
$$
Since all phases tend to the same value, the endpoint tangents of projected paths are almost parallel in this limit. More precisely, the final tangent vector of one path is (nearly) \emph{anti-}parallel to the starting tangent vector of another path.
}
Due to this, the angle between the two trajectories will be acute,
and further looking at the slopes of the $(ii,1)$ trajectory at the theory point using Eq \eqref{eq:spiral}
\begin{equation}
    x'(t_0) = \frac{x_\text{theory} e^{i \vartheta_{a_j}}}{2\pi i}\,,
\end{equation}
we conclude that the phase $\vartheta_{a_j}$ of the translucent green trajectory corresponding to the projection of $a_j$ is less than that of the pink trajectory $\vartheta_{a_k}$ in Figure~\ref{fig:capping}.

Moving on to the actual concatenation, we start by recalling that to concatenate two $(ii,1)$ soliton paths, both the polynomial and logarithmic branch of the end of the first soliton path should match with that of the beginning of the second soliton path. 
Assuming that the phases of the two solitons are different, the redundancy in choosing the base logarithmic branch gives us two distinct possible concatenations which we show in Fig \ref{fig:capping}. 

As it turns out, the correct choice for recovering the quiver is counterclockwise concatenation, which according to above also imposes the \emph{decreasing phase ordering} of concatenation among projected paths. 
Therefore we define the concatenation of two paths of type $(ii,1)$ as follows (oriented from left to right)\footnote{This convention is based on having $\vartheta$ valued in the interval $[-\pi,\pi)$.}
\be
	a_{j} a_{k} \quad \text{if }\vartheta_{a_j}< \vartheta_{a_k}\,.
\ee

The above rule does not address the case of paths with the same phase. 
This situation can happen in two distinct cases: either in the computation of the self-intersection number, when one studies the concatenation of a soliton path with a (shifted, as in \eqref{eq:shift-equiv}) copy of itself, or when moduli of $\Sigma$ are not generic.
In these cases both choices of for the ordering of concatenation give the same intersection pairing.
More details on these cases will be given in actual examples, and further in Appendix \ref{app:intersections2}.

\subsubsection{The intersection pairing matrix}

\begin{figure}
\centering
\begin{subfigure}{.4\textwidth}
  \centering
  \includegraphics[width=0.8\linewidth]{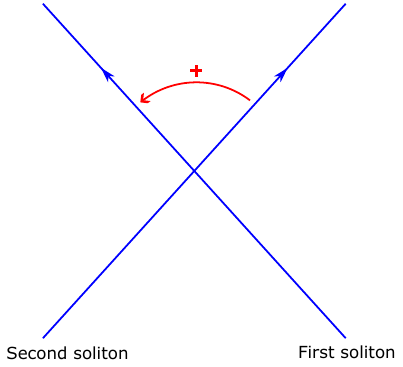}
  \caption{Intersection Number = +1}
  \label{fig:righthandplus}
\end{subfigure}%
\begin{subfigure}{.4\textwidth}
  \centering
  \includegraphics[width=0.8\linewidth]{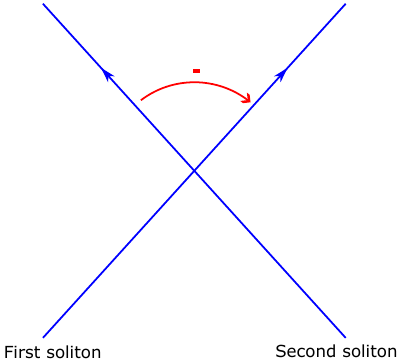}
  \caption{Intersection Number = -1}
  \label{fig:righthandminus}
\end{subfigure}
\caption{Intersection numbers according to the right hand rule}
\label{fig:righthandrule}
\end{figure}

The intersection of two paths $\langle a,b\rangle$ is defined by \eqref{eq:int-pair}, where the write of each path and of their concatenation is obtained by counting double points in the resolved soliton webs $a^{\CW}, b^{\CW}$ and $ab^{\CW}$ defined by choosing appropriate capping paths as explained above.
Each double point of arcs with coincident labels $(j,N) = (j',N')$ contributes $\pm 1$ according to the rule given in Figure \ref{fig:righthandrule}
\be
	\wr(a) = \sum_{\text{double pt.}} \pm1\,.
\ee
This convention is adapted to recover quiver adjacency matrices in examples discussed below.

\section{Examples}\label{sec:examples}

This section illustrates how Conjecture \eqref{conj:QuiverSolitonCorrespondence} is realized through the algorithm described in Section \ref{sec:algorithm} in several examples.
We will consider two examples involving toric branes in Calabi-Yau threefolds with $b_4=0$, respectively $\IC^3$ and the resolved conifold, as well as two examples based on knot conormal Lagrangians in the resolved conifold repectively for the trefoli and figure-eight knots.
In all cases we obtain a quiver from exponential networks, finding perfect agreement with previous literature.

\subsection{Toric brane in $\mathbb{C}^3$}

\subsubsection{Augmentation curve and its $\CQ$-deformation}
The augmentation curve for the framed toric brane in $\IC^3$ has no moduli and is described by a 3-term polynomial \cite{Aganagic:2001nx}
\begin{equation}\label{eq:C3-curve}
    A(x,y) = 1-y-x(-y)^f \,,
\end{equation}
where $f\in \IZ$ is the framing parameter.
The $(ii,n)$ trajectories of the exponential network of this theory were studied extensively in \cite{Gupta:2024ics}.

The curve \eqref{eq:C3-curve} has the structure \eqref{eq:A-i-poly},  which corresponds to a quiver with a single vertex and adjacency matrix
\begin{equation}\label{eq:C3-matrix}
    C = \begin{pmatrix}
            f
        \end{pmatrix}\,.
\end{equation}
The $\CQ$-deformed augmentation polynomial is therefore
\begin{equation}
    A^\CQ(x,y) = 1 - y - c_1 x (-y)^f\,.
\end{equation}
In this case $c_1$ can be absorbed into a rescaling of $x$, so that without loss of generality we can set $c_1=1$ and work directly with \eqref{eq:C3-curve}. 
We have checked that our conjecture \ref{conj:QuiverSolitonCorrespondence} holds for framings $-3\leq f\leq 3$. We illustrate the cases of $f=0,1,2$ below.

\subsubsection{Framing $0$}
\begin{figure}
    \centering
    \includegraphics[width=0.7\linewidth]{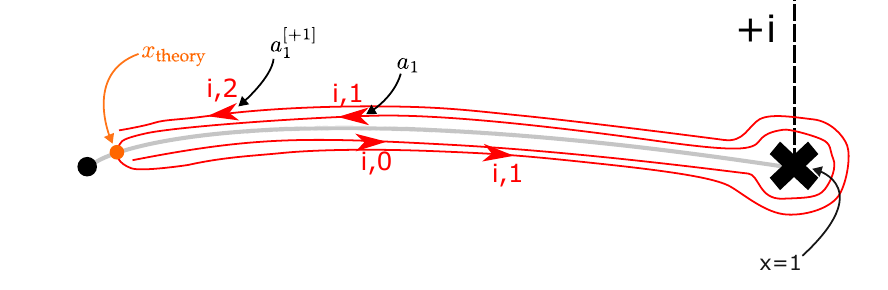}
    \caption{
    The $(ii,1)$ trajectory of the exponential network of $\IC^3$ in framing $f=0$.
    In red: the resolved projections of the soliton path $a_1$ and its shifted copy $a_1^{[+1]}$, concatenated with each other.}
    \label{fig:c3_f0_11}
\end{figure}

The curve is $1-y-x=0$, there is only one sheet $y_i=1-x$ and no polynomial branch points.
The exponential network in this case consists of two trajectories sourced by the logarithmic puncture at $x=1$, and consists of spirals
\be\label{eq:C3-spirals}
	x(t) = e^{\frac{e^{i \vartheta}t}{2\pi i n}}\,.
\ee
For an arbitrary choice of theory point $|x_\theory|<1$,
trajectories of type $(ii,n)$ with $n>0$ will pass through $x_\theory$ whenever $\vartheta = \arg(2\pi i n \log x_\theory+4\pi^2 k)$ for integers $n,k$.
Among these values, only $n=1,k=0$ is relevant because it carries the zero mode kinky vortex in sector $(ii,1)$, see \cite{Gupta:2024ics}. 
There is a unique kinky vortex $a_1$ of type $(ii,1)$, whose corresponding soliton web is shown in Figure~\ref{fig:c3_f0_11}.
The fact that we find a single kinky vortex agrees with the statement that $\CQ$ has a single vertex. 

The $1\times 1$ adjacency matrix can be computed by considering the concatenation of $a_1$ with a shifted copy of itself, denoted here by $a_1^{[+1]}$
\be
	C_{11} = \langle a_1, a_1^{[+1]}\rangle = \wr(a_1 \, a_1^{[+1]}) - \wr(a_1) - \wr(a_1^{[+1]}) = 0\,,
\ee
as can be seen by direct inspection of Figure \ref{fig:c3_f0_11}, and in agreement with \eqref{eq:C3-matrix}.

A subtle but crucial detail is that the projected paths \emph{must circle counterclockwise} around the logarithmic branch point, 
because going around the branch point counterclockwise shifts the logarithmic index as
\be
	\log y +2\pi i N  \to 
	\log y +2\pi i (N+1)\,.
\ee
Correspondingly, the label of the logarithmic branch cut is  $+i$, to denote that the distinguished (unique, in this case) vacuum has a positive logarithmic shift in the counterclockwise direction.

The central charge of this soliton is
\be
	Z_{a_1} = t e^{i \vartheta}\,,
\ee
corresponding to fugacity $c_1=1$, again in agreement with the observation that this is the specialization that takes $A^{\CQ}$ to $A$.

\subsubsection{Framing $1$}

\begin{figure}
    \centering
    \includegraphics[width=0.7\linewidth]{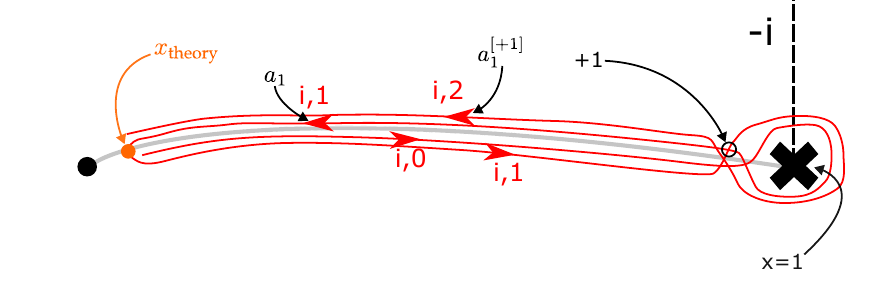}
    \caption{The $(ii,1)$ trajectory of the exponential network of $\IC^3$ in framing $f=1$.
    In red: the resolved projections of the soliton path $a_1$ and its shifted copy $a_1^{[+1]}$, concatenated with each other.}
    \label{fig:c3_f1_11}
\end{figure}

In framing $f=1$ the curve becomes $y = (1-x)^{-1}$. Again there is only one sheet and no polynomial branch points, but there is a logarithmic branch point at $x=1$.
The exponential network again consists of two trajectories sourced by the logarithmic branch point, and consists of spirals
\eqref{eq:C3-spirals}, which will pass through $x_\theory$ when $\vartheta = ...$.
Again there is a unique kinky vortex $a_1$ of type $(ii,1)$ with central charge $...$, with corresponding soliton web is shown in Figure~\ref{fig:c3_f1_11}.

However, a difference with the case $f=0$ is that now $\lim_{x\to 1}y = 0$ instead of diverging, which results in a different type of logarithmic monodromy.
In fact, going around $x=1$ counterclockwise now shifts the logarithmic index of 
\be
	\log y +2\pi i N  \to 
	\log y +2\pi i (N-1)\,.
\ee
For this reason, the projected path for $a_1$ \emph{must circle clockwise} around the logarithmic branch point in this case, see Figure \ref{fig:c3_f1_11}.
This change in the logarithmic shift, and the resulting change of orientation for the soliton path, result in the intersection pairing 
\be
	C_{11} = \langle a_1, a_1^{[+1]}\rangle = 1\,,
\ee
as can be seen by direct inspection. This matches perfectly with the expected quiver adjacency matrix \eqref{eq:C3-matrix}. 
Again, from the central charge $Z_{a_1} = 2\pi i \, \log x_\theory$ we deduce that the quiver fugacity is $c_1 = 1$

\subsubsection{Framing $2$}
In framing $f=2$ the curve has two sheets with polynomial branching at $x=-1/4$, while all logarithmic punctures move to $0$ and $\infty$.
The exponential network contains infinitely many trajectories, all generated by self and mutual intersections of the three trajectories that are sourced by the branch point.
The spectrum of $(ii,1)$ kinky vortices stabilizes if $x_\theory$ is chosen within the MS wall shown in Figure \ref{fig:stabilisation}, and there is a unique soliton $a_1$ with $|\mu_{a_1}|=1$.

\begin{figure}
    \centering
    \includegraphics[width=0.5\linewidth]{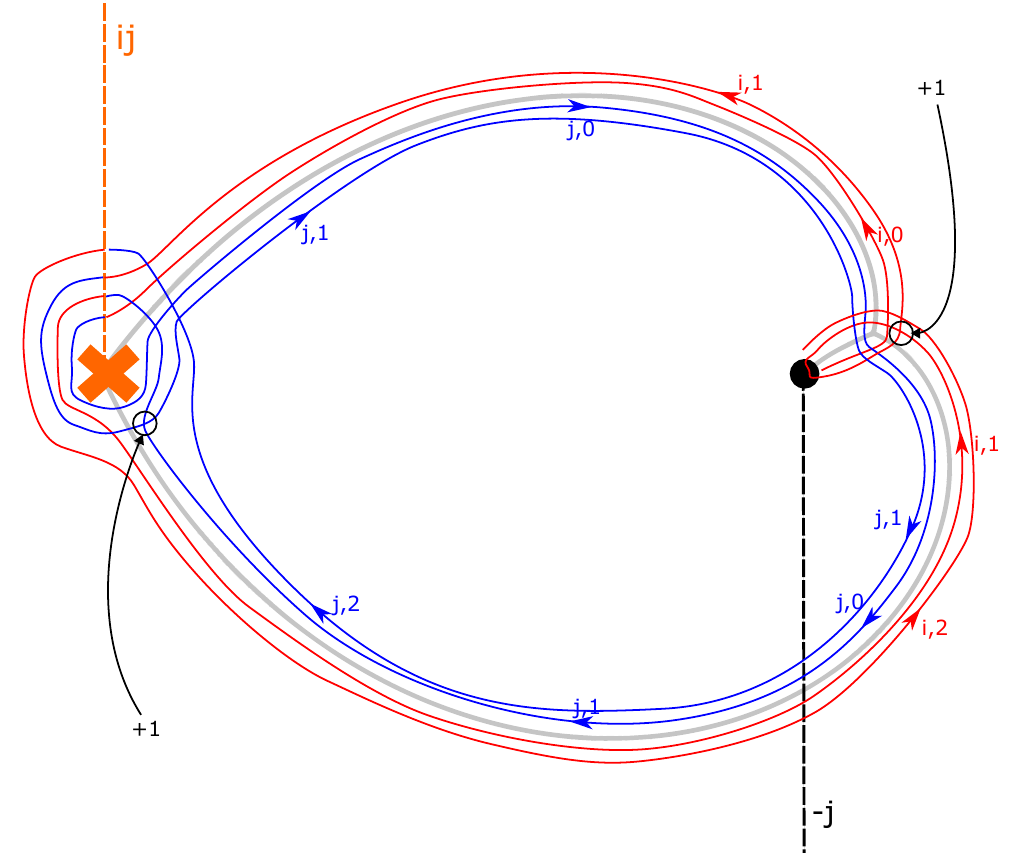}
    \caption{%
    The exponential networks of $\IC^3$ in framing $f=2$.
    Only trajectories that belong to the soliton web $a_1^\CW$ are shown.
    In red and blue: the resolved projections of the soliton path $a_1$ and its shifted copy $a_1^{[+1]}$, concatenated with each other.
    }
    \label{fig:c3_f2_11}
\end{figure}

The adjacency matrix defined by intersections is computed by direct inspection as in Figure \ref{fig:c3_f2_11} and we find
\be
	C_{11} = \langle a_1, a_1^{[+1]}\rangle = 2\,.
\ee
This is agrees with \eqref{eq:C3-matrix}. Again, the quiver fugacity is found to be $c_1 = 1$ according to the fact that $Z_{a_1} = 2\pi i \,\log x_\theory$.

\subsection{Toric brane in the resolved conifold}\label{sec:unknotconormal}

\subsubsection{Augmentation curve and its $\CQ$-deformation}
The augmentation curve for the framed toric brane in $\CO_{\IP^1}(-1)\oplus \CO_{\IP^1}(-1)$ has one complex modulus $Q$, and is described by a 4-term polynomial
\begin{equation}\label{eq:conifold-curve}
    A(x,y,Q) = 1-y -x + Q^2 x y(-y)^f \,,
\end{equation}
where $f\in \IZ$ is the framing parameter. This curve also coincides with the augmentation variety for the unknot conormal brane \cite{Aganagic:2013jpa}.
The underlying quiver has two vertices and adjacency matrix \cite{Kucharski:2017ogk}
\begin{equation}\label{eq:conifold-matrix}
    C = \begin{pmatrix}
            f & f \\
            f & f+1
        \end{pmatrix},
\end{equation}
The $\CQ$-deformed augmentation polynomial is therefore, up to an overall rescaling
\begin{equation}
    A^\CQ(x,y) = (y-1) y-x \left(c_1-c_2 y\right) (-y)^f.\,
\end{equation}
where $c_1, c_2$ are the quiver fugacities.
One of them, say $c_1$ can be absorbed into a rescaling of $x$ without loss of generality, while $c_2/c_1=Q$ can be identified with the complex parameter in \eqref{eq:conifold-curve}. We can therefore work directly with \eqref{eq:conifold-curve}. 
For definiteness, we choose the complex modulus and the theory point as follows
\begin{equation}
    Q = \frac{11}{10}+\frac{2}{10}i\,,\qquad
    x_\text{theory} = \frac{e^{i \pi/3}}{1000}\,.
\end{equation}
We have checked that our conjecture \ref{conj:QuiverSolitonCorrespondence} holds for framings $-3\leq f\leq 3$. We illustrate the cases of $f=0,-1$ below.

\subsubsection{Framing $0$}

\begin{figure}
\centering
\begin{subfigure}{.4\textwidth}
  \centering
  \includegraphics[width=1\linewidth]{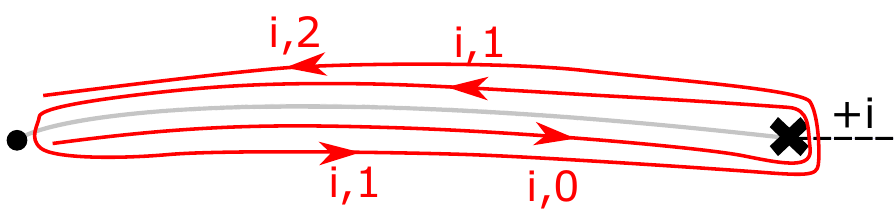}
  \caption{The soliton $a_1$ and computation of $C_{11}$.}
  \label{fig:conifold_f0_11}
\end{subfigure}%
\hfill
\begin{subfigure}{.4\textwidth}
  \centering
  \includegraphics[width=1\linewidth]{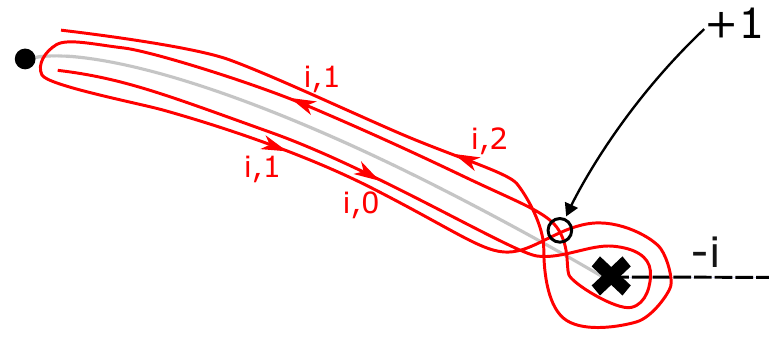}
  \caption{The soliton $a_2$ and computation of $C_{22}$.}
  \label{fig:conifold_f0_22}
\end{subfigure}
\begin{subfigure}{.48\textwidth}
  \centering
  \includegraphics[width=1\linewidth]{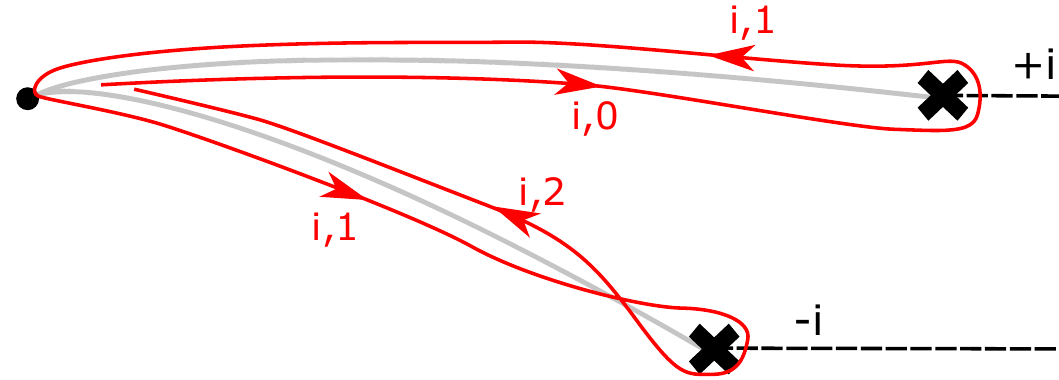}
  \caption{Concatenation $a_1 a_2^{[+1]}$ and computation of $C_{12}$.}
\end{subfigure}
\caption{Soliton webs $a_i^\CW$ for the resolved conifold in framing $f=0$ shown as grey network trajectories.
Resolved projections of soliton paths, and the concatenation with their own their logarithmic shifts are also shown. The black dot denotes the puncture at $x=0$.}
\label{fig:conifold-f0}
\end{figure}

In framing $f=0$ the curve becomes $1-y-x+Q^2 xy=0$, which is a one-sheeted covering of the $x$-plane with logarithmic branch points at $x=1$ and at $x=Q^{-2}$. The exponential network therefore consists of two sets of spirals, one for each logarithmic branch point, which never intersect.
Since 
\be
	\lim_{x\to 1} y(x) = 0\,,\qquad
	\lim_{x\to Q^{-2}} y(x) = \infty\,,
\ee
it follows that each of the logarithmic punctures sources a soliton with the same exact properties as in the cases of $\IC^3$ with framings $f=0$ and $f=1$ respectively. This is confirmed by direct inspection, see Figure \ref{fig:conifold-f0}.
Since there are no intersections, we immediately deduce that the two basic solitons have pairings
\be
	C_{ij} = \langle a_i, a_j^{[+1]}\rangle = 
	 \begin{pmatrix}
            0 & 0 \\
            0 & 1
        \end{pmatrix}
        \,.
\ee

The central charges are obtained by direct integration to be 
\be
	Z_{a_1} = 2\pi i \, \log x_\theory\,,
	\qquad
	Z_{a_2} = 2\pi i \, \log (Q^2 x_\theory)\,,
\ee
which confirms that $c_1=1$ and $C_2 = Q^2$.

\subsubsection{Framing $-1$}
\begin{figure}
\centering
\begin{subfigure}{.48\textwidth}
  \centering
  \includegraphics[width=1\linewidth]{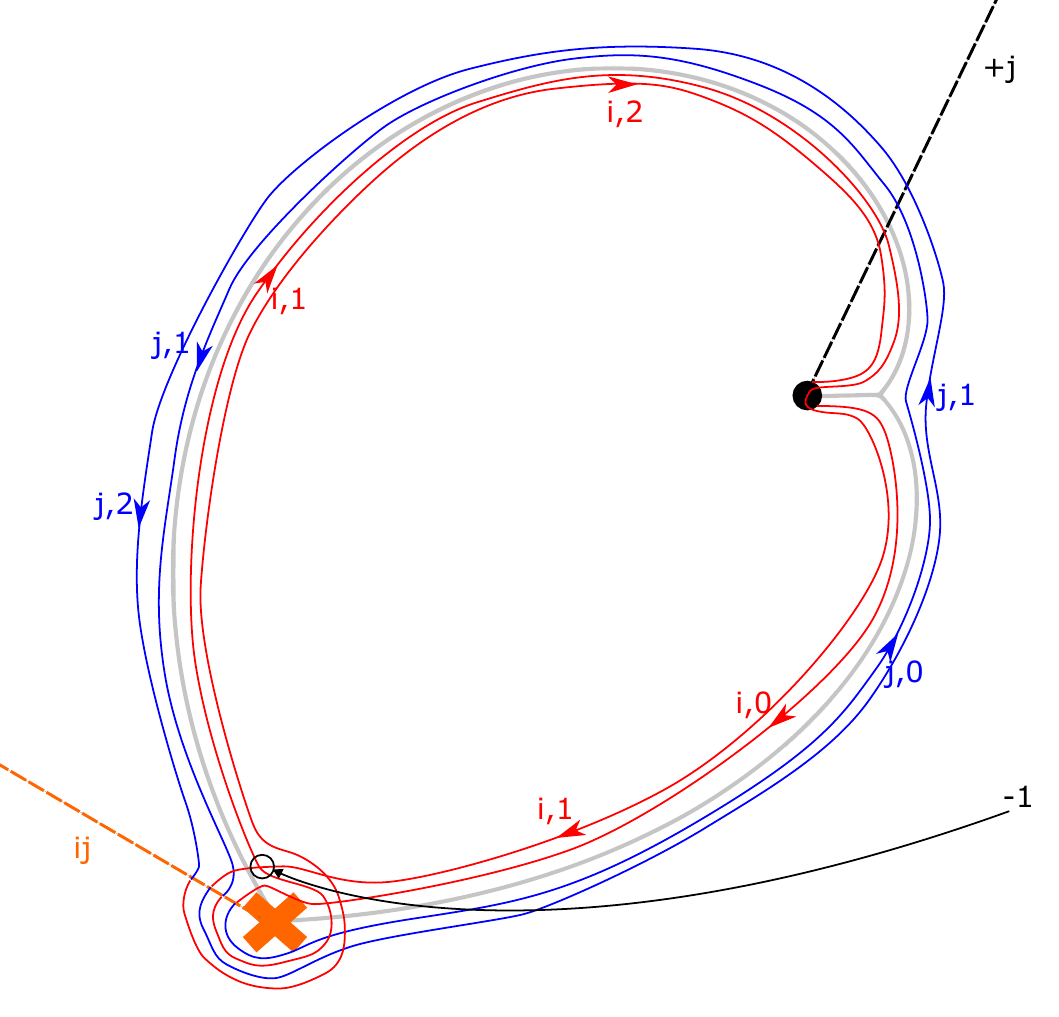}
  \caption{The soliton $a_1$ and computation of $C_{11}$.}
  \label{fig:conifold_fm1_11}
\end{subfigure}%
\hfill
\begin{subfigure}{.48\textwidth}
  \centering
  \includegraphics[width=1\linewidth]{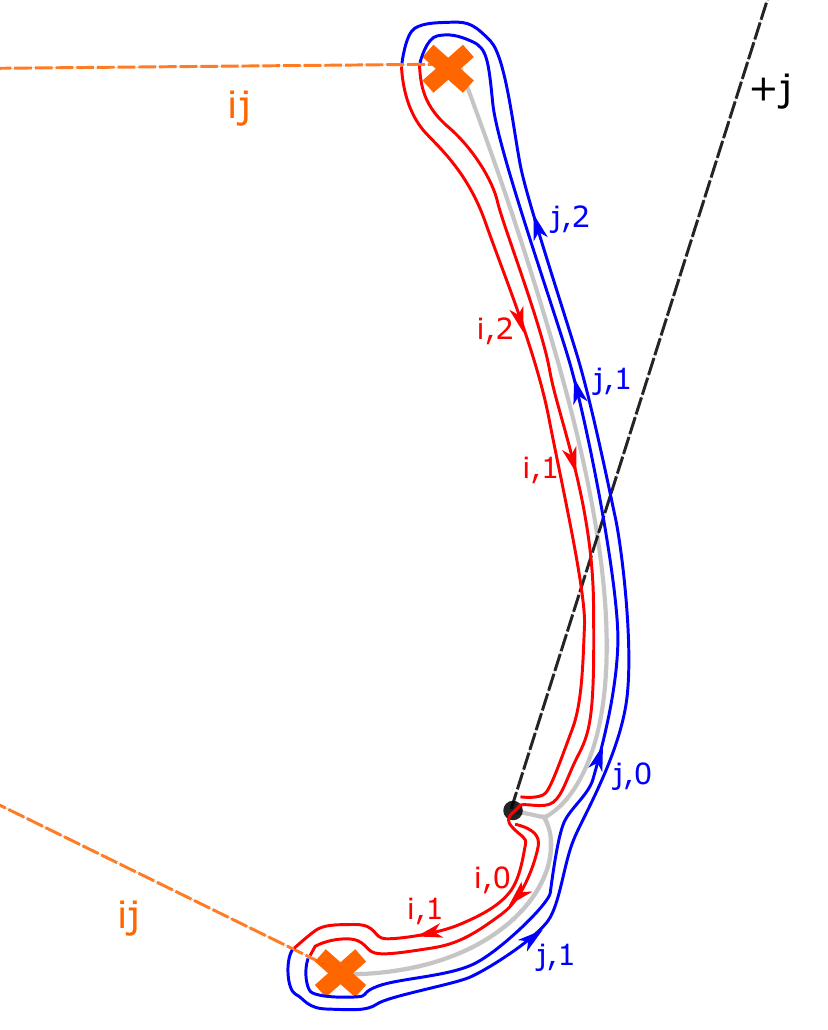}
  \caption{The soliton $a_2$ and computation of $C_{22}$.}
  \label{fig:conifold_fm1_22}
\end{subfigure}
\caption{Soliton webs $a_i^\CW$ for the resolved conifold in framing $f=-1$ shown as grey network trajectories.
Resolved projections of soliton paths, and the concatenation with their own their logarithmic shifts are also shown (in red and in blue).}
\label{fig:conifold-basic-solitons}
\end{figure}

\begin{figure}
    \centering
    \includegraphics[width=1\linewidth]{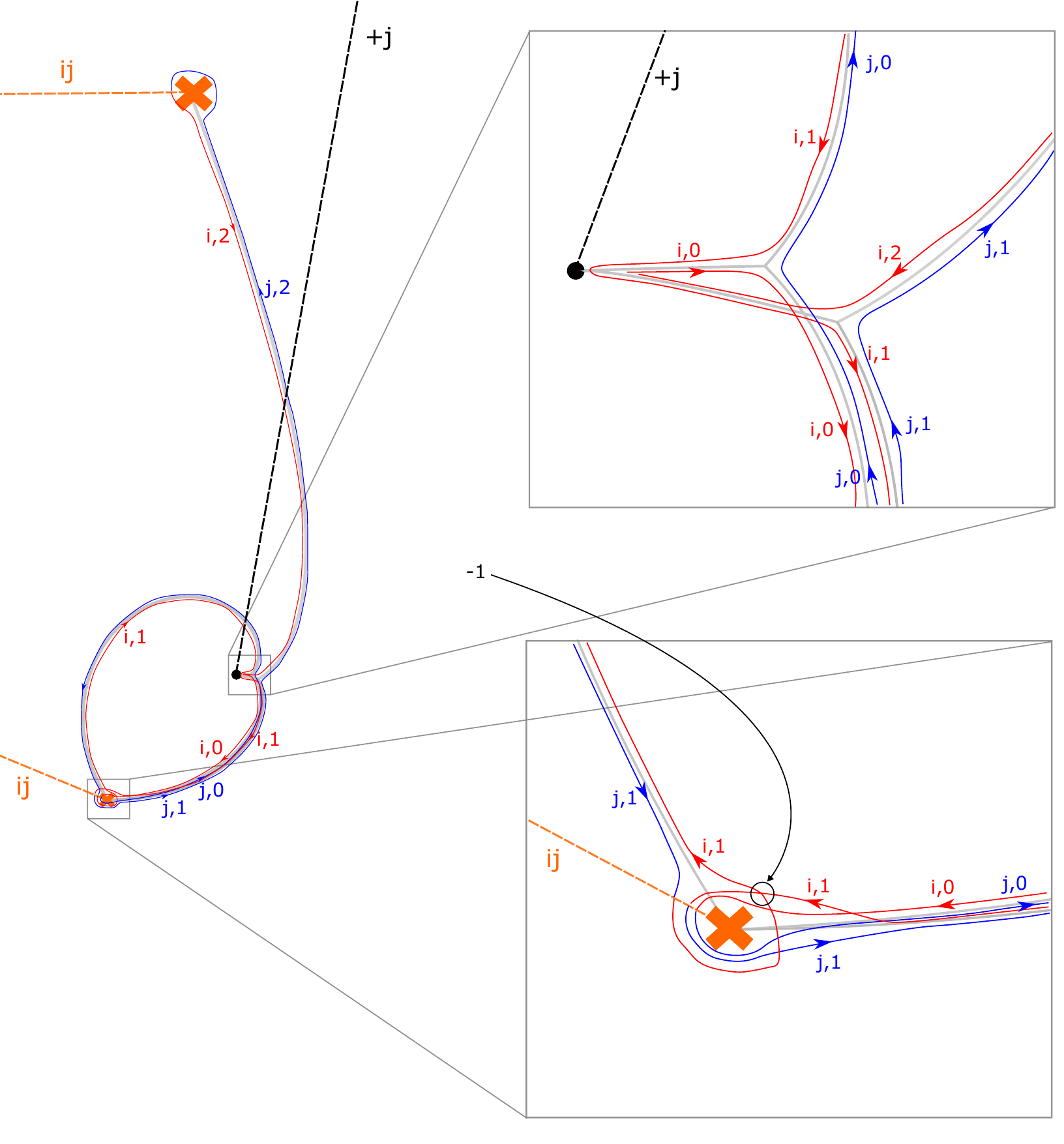}
    \caption{Concatenation of $a_1$ and $a_2^{[+1]}$ for the computation of $C_{12}$. }
    \label{fig:conifold_fm1_12}
\end{figure}

In framing $f=0$ the curve becomes $1-y-x- Q^2 xy^{-1}=0$, which is a two-sheeted covering of the $x$-plane with two polynomial branch points.
All logarithmic branch points are now at $0,\infty$ and do not play a role in sourcing solitons.
After fixing the theory point sufficiently close to $0$, we find two kinky vortices of type $(ii,1)$ with central charges 
\be
	Z_{a_1} = 2\pi i\, \log x_\theory\,,
	\qquad
	Z_{a_2} = 2\pi i\, \log (Q^2 x_\theory)\,.
\ee
which correspond to a quiver with two vertices, and fugacities
\be
	c_1 = 1\,,\qquad
	c_2 = Q^2\,.
\ee
The corresponding soliton paths, and the computation of their self- and mutual intersections are shown in Figures \ref{fig:conifold-basic-solitons} and \ref{fig:conifold_fm1_12} respectively. 
The result is the quiver adjacency matrix
\be
	C_{ij} = \langle a_i, a_j^{[+1]}\rangle = 
	 \begin{pmatrix}
            -1 & -1 \\
            -1 & 0
        \end{pmatrix}
        \,.
\ee
in perfect agreement with \eqref{eq:conifold-matrix} for $f=-1$.

\subsection{Trefoil conormal}

\subsubsection{Augmentation curve and its $\CQ$-deformation}

The augmentation curve for the conormal brane of the trefoil knot is 
\begin{equation}\label{eq:trefoil-curve}
    A(x,y,Q) = Q^6 x y^4+Q^2 y^3 \left(1-Q^2 x\right)+y \left(1-Q^2 x\right)+Q^2 x+y^2 \left(Q^4 \left(-x^2\right)-2 Q^4 x+2 Q^2 x-Q^2-1\right)\,,
\end{equation}
where $Q^2=a^2$ is the HOMFLYPT variable that corresponds to the K\"ahler parameter of the resolved conifold.
The framed version can be obtained by the change of variables $x\to x(-y)^f$ with $f\in \IZ$.\footnote{More precisely, this curve corresponds to the \emph{normalized} trefoil curve, which corresponds to the conventions in which the unknot HOMFLYPT polynomial is $1$.} The underlying quiver has three vertices and adjacency matrix \cite{Kucharski:2017ogk}
\begin{equation}\label{eq:trefoil-matrix}
   C = \begin{pmatrix}
            f & f + 1 & f + 1 \\
            f + 1 & f+2 & f + 2\\
            f + 1 & f + 2 & f + 3
        \end{pmatrix},
\end{equation}
The $\CQ$-deformed augmentation polynomial is therefore, up to an overall rescaling
\begin{equation}
\begin{split}
    A^\CQ(x,y) 
    & = c_3 y^2 \left(x y \left(c_3 y-c_2\right) (-y)^f+y-1\right)+c_1^2 x (-y)^f\\
    &-c_1 y \left(x (-y)^f \left(c_2^2 x y (-y)^f+c_2 (1-2 y)+2 c_3 y\right)+y-1\right)\,.
\end{split}
\end{equation}
One of the quiver fugacities, say $c_1$ can be absorbed into a rescaling of $x$ without loss of generality, while $c_2$ and $c_3$ are free parameters. 
The $\CQ$-deformed curve $\Sigma^\CQ$ therefore has a two-dimensional parameter space (up to rescaling of $x$) while $\Sigma$ has only one parameter $Q$.
The original augmentation curve is obtained after specializing 
\begin{equation}\label{eq:trefoil-KQ-fugacities}
    c_1 = c_2 = Q^2, \hspace{0.2cm} c_3 = Q^4\,.
\end{equation}
We have checked that our conjecture \ref{conj:QuiverSolitonCorrespondence} holds for framings $-1\leq f\leq 1$. We illustrate the case $f=-1$ below.

\subsubsection{Framing $-1$}

For definiteness, we choose the complex modulus and the theory point as follows
\begin{equation}
    Q = \frac{11}{10} + \frac{3}{5}i
    \,,\qquad
    x_\text{theory} =  \frac{e^{i \pi/3}}{1000}\,.
\end{equation}

\begin{figure}
\centering
\begin{subfigure}{.48\textwidth}
  \centering
  \includegraphics[width=1\linewidth]{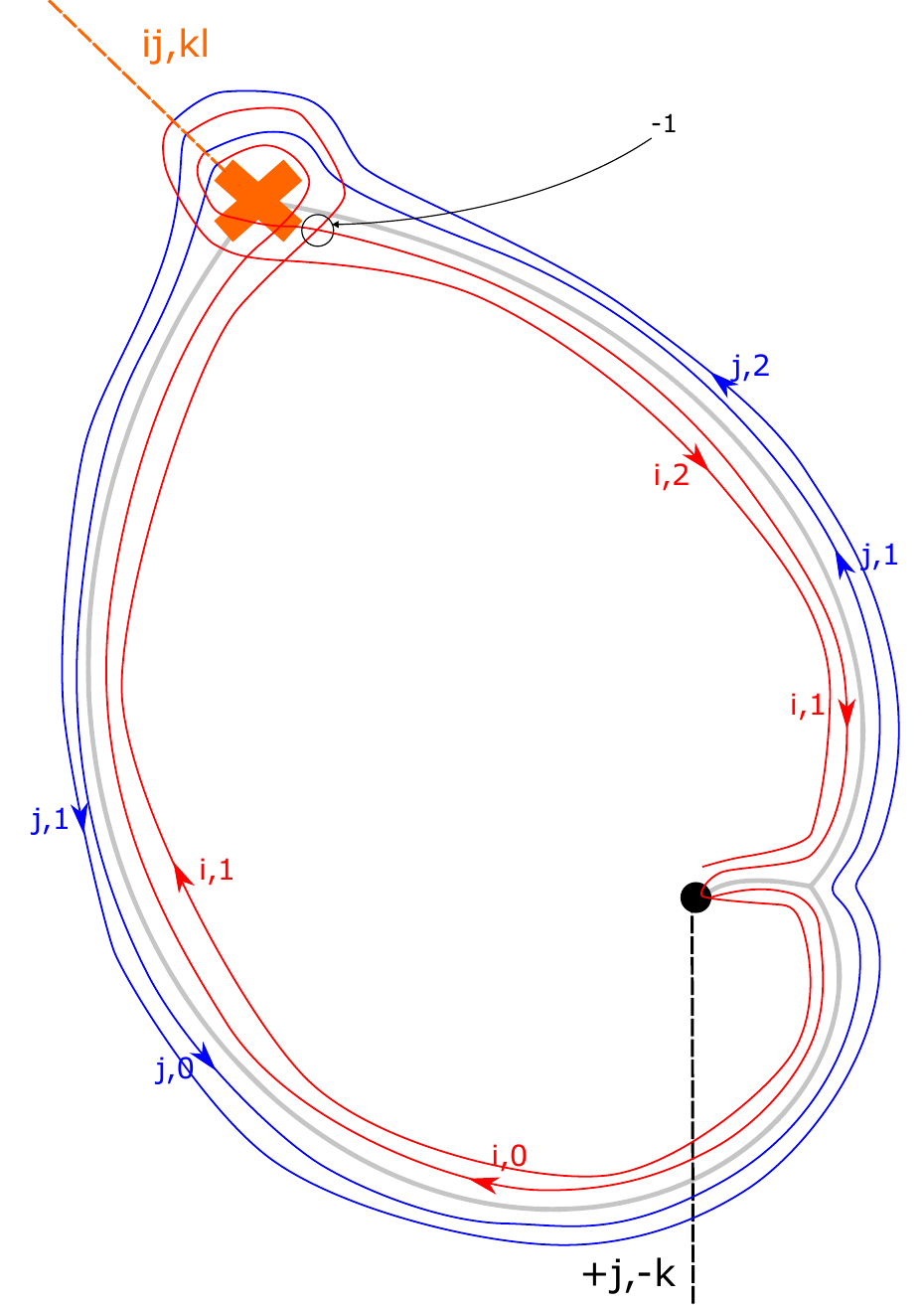}
  \caption{The soliton $a_1$ and computation of $C_{11}$.}
  \label{fig:trefoil_fm1_11}
\end{subfigure}%
\hfill
\begin{subfigure}{.48\textwidth}
  \centering
  \includegraphics[width=1\linewidth]{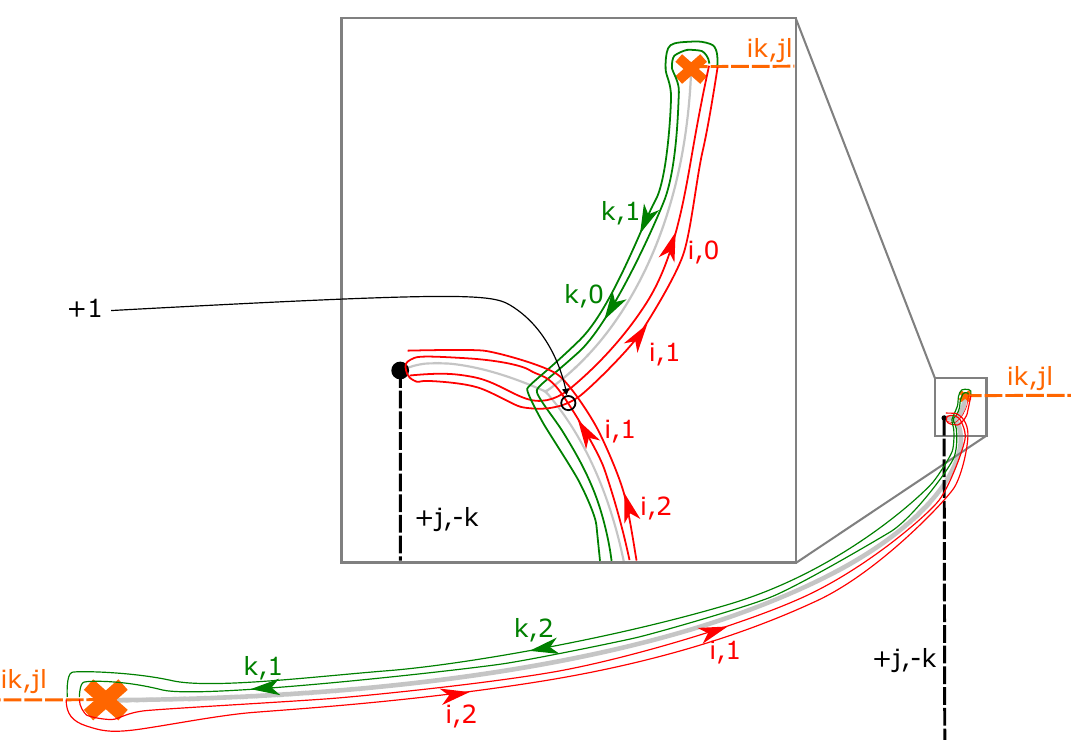}
  \caption{The soliton $a_2$ and computation of $C_{22}$.}
  \label{fig:trefoil_fm1_22}
\end{subfigure}
\begin{subfigure}{.48\textwidth}
  \centering
  \includegraphics[width=1\linewidth]{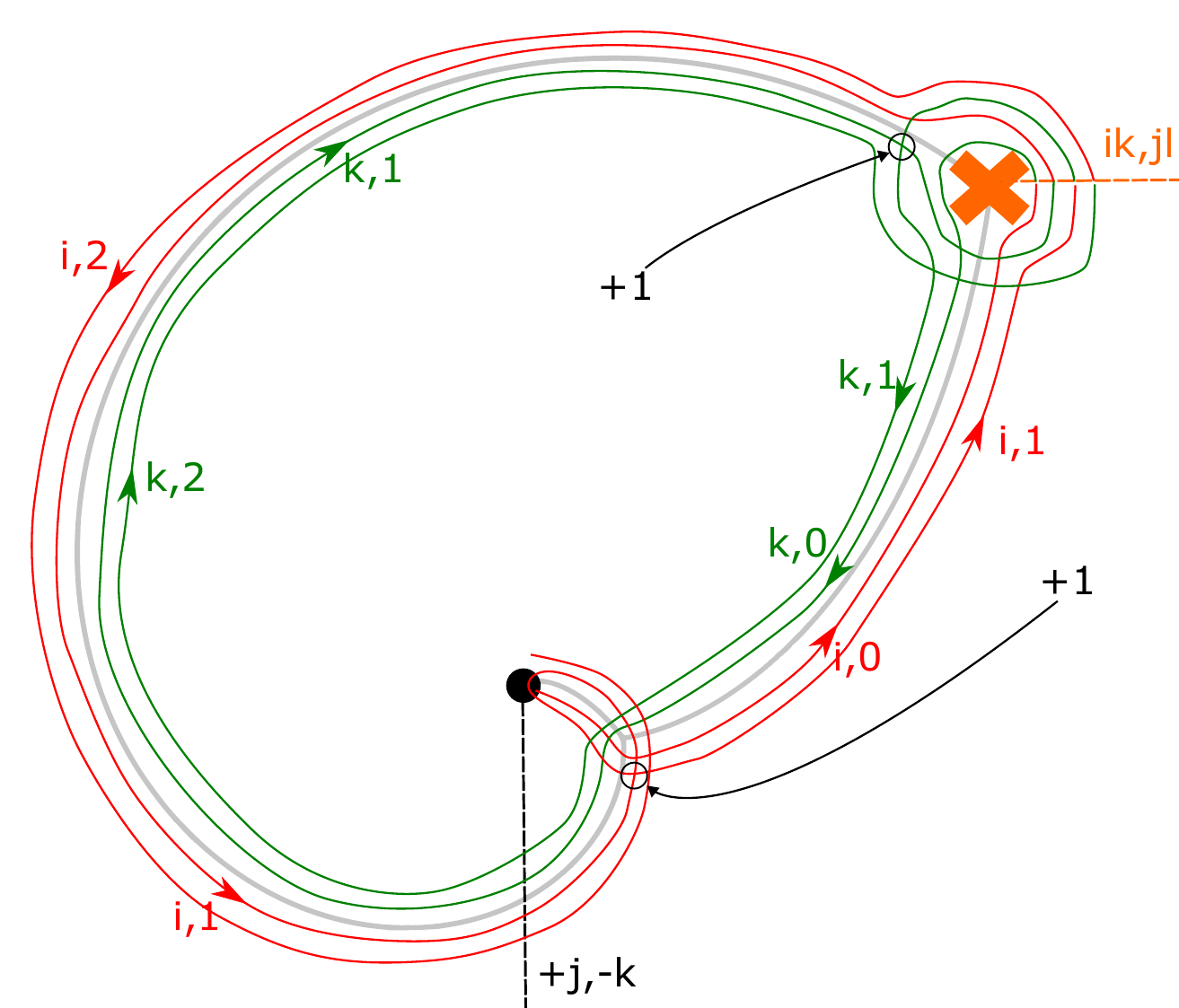}
  \caption{The soliton $a_3$ and computation of $C_{33}$.}
  \label{fig:trefoil_fm1_33}
\end{subfigure}
\caption{Soliton webs $a_i^\CW$ for the conormal lagrangian of the trefoil knot in framing $f=-1$ shown as grey network trajectories.
Resolved projections of soliton paths, and the concatenation with their own their logarithmic shifts are also shown. The black dot denotes the puncture at $x=0$.}
\label{fig:trefoil-basic-solitons}
\end{figure}

The curve \eqref{eq:trefoil-curve} has four sheets, with three quadratic polynomial branch points at $0<|x|<\infty$ and no logarithmic branch points at finite $x$. 
As we vary $\vartheta$ between $-\pi$ and $0$, we find two critical phases, where three distinct $(ii,1)$ solitons appear. The respective central charges are as follows 
\be
\begin{split}
	\vartheta_1 = \vartheta_2 \ &: \qquad Z_{a_1} = Z_{a_2} = 2\pi i \log (Q^2 x_\theory) \,, \\
	\vartheta_3 \ &: \qquad Z_{a_3} = 2\pi i \log (Q^4 x_\theory) \,. \\
\end{split}	
\ee
This is compatible with a three-vertex quiver with flavour fugacities exactly as in \eqref{eq:trefoil-KQ-fugacities}.
The three soliton paths are shown in Figure \ref{fig:trefoil-basic-solitons}.

\begin{figure}
\centering
\begin{subfigure}{.48\textwidth}
  \centering
  \includegraphics[width=1\linewidth]{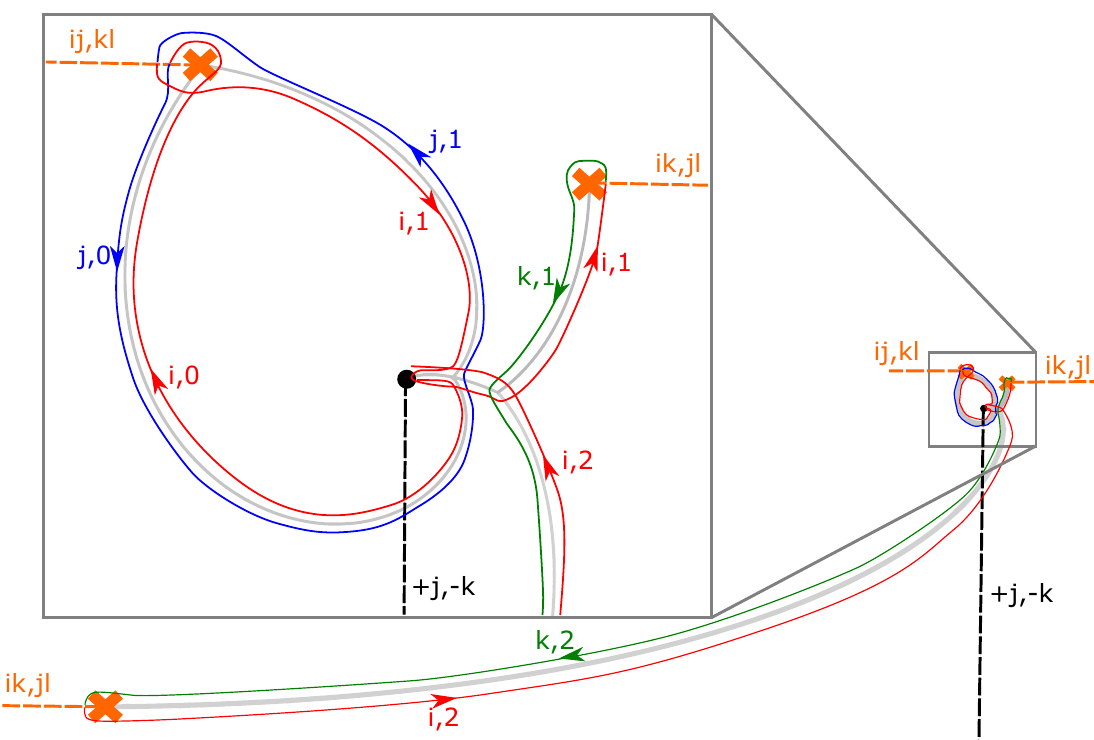}
  \caption{Computation of $C_{12}$.}
  \label{fig:trefoil_fm1_12}
\end{subfigure}%
\hfill
\begin{subfigure}{.48\textwidth}
  \centering
  \includegraphics[width=1\linewidth]{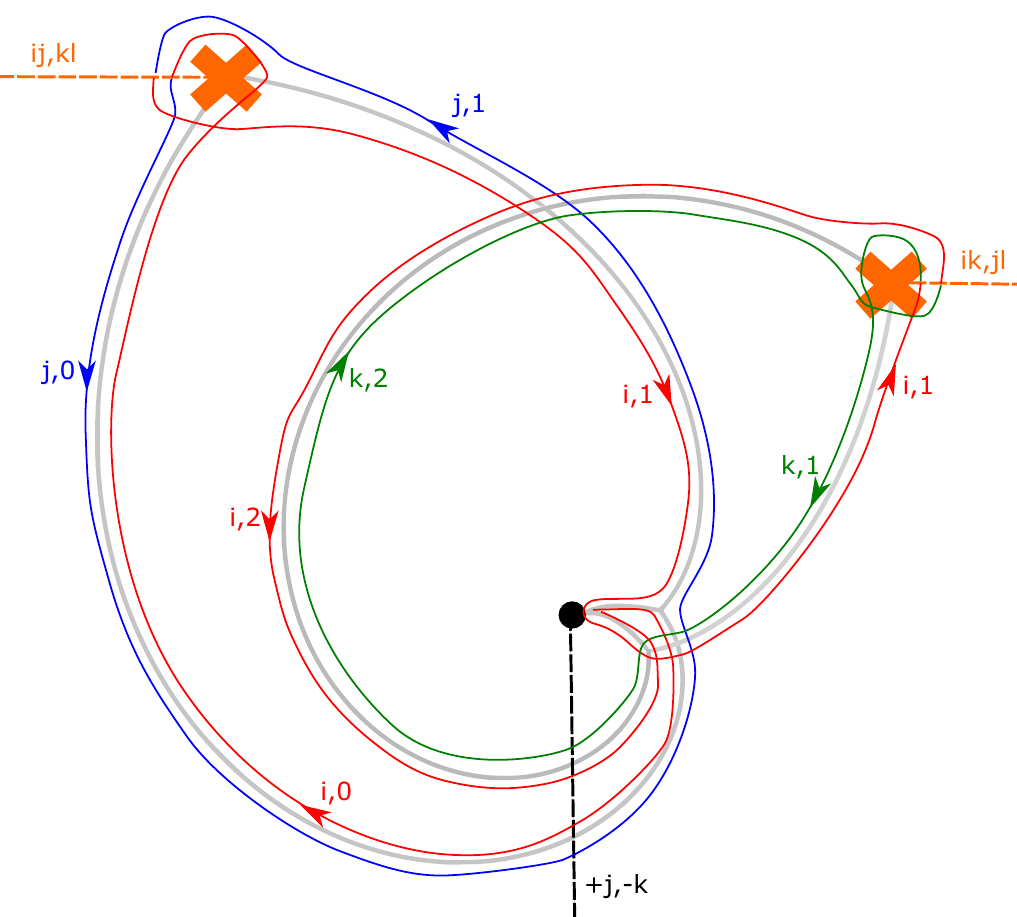}
  \caption{Computation of $C_{13}$.}
  \label{fig:trefoil_fm1_13}
\end{subfigure}
\begin{subfigure}{.48\textwidth}
  \centering
  \includegraphics[width=1\linewidth]{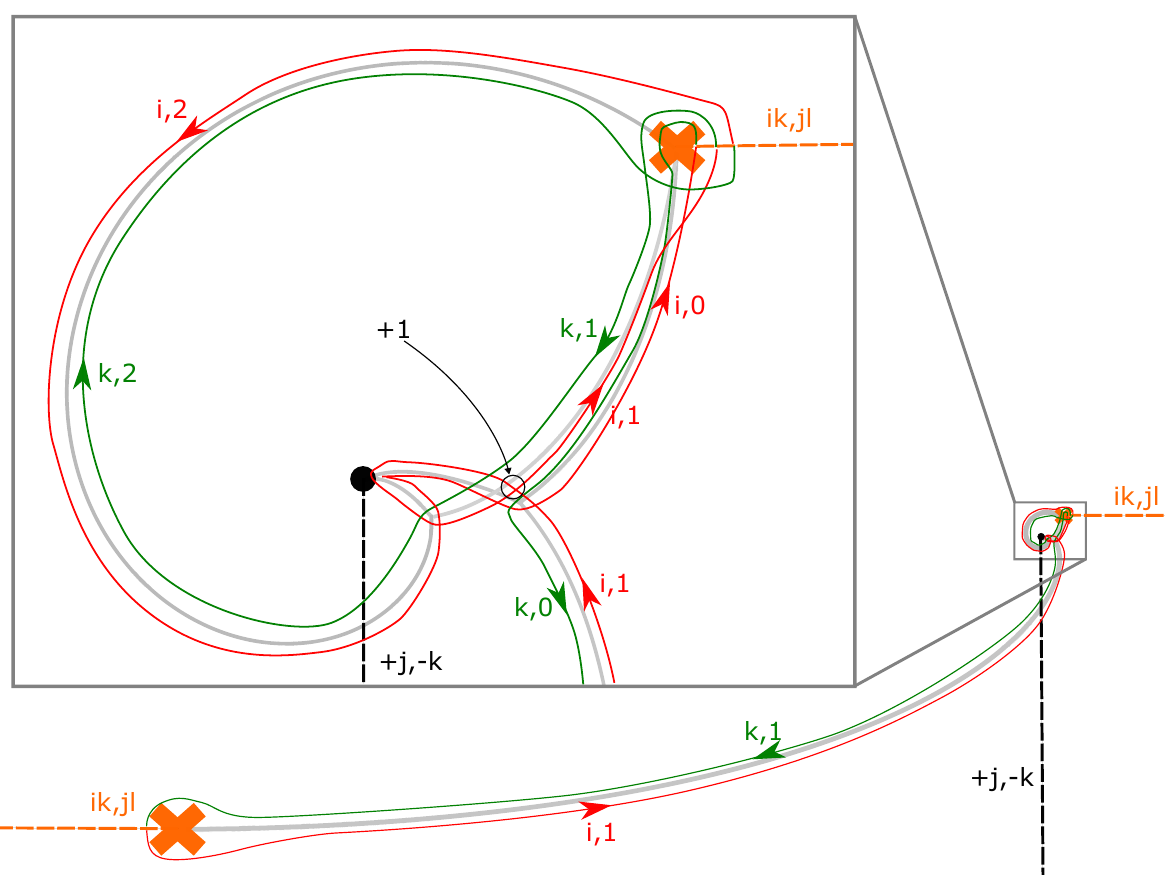}
  \caption{Computation of $C_{23}$.}
  \label{fig:trefoil_fm1_23}
\end{subfigure}
\caption{Computation of off-diagonal entries of $C_{ij}$ for the trefoil augmentation curve in framing $f=-1$. shown as grey network trajectories.}
\label{fig:trefoil-mutual-intersections}
\end{figure}

Using these, it is a tedious but straightforward exercise to compute the intersection matrix. Details of the computation are given in Figure \ref{fig:trefoil-basic-solitons} for the diagonal entries, and in Figure \ref{fig:trefoil-mutual-intersections} for the off-diagonal ones. The result is
\be
	C_{ij} = \langle a_i, a_j^{[+1]}\rangle = 
	 \begin{pmatrix}
            -1 & 0 & 0 \\
            0 & 1 & 1 \\
            0 & 1 & 2
        \end{pmatrix}
        \,.
\ee
in perfect agreement with \eqref{eq:trefoil-matrix} for $f=-1$.

\subsection{Figure-eight conormal}\label{sec:figureeight}

\subsubsection{Augmentation curve and its $\CQ$-deformation}

The augmentation curve for the conormal brane of the trefoil knot is 
\begin{equation}\label{eq:f8-curve}
\begin{split}
    A(x,y,Q)=&\frac{\left(Q^4 y^2+x \left(Q^2 y-1\right) (-y)^f\right) }{Q^6} \Big(Q^4 x^3 y^2 (-y)^{3 f}+Q^2 (y-1) y^2 \left(Q^2 y-1\right)^2\\&+Q^2 x^2 \left(Q^4 (y-2) y^4+2 y-1\right) (-y)^{2 f}+x \left(Q^2 y-1\right) \left(Q^6 y^5-2 Q^4 y^4+2 Q^2 y-1\right) (-y)^f\Big)
    \,,
\end{split}
\end{equation}
where $Q^2=a^2$ is the HOMFLYPT variable that corresponds to the K\"ahler parameter of the resolved conifold.
Note that the curve has two factors, and the distinguished vacuum $y_i(x)$ belongs to the second one, however for reasons that will become clear shortly, it is important to keep both to obtain the full quiver description.\footnote{This is also natural in view of the fact that  quantum effects, such as string instantons with Euler characteristic $\chi<1$, spoil the factorization property at the level of the quantum curve $\hat A(\hat x, \hat y, Q)$.}
The framed version can be obtained by the change of variables $x\to x(-y)^f$ with $f\in \IZ$.\footnote{As for the trefoil knot, this corresponds to the \emph{normalized} augmentation curve.} The underlying quiver has five vertices and adjacency matrix \cite{Kucharski:2017ogk}
\begin{equation}\label{eq:f8-matrix}
   C = \begin{pmatrix}
            f & f & f-1 & f & f-1 \\
            f & f+2 & f & f+1 & f-1\\
            f-1 & f & f-1 & f & f-2\\
            f & f+1 & f & f+1 & f-1\\
            f-1 & f-1 & f-2 & f-1 & f-2
        \end{pmatrix},
\end{equation}
The $\CQ$-deformed augmentation polynomial is therefore, up to an overall rescaling 
\begin{equation}\label{eq:f8-Q-curve}
\begin{split}
    A^\CQ(x,y) 
    & = 
	x c_2^3 y^8+c_2^2 (2 c_1 c_4 x^2-(c_1+c_3+c_4) x+1) y^7-c_2 ((x c_4-1) ((x c_1-2) c_3\\&+x (-x c_1^2+c_1+c_3) c_4)+c_2 (x c_1 (x (c_1+c_3)-2)+2 x (c_3+c_5)+1)) y^6\\&-(x (4 x c_1-3) c_5 c_2^2+(x c_1 c_4 (x c_1-1){}^2-2 x c_3^2+x (-x c_1^2+c_1-x c_4^2+c_4) c_5\\&+c_3 (x (c_4+c_1 (x c_1 (x c_4-2)+4)-2 c_5)-2)) c_2+c_3 (x c_4-1){}^2 (x c_1 c_4-c_3)) y^5\\&+(x c_2 c_5^2+x (2 x^2 c_2 c_1^3-x c_2 (2 x c_4+3) c_1^2+(5 x c_3 c_2+c_2+c_3+x (c_2-c_3) c_4) c_1\\&+c_3 c_4 (1-x c_4)+c_2 (c_4+c_3 (2 x c_4-3))) c_5+c_3 (x (x^2 c_2 c_1^3+x (c_4 (x c_4-1)\\&-2 c_2) c_1^2+(c_2 (2 x c_3+1)+(x c_4-1) ((x c_3-1) c_4-2 c_3)) c_1+c_3 (c_2+c_4))-c_3)) y^4\\&+x (-((2 x (c_4-c_1) c_2+c_2+c_3) c_5^2)-c_3^2 (c_3+c_1 (-x c_1+x (x c_1-2) c_4+1))\\&+(x^3 c_2 c_1^4+x (c_2+2 c_3) c_1^2+x^2 (-2 c_2 c_1^2+c_3 (3 c_2-2 c_4) c_1+2 c_3 c_4^2) c_1-2 x c_3^2 c_4\\&-c_3 (c_1+c_2+c_4)) c_5) y^3-x (x c_1 c_3^3+(x c_1 (-x c_1^2+c_1+c_3+x (c_1 (x c_1-1)\\&+2 c_3) c_4)-c_3) c_5 c_3+(x (-2 x c_2 c_1^2+2 (c_2+c_3) c_1+c_3 (c_2-2 c_4))-c_3) c_5^2) y^2\\&+x^2 c_5 (c_5 (c_3^2+2 c_1 c_3+c_2 c_5)-x c_1 c_3 (c_4 c_5+c_1 (c_3+c_5))) y\\&+x^2 c_3 (x c_1 c_3-c_5) c_5^2.
\end{split}
\end{equation}

One of the quiver fugacities, say $c_1$ can be absorbed into a rescaling of $x$ without loss of generality, while $c_2,\dots, c_5$ are free parameters. 
The $\CQ$-deformed curve $\Sigma^\CQ$ therefore has a four-dimensional parameter space (up to rescaling of $x$) while $\Sigma$ has only one parameter $Q$.
The original augmentation curve is obtained after specializing 
\begin{equation}\label{eq:f8-KQ-fugacities}
    c_1 = c_3 = c_4 = 1\,,
    \qquad
    c_2 = (c_5 )^{-1} =  Q^2\,.
\end{equation}

\subsubsection{Framing $0$}

For illustration purposes we will focus on the case of framing $f=0$.
As it turns out, the original curve is degenerate as can be seen from the factorization in \eqref{eq:f8-curve}, and it will be convenient to work instead with the $\CQ$-deformed curve \eqref{eq:f8-Q-curve} with the following choice of moduli
\be\label{eq:f8-fugacities-deformed}
	c_1 = 1+\epsilon_1,\hspace{0.2cm} c_2 = Q^2 + \epsilon_2, \hspace{0.2cm} c_3 = 1+\epsilon_3, c_4 = 1, c_5 = Q^{-2} + \epsilon_5\,,
\ee
where for definitess we fix
\be
	Q = \frac{2}{3}+\frac{4}{5}i,\hspace{0.2cm}\epsilon_1 = -\frac{1+i}{1000},\hspace{0.2cm}\epsilon_2 = -\frac{1769+9600i}{900000},\hspace{0.2cm}\epsilon_3 = \frac{1-2i}{1000},\hspace{0.2cm}\epsilon_5 = -\frac{618750+3378721 i}{372100000}\,.
\ee
We moreover fix the theory point as follows
\begin{equation}
    x_\text{theory} = \frac{e^{i\pi/3}}{1000000}.
\end{equation}

\begin{figure}
\centering
\begin{subfigure}{.28\textwidth}
  \centering
  \includegraphics[width=1\linewidth]{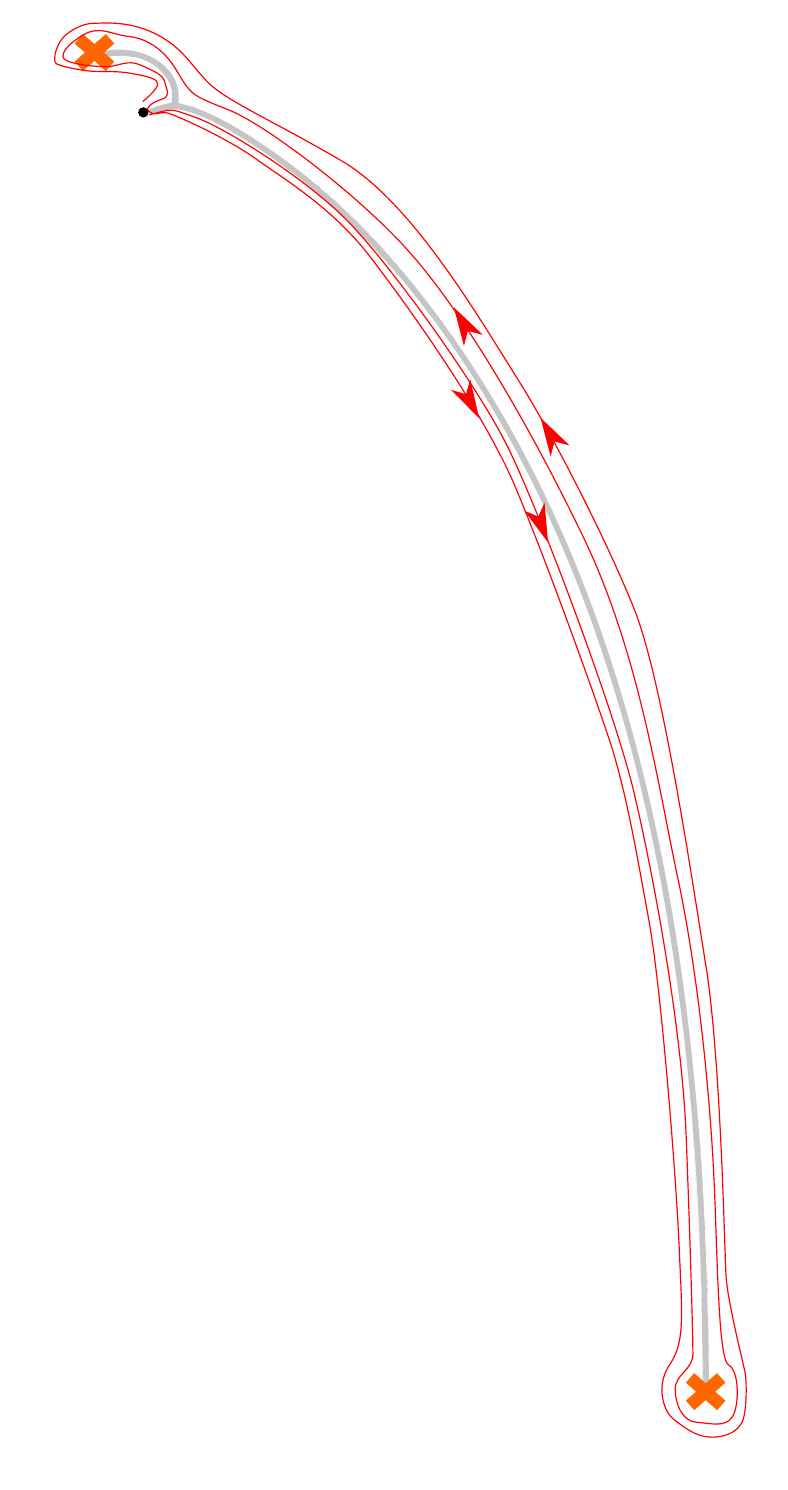}
  \caption{The soliton $a_1$ and computation of $C_{11}$.}
  \label{fig:figureeight_f0_11}
\end{subfigure}%
\hfill
\begin{subfigure}{.28\textwidth}
  \centering
  \includegraphics[width=1\linewidth]{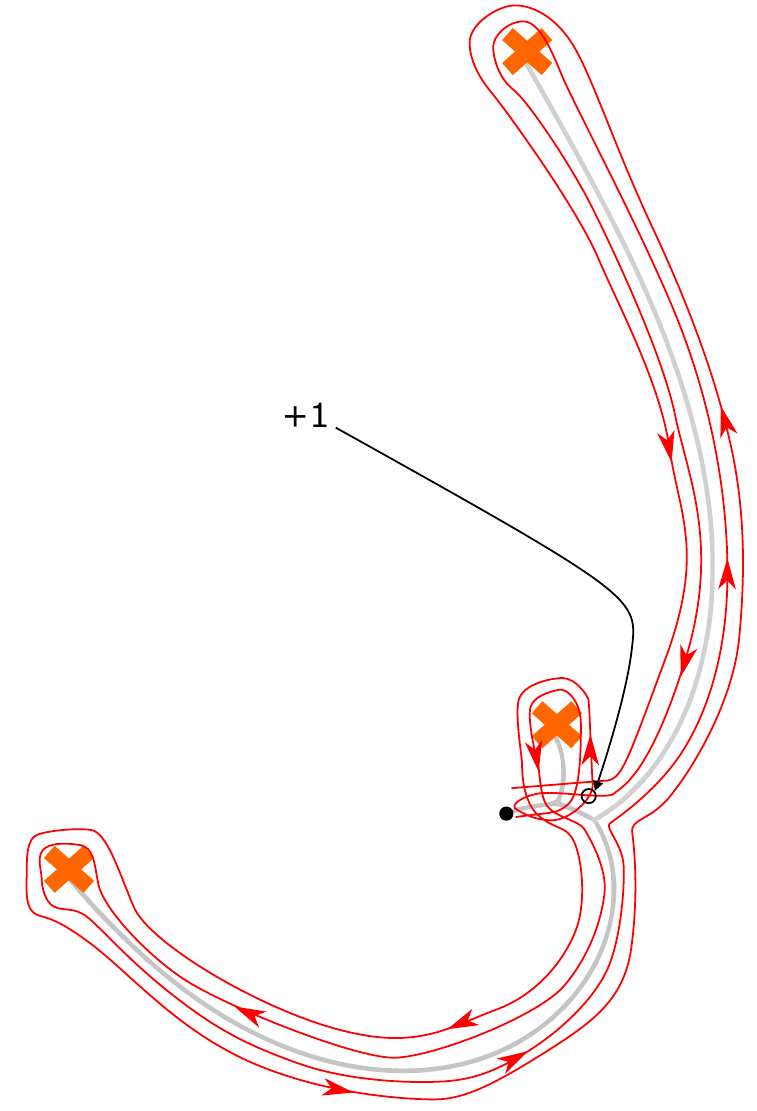}
  \caption{The soliton $a_4$ and computation of $C_{44}$.}
  \label{fig:figureeight_f0_44}
\end{subfigure}
\hfill
\begin{subfigure}{.35\textwidth}
  \centering
  \includegraphics[width=1\linewidth]{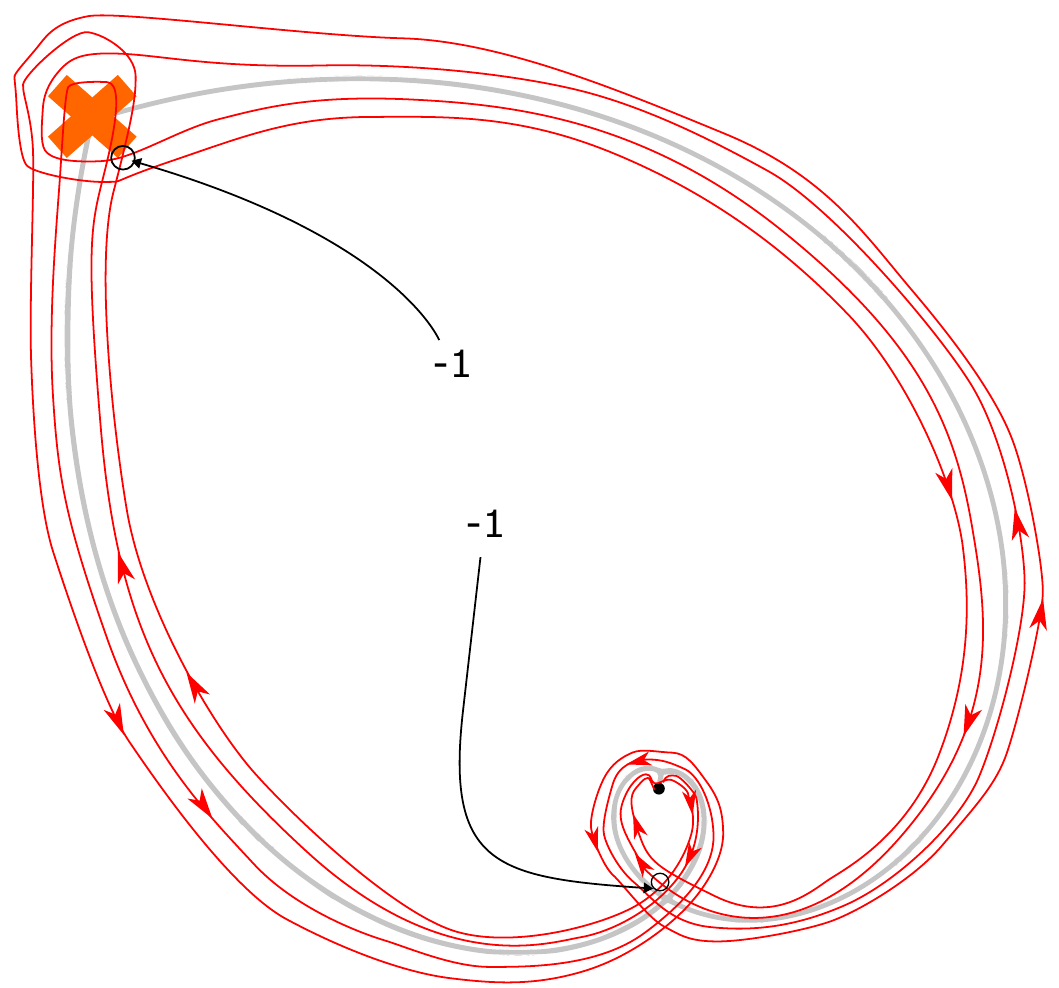}
  \caption{The soliton $a_5$ and computation of $C_{55}$.}
  \label{fig:figureeight_f0_55}
\end{subfigure}
\\
\begin{subfigure}{.48\textwidth}
  \centering
  \includegraphics[width=1\linewidth]{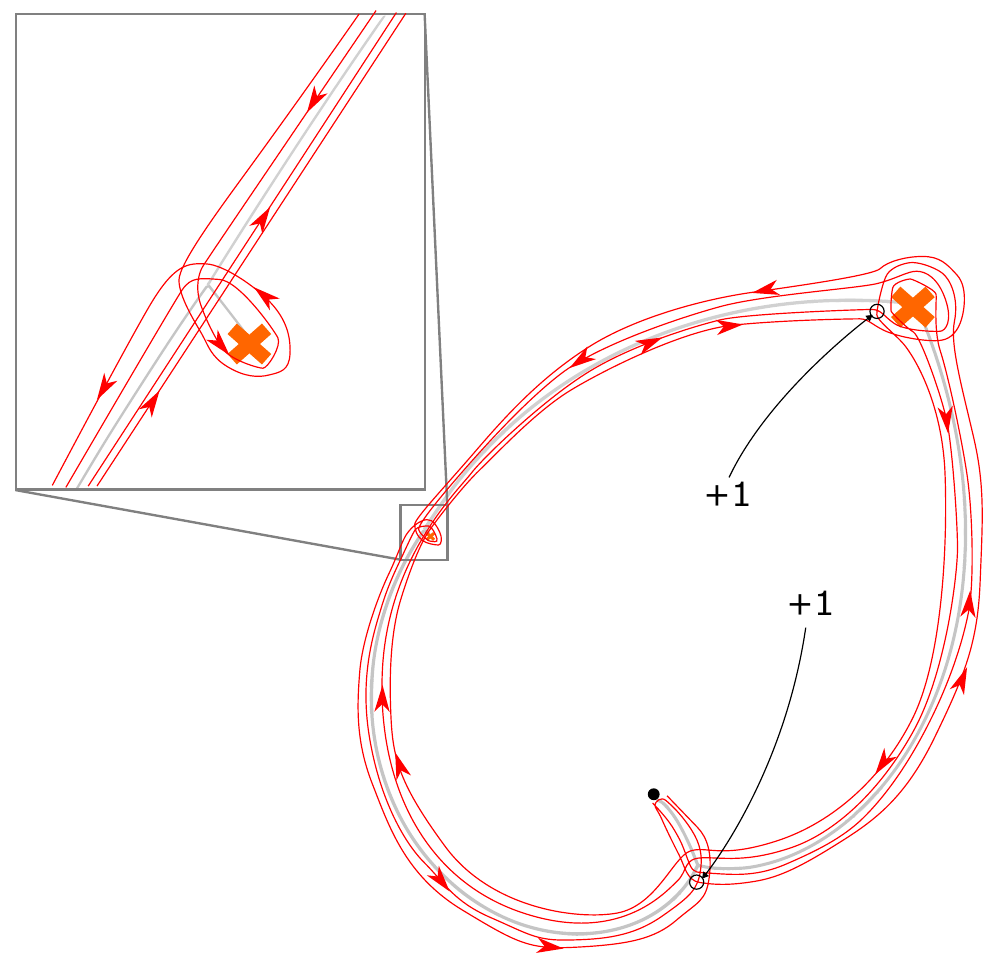}
  \caption{The soliton $a_2$ and computation of $C_{22}$.}
  \label{fig:figureeight_f0_22}
\end{subfigure}
\hfill
\begin{subfigure}{.48\textwidth}
  \centering
  \includegraphics[width=1\linewidth]{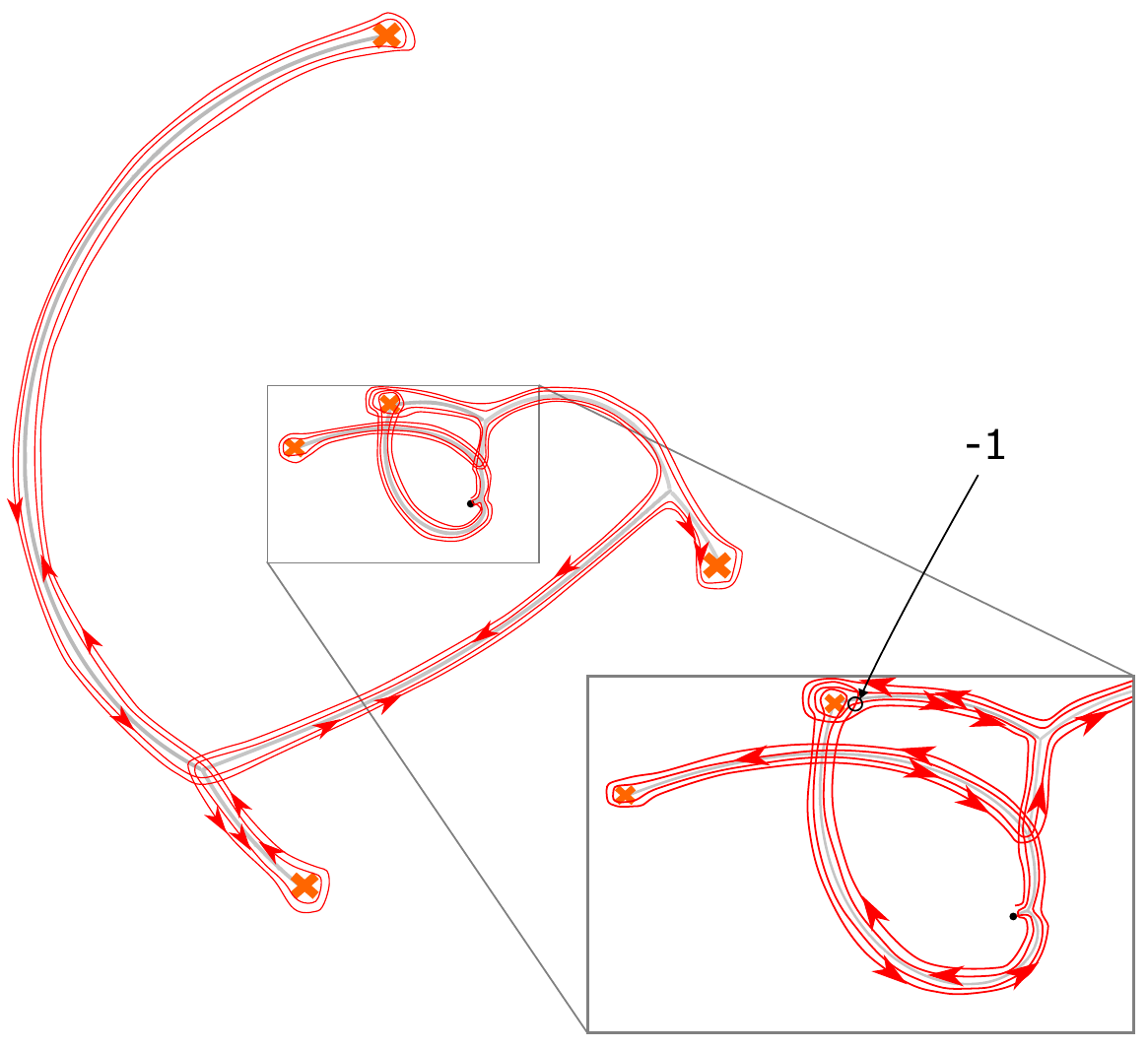}
  \caption{The soliton $a_3$ and computation of $C_{33}$.}
  \label{fig:figureeight_f0_33}
\end{subfigure}
\caption{Soliton webs $a_i^\CW$ for the conormal lagrangian of the figure eight knot in framing $f=0$ shown as grey network trajectories.
Resolved projections of soliton paths, and the concatenation with their own their logarithmic shifts are also shown. The black dot denotes the puncture at $x=0$.}
\label{fig:f8-basic-solitons}
\end{figure}

The corresponding augmentation curve is rather complicated: it has 8 sheets, with 15 polynomial branch points (all of which are quadratic) and one logarithmic branch point at finite $x$. 
As we vary $\vartheta$ between $-\pi$ and $0$, we find five critical phases, where five distinct $(ii,1)$ solitons appear. 
The respective central charges are as follows
\be
\begin{split}
	\vartheta_k =  2\pi i \log (c_k\, x_\theory) \,, \\
\end{split}	
\ee
with $c_k$ precisely as given in \eqref{eq:f8-fugacities-deformed}. 
Thus our algorithm associates to $\Sigma^\CQ$ a five-vertex quiver with flavour fugacities exactly as in \eqref{eq:trefoil-KQ-fugacities}, providing another highly nontrivial test of Conjecture \ref{conj:QuiverSolitonCorrespondence}.
The five soliton paths are shown in Figure \ref{fig:f8-basic-solitons}.

Using these, it is a tedious but straightforward exercise to compute the intersection matrix. 
Details of the computation\footnote{%
Rather than fixing a global choice of trivialization for the logarithmic covering $\tSigma^\CQ$
 of the exponential curve and then analytically tracking the polynomial and logarithmic branch labels of trajectories as we have done for the unknot and trefoil, we used here the numerical solutions to the differential equation for trajectory evolution given by \eqref{eq:E-walls} to check where the soliton paths intersect. 
For this reason we do not include in the figures the branch cuts and we do not need to label trajectories explicitly (as we have been doing till now), but we can still compute intersections of soliton paths. 
} are given in Figure \ref{fig:f8-basic-solitons} for the diagonal entries, and in Figures
\ref{fig:figureeight_f0_12}, \ref{fig:figureeight_f0_13}, \ref{fig:figureeight_f0_14}, \ref{fig:figureeight_f0_15}, \ref{fig:figureeight_f0_23}, \ref{fig:figureeight_f0_24}, \ref{fig:figureeight_f0_25}, \ref{fig:figureeight_f0_34}, \ref{fig:figureeight_f0_35} and \ref{fig:figureeight_f0_45}.
for the off-diagonal ones. The result is
\be
	C_{ij} = \langle a_i, a_j^{[+1]}\rangle = 
	 \begin{pmatrix}
            0 & 0 & -1 & 0 & -1 \\
            0 & 2 & 0 & 1 & -1\\
            -1 & 0 & -1 & 0 & -2\\
            0 & 1 & 0 & 1 & -1\\
            -1 & -1 & -2 & -1 & -2
        \end{pmatrix}
        \,.
\ee
in perfect agreement with \eqref{eq:f8-matrix} for $f=0$.

\appendix

\section{Orbital angular momentum of Chern-Simons vortices}\label{app:interactionspin}
In this section, we will explain how mixed Chern-Simons coupling contributes to the orbital angular momentum of the 2-vortex bound state. Our focus will be on Maxwell-Chern-Simons theories with gauge group $U(1)_1\times U(1)_2 \times \dots \times U(1)_m$ defined by the classical Lagrangian

\begin{equation}
    \mathcal{L} = -\frac{1}{4e^2}F^{(i)}_{\mu \nu}F^{(i)\mu \nu} + \frac{\kappa_{ij}}{4\pi}\epsilon^{\mu \nu \rho} A^{(i)}_\mu \partial_\nu A^{(j)}_\rho - A^{(i)}_\mu \mathcal{J}^{(i)\mu}\,,
\end{equation}

where $i,j$ denote the index of the gauge group, $e$ is the electric charge, $\kappa_{ij}$ is the mixed Chern-Simons coupling, $A_
\mu^{(i)}$ is the gauge field, $F_{\mu \nu}^{(i)}$ is the corresponding gauge field strength and $\mathcal{J}^{(i)\mu}$ is a non-dynamical current. We will use $\Vec{E}_i = (F^{(i)}_{10},F^{(i)}_{20})$ and $B_i = F^{(i)}_{12}$ to denote the electric and magnetic field respectively of the gauge group $U(1)_i$. Then, the corresponding equations of motion are given by

\begin{equation}
    -\frac{1}{e^2} \partial_\nu F^{(i)\nu \mu} = - \mathcal{J}^{(i)\mu} + \frac{\kappa_{ij}}{2\pi}\epsilon^{\mu\nu\rho}\partial_\nu A_\rho^{(j)}\,.
\end{equation}

Setting $\mu = 0$, and focussing on the case of vortices, ie. $\mathcal{J}^{(i)0} = 0$, we recover the well-known mixing of electric and magnetic fluxes induced by Chern-Simons couplings
\begin{equation}
    \Vec{\nabla}\cdot\Vec{E}_i = -\frac{e^2\kappa_{ij}}{2\pi}B_j\,.
\end{equation}

More specifically, we want to consider background currents that model vortices $V_i$ with unit vorticity for the gauge group $U(1)_i$, since these serve as models for the basic vortices considered in this paper.\footnote{Indeed, in Chern-Simons-Matter theories where vortices are dynamical, taking the limit $x_\theory \to 0$ causes vortices to become point-like \cite{Gupta:2024ics}.} 
A vortex $V_i$ located at the origin $\Vec{x} = 0$ therefore sources both magnetic and electric fields 
\begin{equation}
    B_{j}(\Vec{x}) = 2\pi\delta_{ij}\delta(\Vec{x}) \implies \Vec{E}_j(\Vec{x}) = -\frac{e^2 \kappa_{ij}}{2\pi |\Vec{x}|^2} \Vec{x}\,.
\end{equation}

Now, we want to determine the contribution of mixed Chern-Simons coupling to the orbital angular momentum of the 2-vortex system. Hence, we consider the situation where we have two fundamental vortices $V_i$ and $V_j$ at $\Vec{x}_1$ and $\Vec{x}_2$ respectively. Then

\begin{equation}
    B_k(\Vec{x}) = 2\pi\left(\delta_{ki} \delta(\Vec{x}-\Vec{x}_1)+\delta_{kj} \delta(\Vec{x}-\Vec{x}_2)\right),\hspace{0.2cm} \Vec{E}_k = -\frac{e^2}{2\pi} \left(\frac{\kappa_{ki}(\Vec{x}-\Vec{x_1})}{|\Vec{x}-\Vec{x_1}|^2}+\frac{\kappa_{kj}(\Vec{x}-\Vec{x_2})}{|\Vec{x}-\Vec{x_2}|^2}\right)\,.
\end{equation}

\begin{figure}[h!]
\centering
\begin{subfigure}{.5\textwidth}
  \centering
  \includegraphics[width=0.9\linewidth]{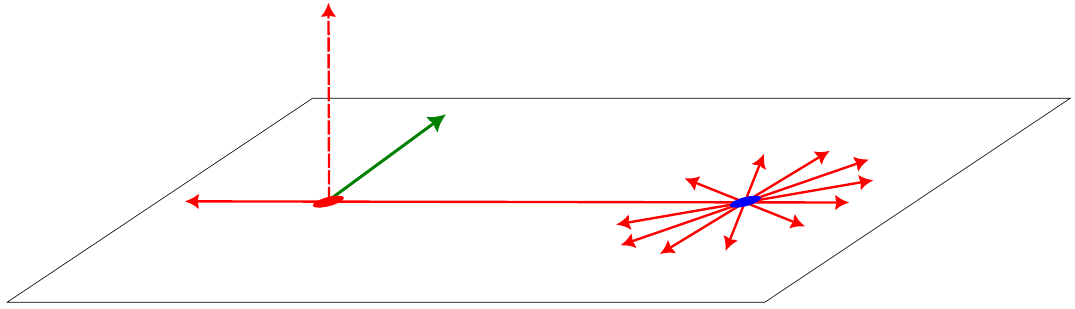}
  \caption{Orbital angular momentum contribution from $U(1)_i$}
  \label{fig:interactionspin2}
\end{subfigure}%
\begin{subfigure}{.5\textwidth}
  \centering
  \includegraphics[width=0.9\linewidth]{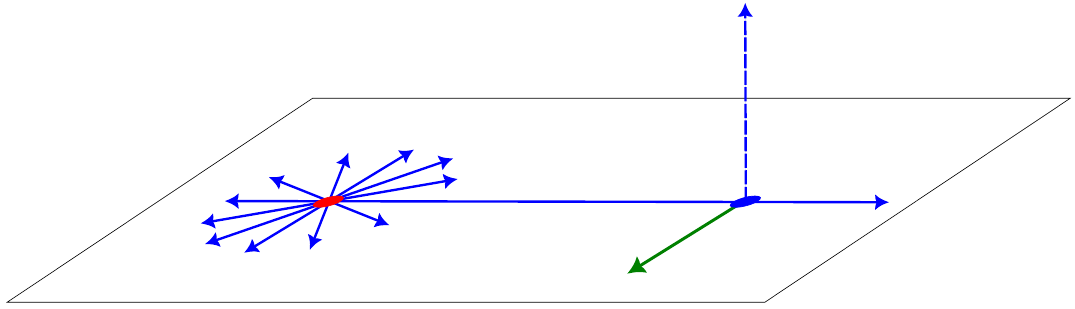}
  \caption{Orbital angular momentum contribution from $U(1)_j$}
  \label{fig:interactionspin1}
\end{subfigure}
\caption{$V_i$ and $V_j$ are shown as red and blue dots respectively. The electric field lines for $U(1)_i$ ($U(1)_j$) are shown in solid red (blue) and the magnetic field lines in dashed red (blue). The green arrow indicates the Poynting vector induced by the interaction between the two vortices.}
\label{fig:orbitalcontribution}
\end{figure}

Since we are in two spatial dimensions, the contribution of electro-magnetic fields to the orbital angular momentum is given by

\begin{equation}
    J_{\text{orbital}} = -\frac{1}{e^2}\int_{\mathbb{R}^2} d^2 x \left[(\Vec{x}\cdot\Vec{E}_k) B_k - \text{self-interactions}\right] \,,
\end{equation}

where we have to get rid of the self-interaction contributions because they do not contribute to the orbital angular momentum. The remaining contributions are due to the interactions between magnetic field of one vortex with the induced electric field of the other vortex for both $U(1)_i$ and $U(1)_j$, and we show these contributions in Figure \ref{fig:orbitalcontribution}.

Then, a simple substitution shows that the orbital angular momentum of the 2-vortex bound state is given precisely by the mixed Chern-Simons coupling
\begin{equation}\label{eq:J-orbit}
    J_{\text{orbital}} = \kappa_{ij}\,.
\end{equation}

\section{Intersections of projected paths}\label{app:intersections}

Finding intersections of concatenated projected paths corresponding to different phases and away from branch points is fairly simple to do as should be clear from the main text. However, when either of these assumptions is not true, then finding intersections can be subtle, and we explain both of these cases here.

\subsection{Around branch points}\label{app:intersections1}
Branch points can be of either logarithmic type or polynomial type.

\subsubsection{Logarithmic branch points}
\begin{figure}[h!]
\centering
\begin{subfigure}{.4\textwidth}
  \centering
  \includegraphics[width=0.7\linewidth]{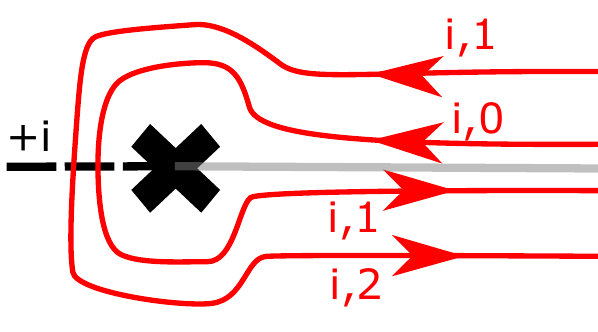}
  \caption{Positive counterclockwise Monodromy}
  \label{fig:positivelog}
\end{subfigure}%
\begin{subfigure}{.4\textwidth}
  \centering
  \includegraphics[width=0.7\linewidth]{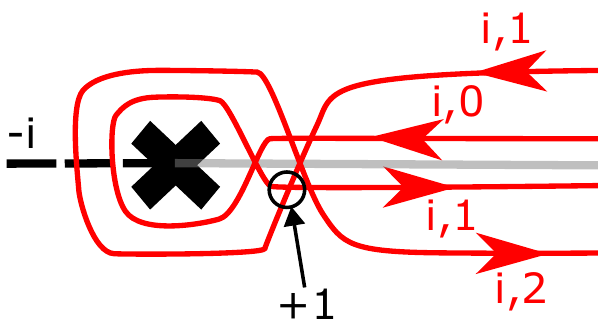}
  \caption{Negative counterclockwise Monodromy}
  \label{fig:negativelog}
\end{subfigure}
\caption{Two possible monodromies and the corresponding concatenated soliton paths around a logarithmic branch point.}
\label{fig:logarithmicBranchPoint}
\end{figure}

Since we want to focus on the case of $x_\text{theory} \to 0$, or more specifically, $\vartheta \to -\frac{\pi}{2}$, it suffices to consider here just the case of coincident trajectories originating from a logarithmic branch point. Then, we have two cases depending on the counterclockwise monodromy around the logarithmic branch point, and we show these in Fig \ref{fig:logarithmicBranchPoint}. We observe that in the case of positive monodromy, we get $0$ intersection, while in the case of negative monodromy, we get $+1$ intersection number. This is because we have to change the direction of the soliton path around the logarithmic branch point as we go from positive monodromy to negative monodromy, but this still needs to be compatible with our rule of resolving soliton paths so that for a trajectory of type $(ii,1)$, the corresponding $(i,n+1)$ projected path is on the right and $(i,n)$ projected path is on the left with respect to the direction of the trajectory.

\subsubsection{Polynomial branch points}

\begin{figure}[h!]
\centering
\begin{subfigure}{.25\textwidth}
  \centering \includegraphics[width=0.8\linewidth]{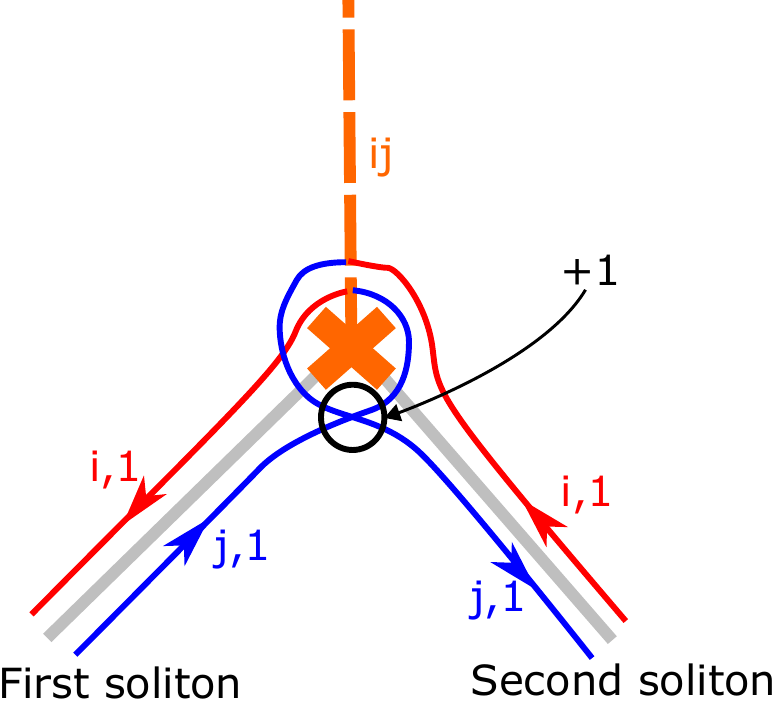}
\end{subfigure}%
\begin{subfigure}{.25\textwidth}
  \centering \includegraphics[width=0.8\linewidth]{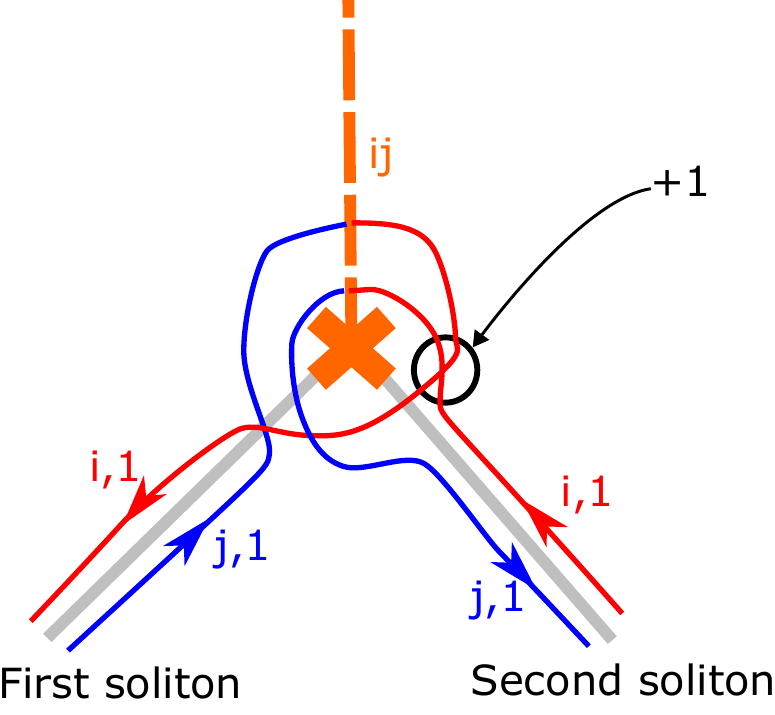}
\end{subfigure}%
\begin{subfigure}{.25\textwidth}
  \centering \includegraphics[width=0.8\linewidth]{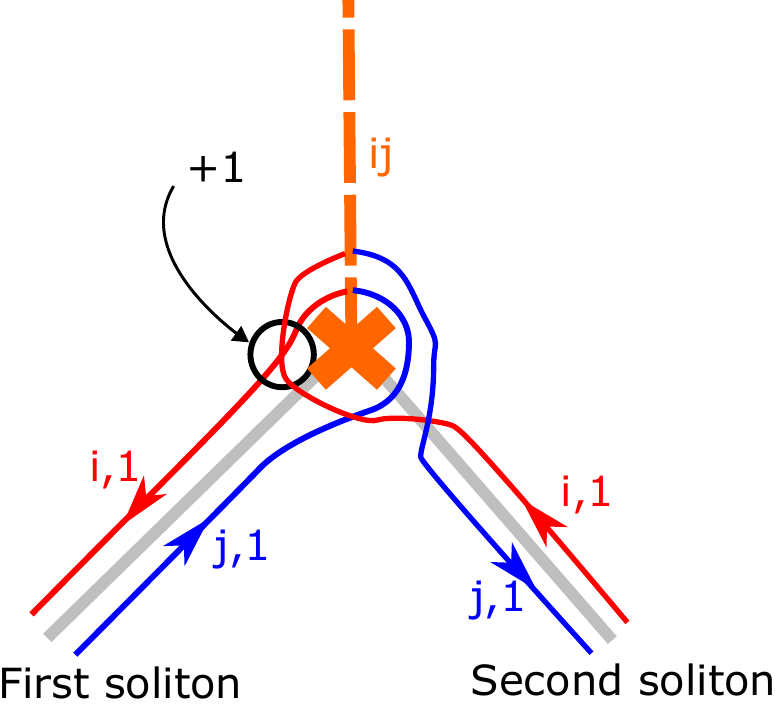}
\end{subfigure}%
\begin{subfigure}{.25\textwidth}
  \centering \includegraphics[width=0.8\linewidth]{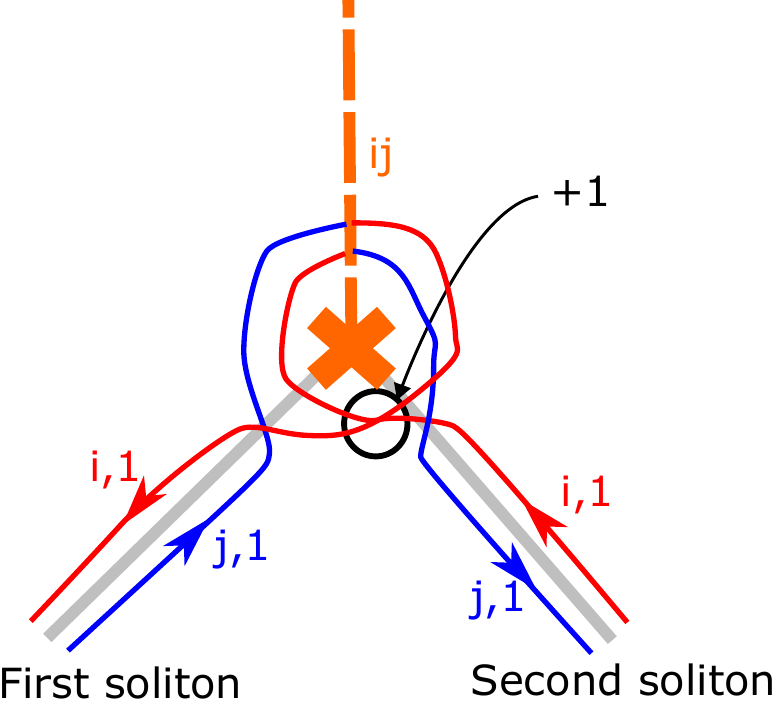}
\end{subfigure}

\caption{Different cases for intersection number $+1$ around a branch point}
\label{fig:poly1}
\end{figure}

Polynomial branch points are characterised by their degree. In all the examples we have considered till now, we only encountered polynomial branch points of degree two, so we will focus on these. Because these branch points are of degree two, the order in which we go around them shouldn't matter, hence for the intersection number of projected paths to be well defined, it should also be invariant under changing the monodromy direction around degree two branch points. Around a branch point, we can have two possible non-trivial intersections: $+1$ or $-1$. We show the first case in Fig \ref{fig:poly1}. The second case is the same as the first case, with the ``First'' and ``Second'' solitons interchanged. Since for any direction around the branch point, the intersection number is invariant, we conclude that the direction in which we go around degree two branch points is immaterial, which is what we wanted. 

The generalisation to calculating intersections around higher degree branch points is straighforward, although computationally more challenging since we would now have to take care of the monodromy direction around the branch point, and also because for an $n$-degree branch point, the number of trajectories generated from it are $n^2-1$, i.e. their number grows quadratically.

\subsection{Solitons with the same phase}\label{app:intersections2}

\begin{figure}[h!]
\centering
\boxed{
\begin{subfigure}{.48\textwidth}
\includegraphics[width=0.44\linewidth]{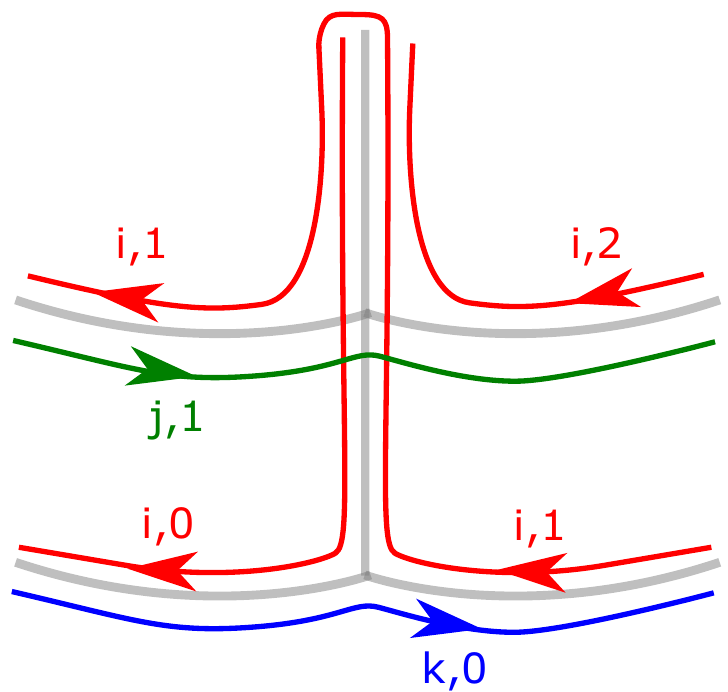}
\hfill
\includegraphics[width=0.44\linewidth]{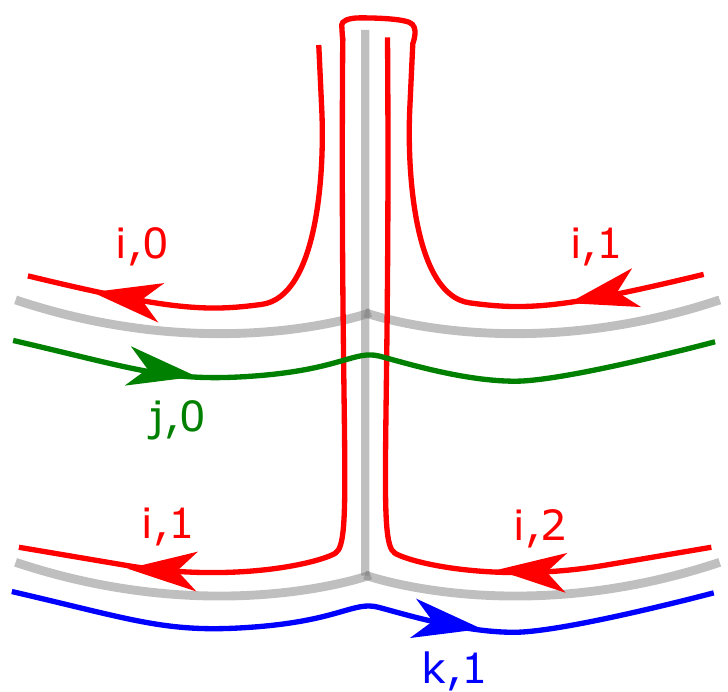}
\caption{}
\label{fig:nn}
\end{subfigure}
}
\hfill
\boxed{
\begin{subfigure}{.48\textwidth}
\includegraphics[width=0.44\linewidth]{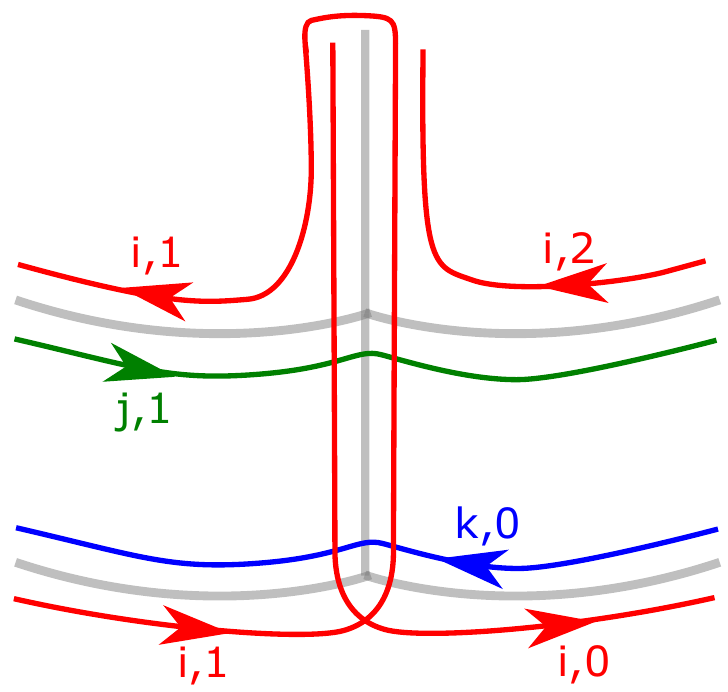}
\hfill
\includegraphics[width=0.44\linewidth]{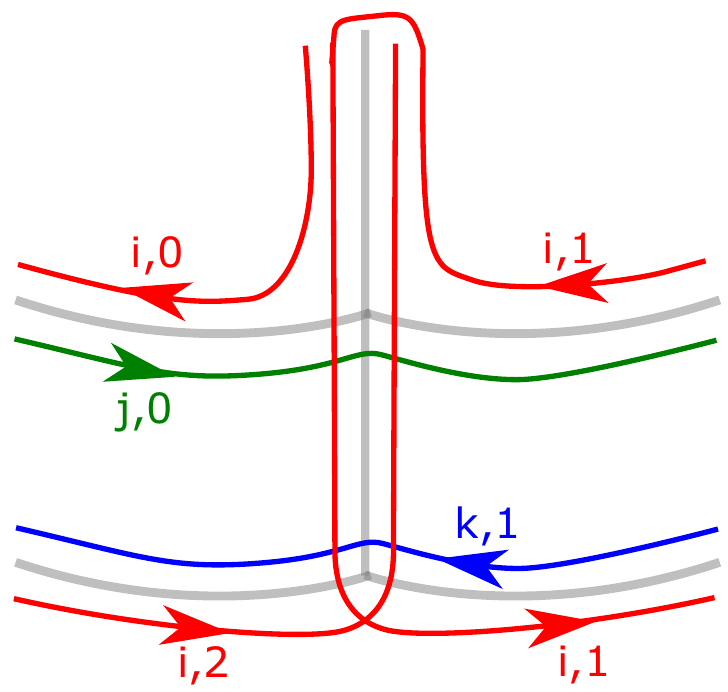}
\caption{}
\label{fig:np}
\end{subfigure}
}
\\
\boxed{
\begin{subfigure}{.48\textwidth}
\includegraphics[width=0.44\linewidth]{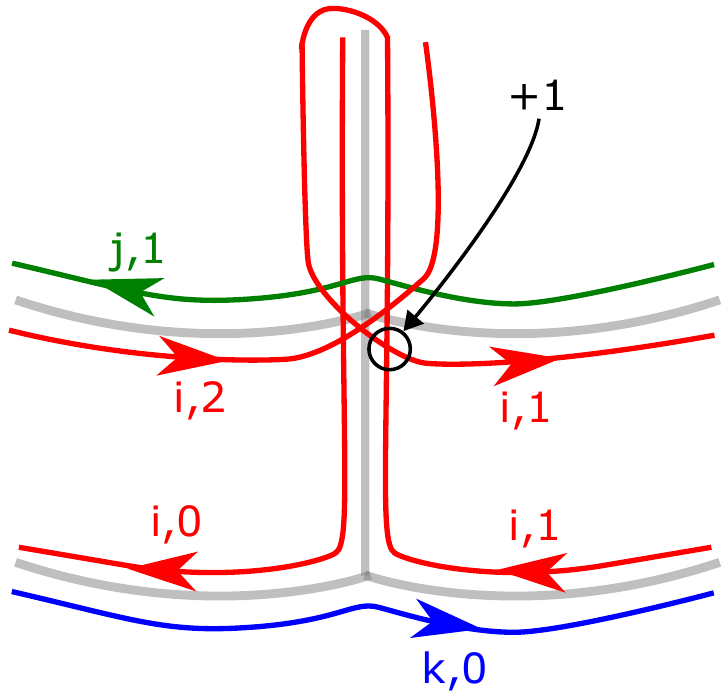}
\hfill
\includegraphics[width=0.44\linewidth]{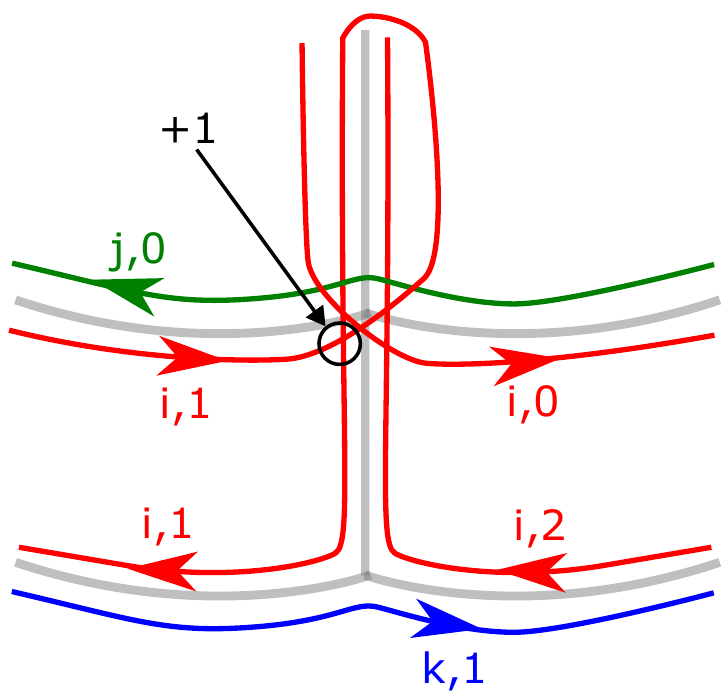}
\caption{}
\label{fig:pn}
\end{subfigure}
}
\hfill
\boxed{
\begin{subfigure}{.48\textwidth}
\includegraphics[width=0.44\linewidth]{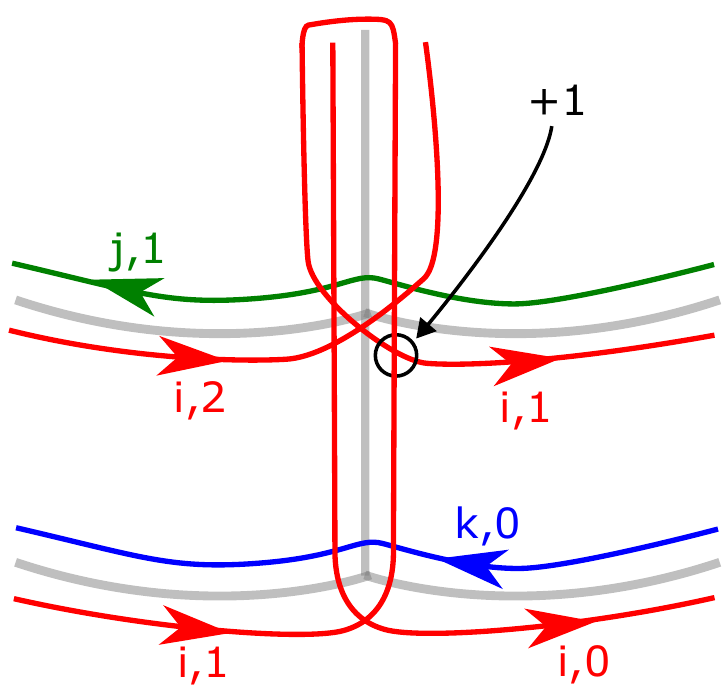}
\hfill
\includegraphics[width=0.44\linewidth]{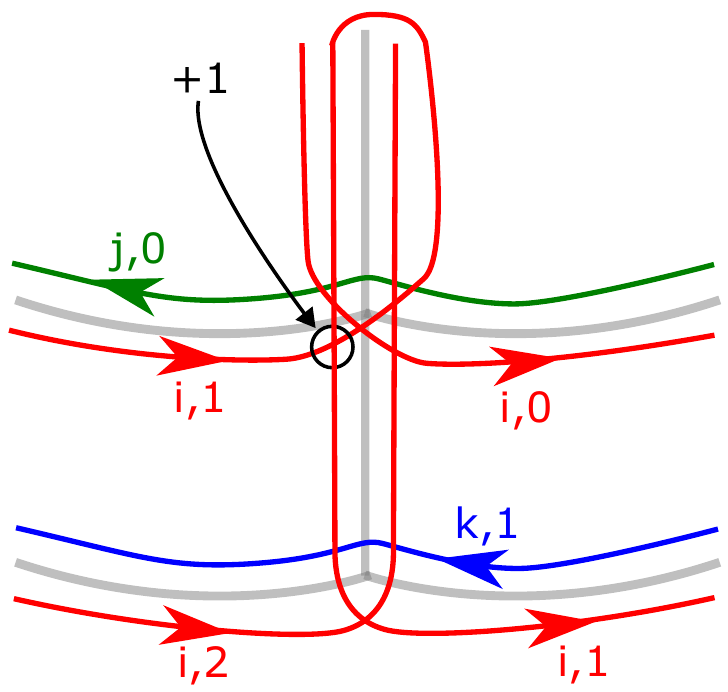}
\caption{}
\label{fig:pp}
\end{subfigure}
}
\caption{Different concatenations for solitons with the same phase.}
\end{figure}

Another ambiguity arises in the case of coincident trajectories, or equivalently when we combine two solitons with the same phase. In this case, we need the intersection number to be invariant under the change of ordering of solitons while concatenating. Generically, there are four cases with two possible concatenation orderings in each case, and we show them and the corresponding intersections in Figures \ref{fig:nn}, \ref{fig:np}, \ref{fig:pn} and \ref{fig:pp}. Clearly, in all these cases, the ordering doesn't change the intersection number. 

\section{Boson-fermion cancellations of CFIV indices}\label{sec:CFIVcancellations}
We observe CFIV cancellations among the trajectories of the network that ensure that corresponding to each vertex of a quiver corresponding to some augmentation curve, there exists only a single $(ii,1)$ kinky vortex. Hence, it is important to not just plot the network but also to keep track of the corresponding soliton data. In this section, we give an explicit example of this phenomenon in the case of figure eight. For presentation purposes, we will work with a different example than Section \ref{sec:figureeight}.

Consider the figure eight augmentation curve with the KQ change of variables as given in Eq \eqref{eq:f8-curve}. We fix

\begin{equation}
    Q = \frac{4}{9}+\frac{4}{15}i, \hspace{0.2cm} x_\theory = \frac{e^{i \pi/3}}{1000000}\,,
\end{equation}

and focus on the exponential network at phase 

\begin{equation}
    \vartheta_2 = \arg (2\pi i \log (c_2 x_\theory))\,.
\end{equation}

\begin{figure}[h!]
\centering
\begin{subfigure}{.5\textwidth}
  \centering \includegraphics[width=0.9\linewidth]{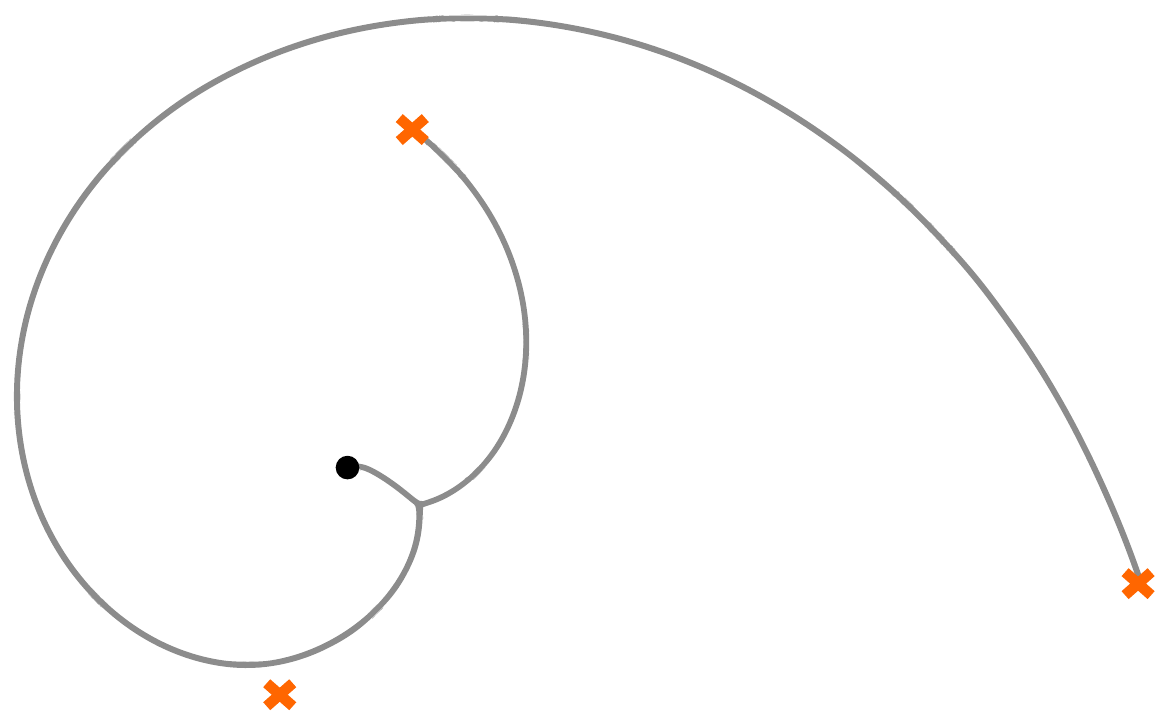}
  \label{fig:cancel1}
\end{subfigure}%
\begin{subfigure}{.5\textwidth}
  \centering \includegraphics[width=0.9\linewidth]{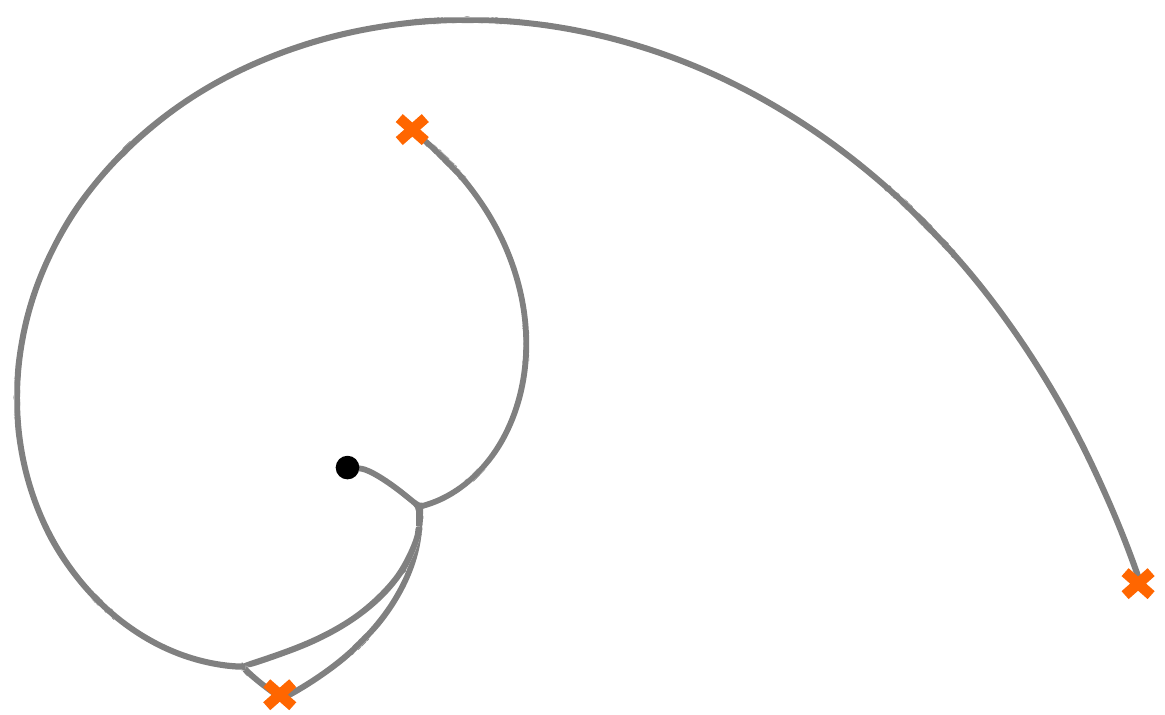}
  \label{fig:cancel2}
\end{subfigure}
\caption{Soliton-antisoliton webs.}
\label{fig:solitonantisolitonwebs}
\end{figure}

Then, we observe that along with the usual soliton $a_2$ corresponding to this phase, we get a soliton-antisoliton pair with the corresponding webs shown in Fig \ref{fig:solitonantisolitonwebs}. 

\section{Off-diagonal intersections of basic solitons for the figure-eight augmentation curve}

\begin{figure}
    \centering   \includegraphics[width=0.4\linewidth]{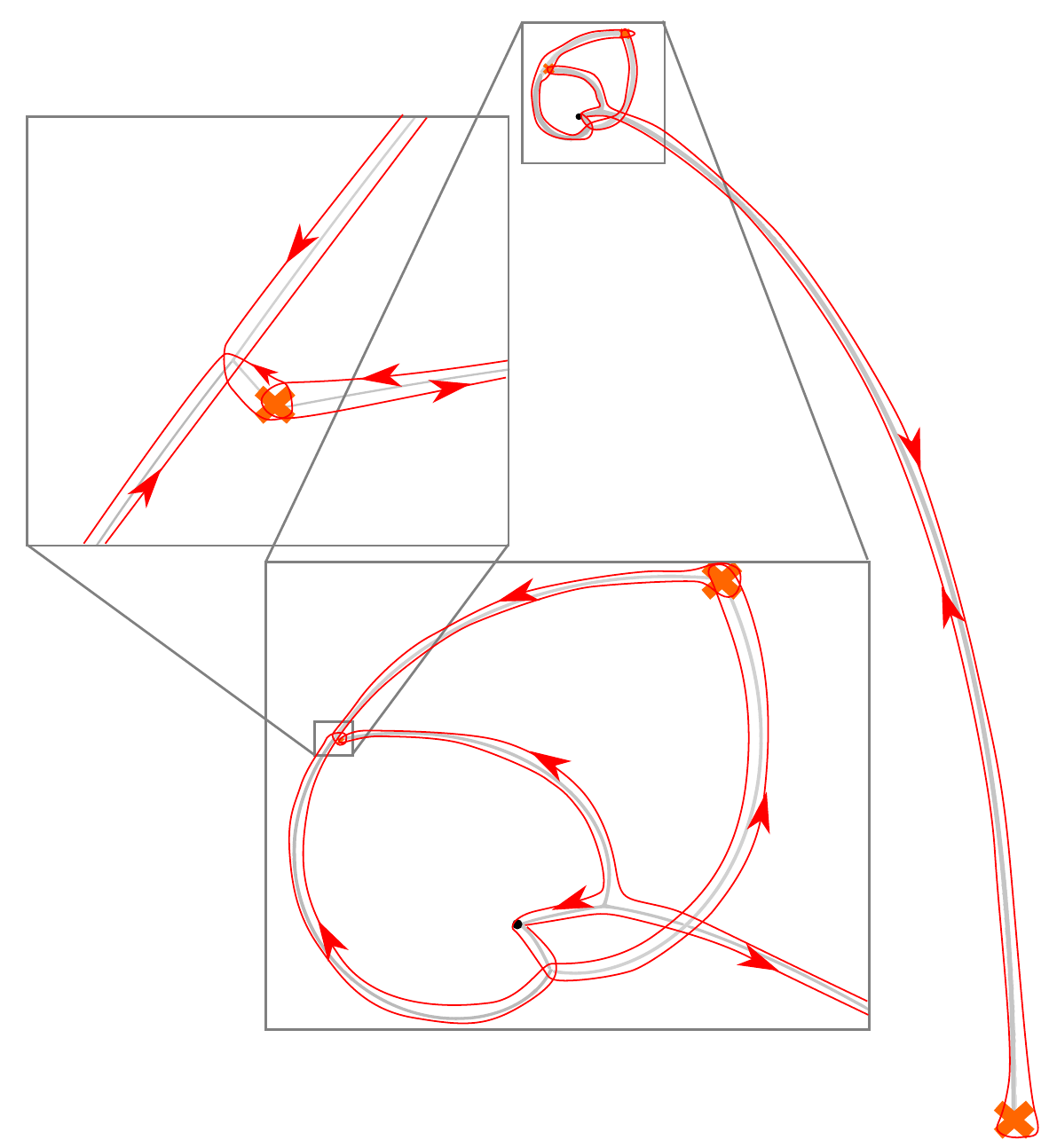}
    \caption{Concatenation of $a_1$ and $a_2^{[+1]}$ for the computation of $C_{12}$.}
    \label{fig:figureeight_f0_12}
\end{figure}

\begin{figure}
    \centering   \includegraphics[width=0.7\linewidth]{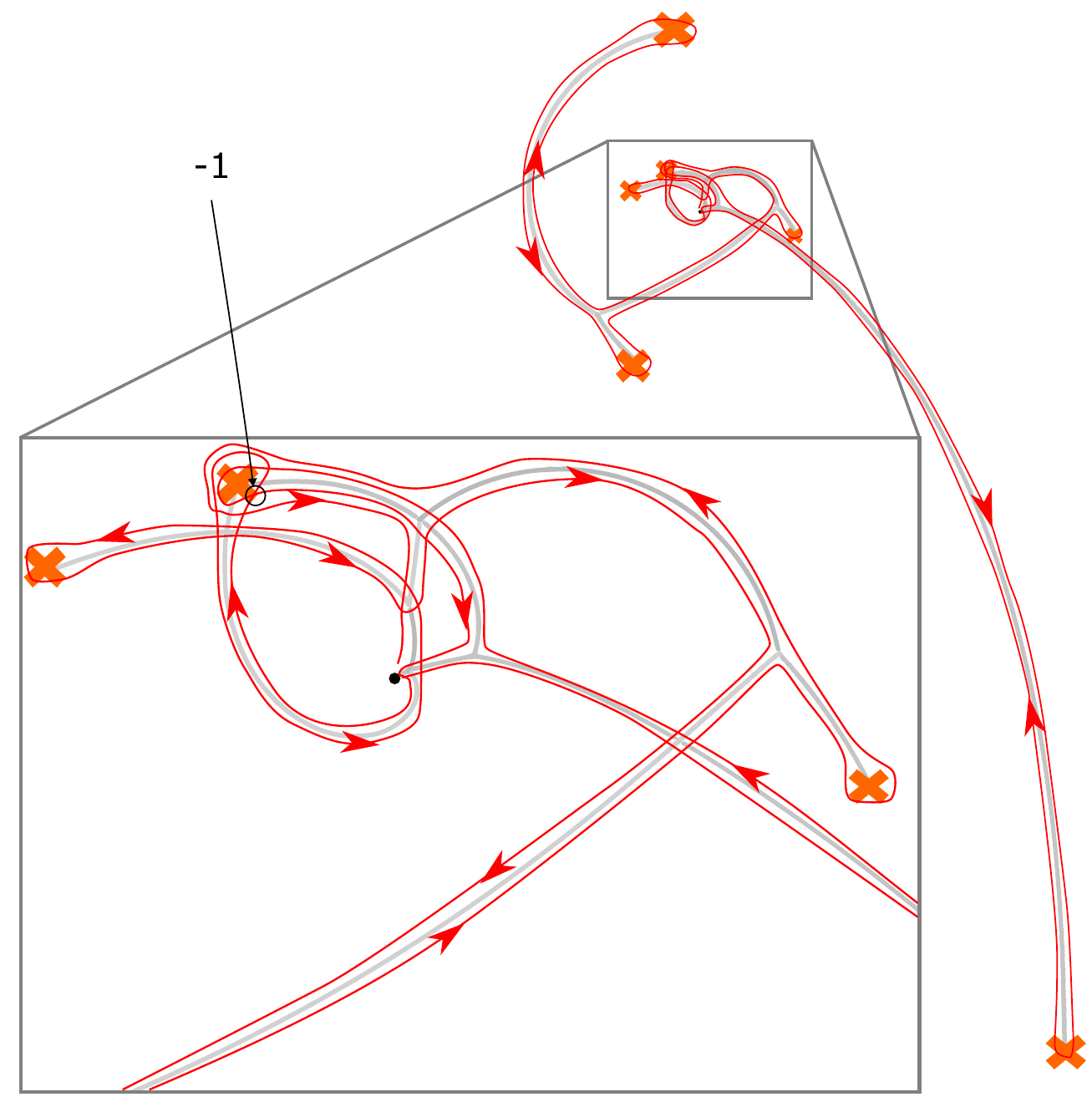}
    \caption{Concatenation of $a_1$ and $a_3^{[+1]}$ for the computation of $C_{13}$.}
    \label{fig:figureeight_f0_13}
\end{figure}

\begin{figure}
    \centering   \includegraphics[width=0.5\linewidth]{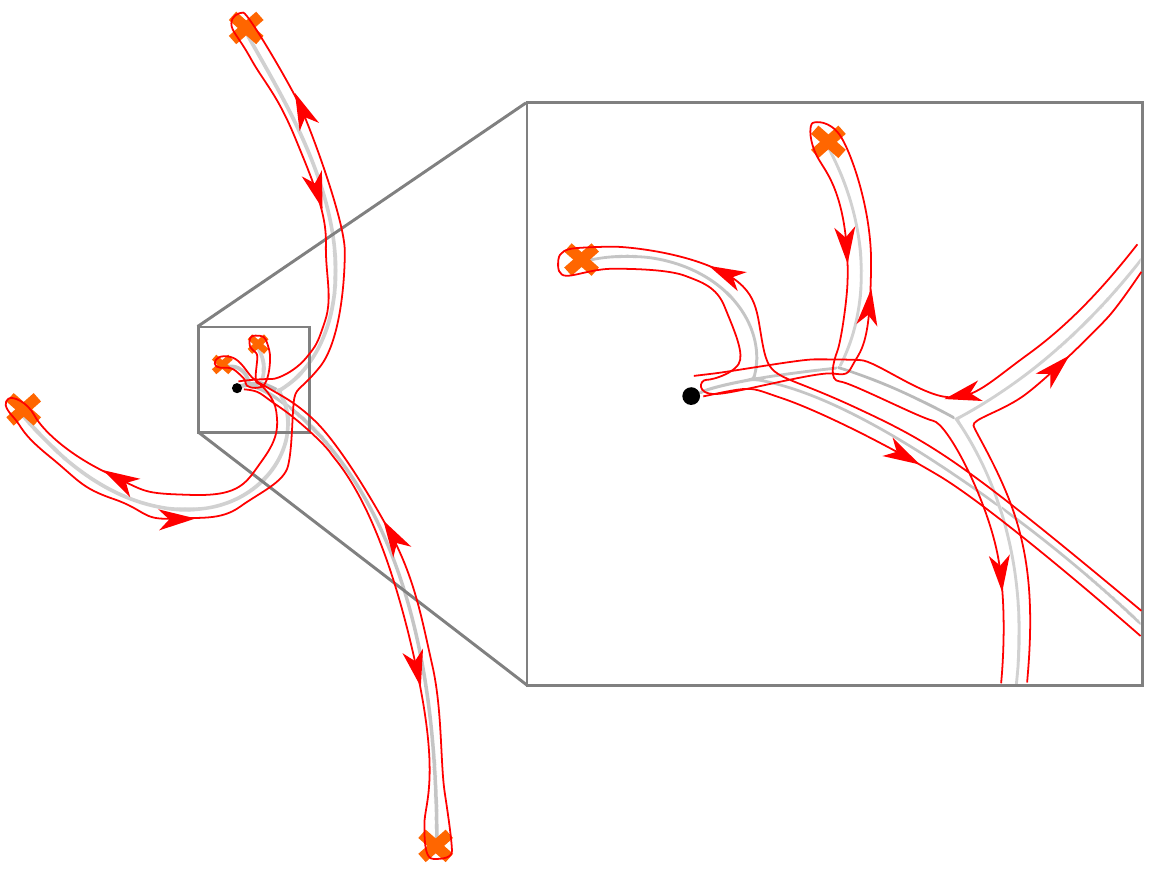}
    \caption{Concatenation of $a_1$ and $a_4^{[+1]}$ for the computation of $C_{14}$. }
    \label{fig:figureeight_f0_14}
\end{figure}

\begin{figure}
    \centering   \includegraphics[width=0.5\linewidth]{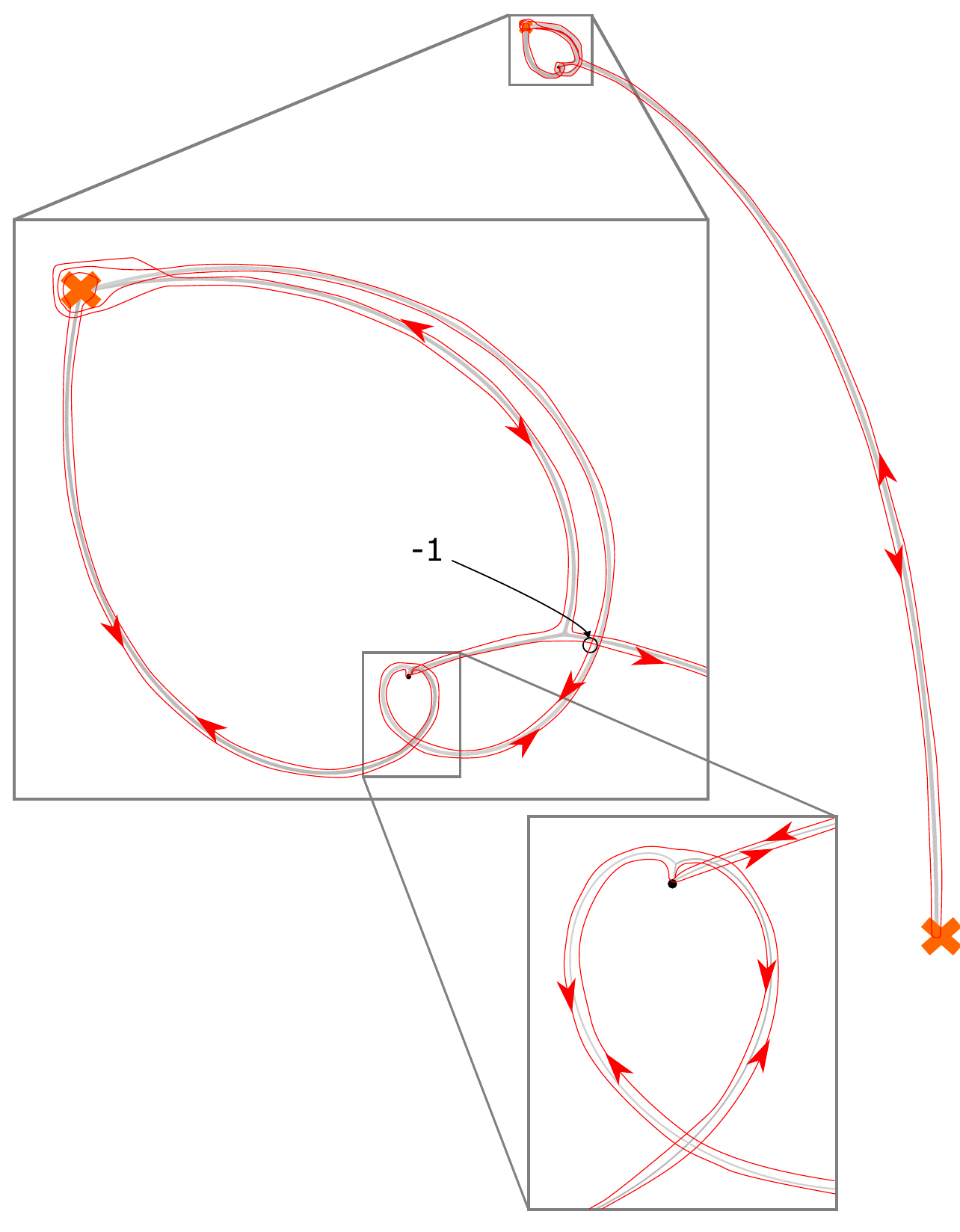}
    \caption{Concatenation of $a_5$ and $a_1^{[+1]}$ for the computation of $C_{15}$.}
    \label{fig:figureeight_f0_15}
\end{figure}

\begin{figure}
    \centering   \includegraphics[width=0.5\linewidth]{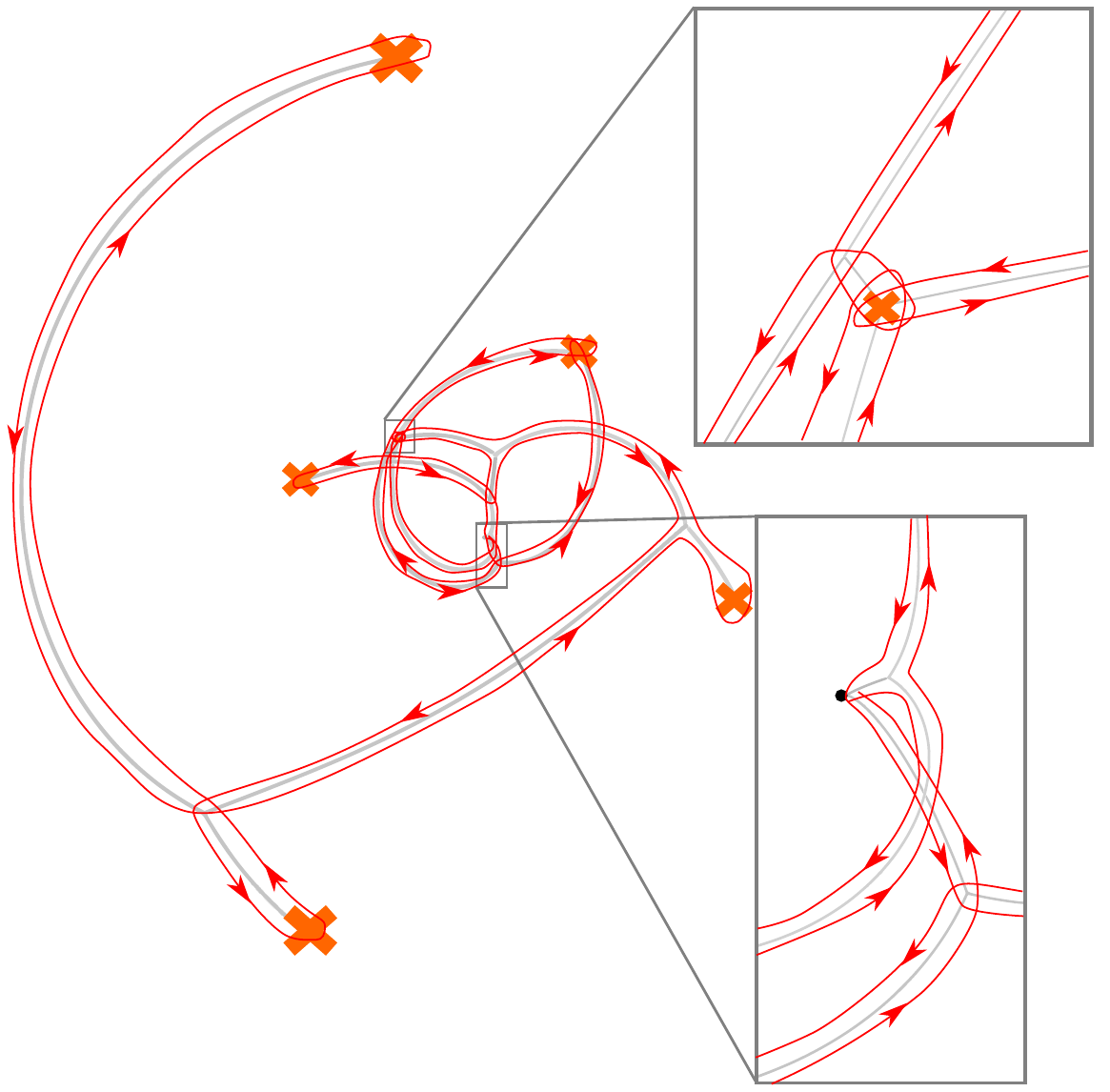}
    \caption{Concatenation of $a_3$ and $a_2^{[+1]}$ for the computation of $C_{23}$.}
    \label{fig:figureeight_f0_23}
\end{figure}

\begin{figure}
    \centering   \includegraphics[width=0.6\linewidth]{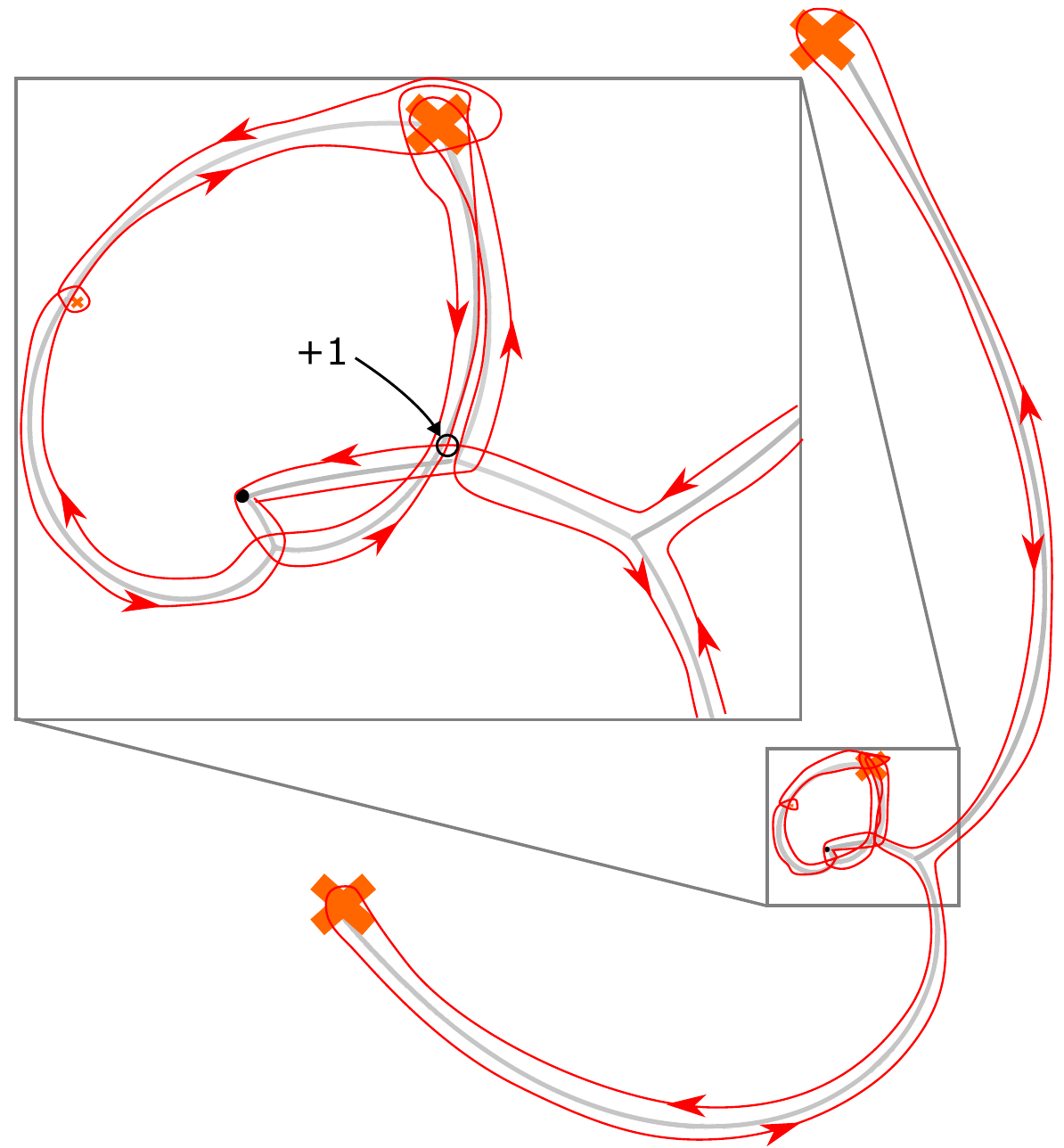}
    \caption{Concatenation of $a_4$ and $a_2^{[+1]}$ for the computation of $C_{24}$.}
    \label{fig:figureeight_f0_24}
\end{figure}

\begin{figure}
    \centering   \includegraphics[width=0.6\linewidth]{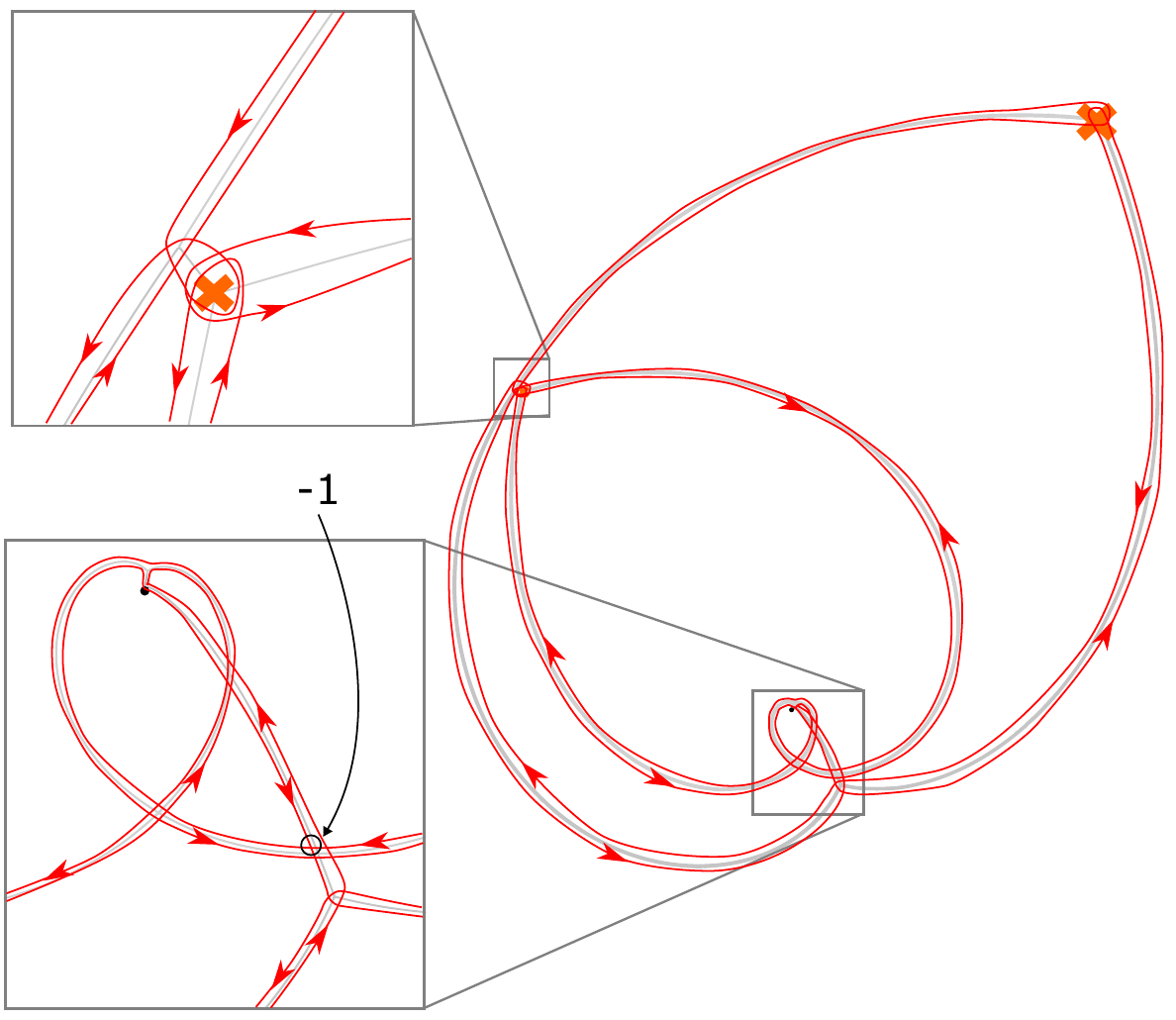}
    \caption{Concatenation of $a_5$ and $a_2^{[+1]}$ for the computation of $C_{25}$.}
    \label{fig:figureeight_f0_25}
\end{figure}

\begin{figure}
    \centering   \includegraphics[width=0.6\linewidth]{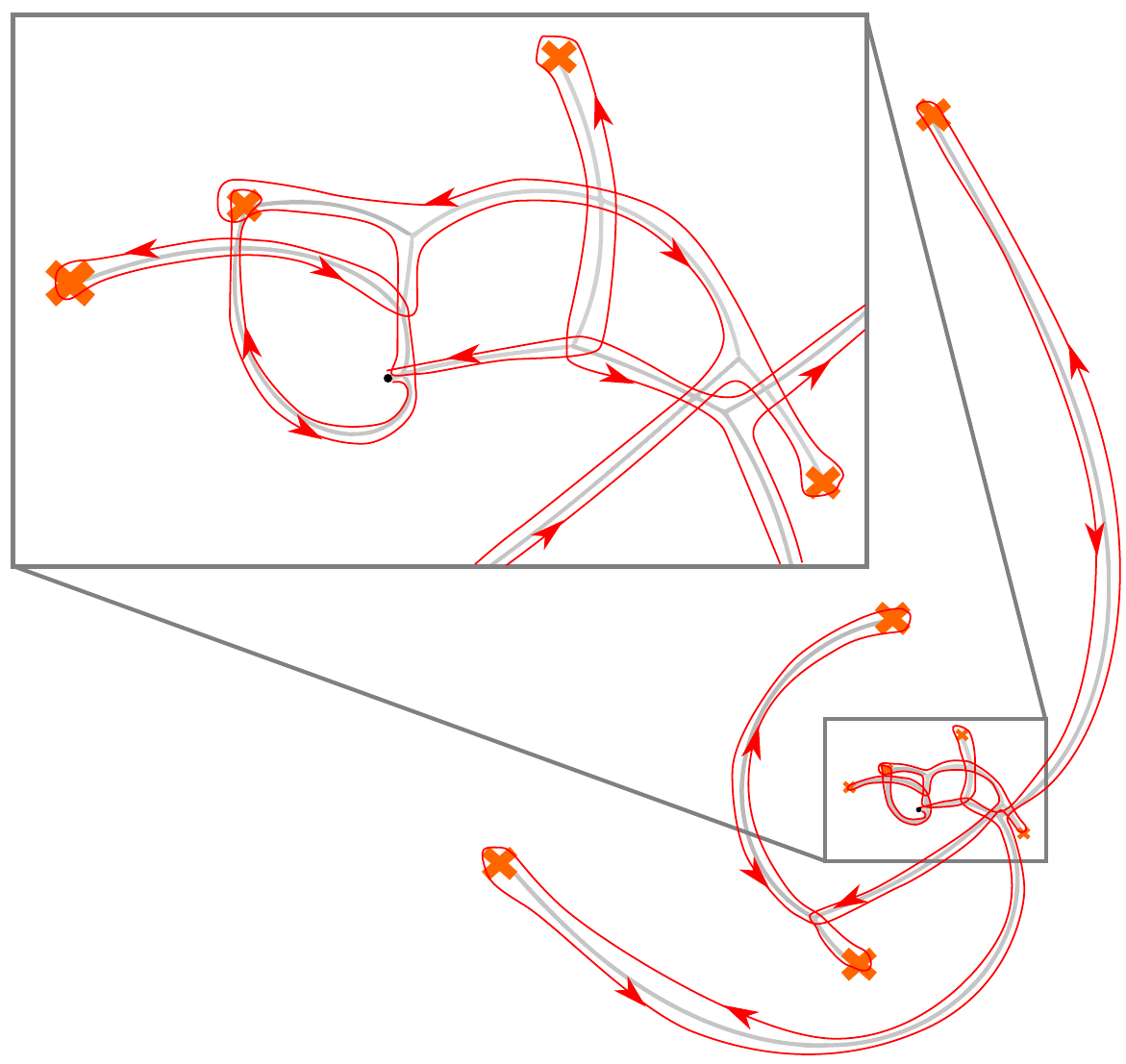}
    \caption{Concatenation of $a_3$ and $a_4^{[+1]}$ for the computation of $C_{34}$.}
    \label{fig:figureeight_f0_34}
\end{figure}

\begin{figure}
    \centering   \includegraphics[width=0.4\linewidth]{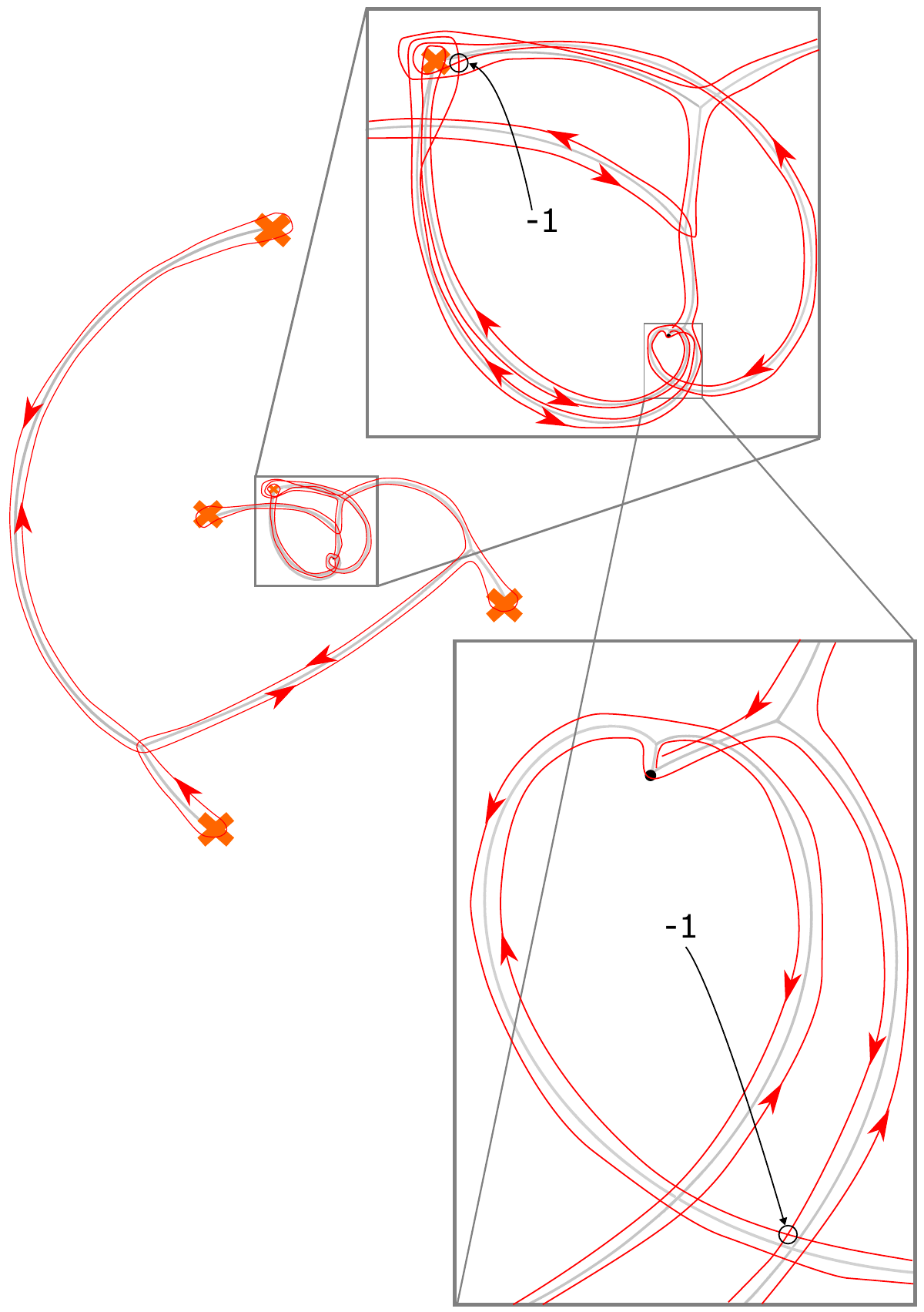}
    \caption{Concatenation of $a_5$ and $a_3^{[+1]}$ for the computation of $C_{35}$.}
    \label{fig:figureeight_f0_35}
\end{figure}

\begin{figure}
    \centering   \includegraphics[width=0.5\linewidth]{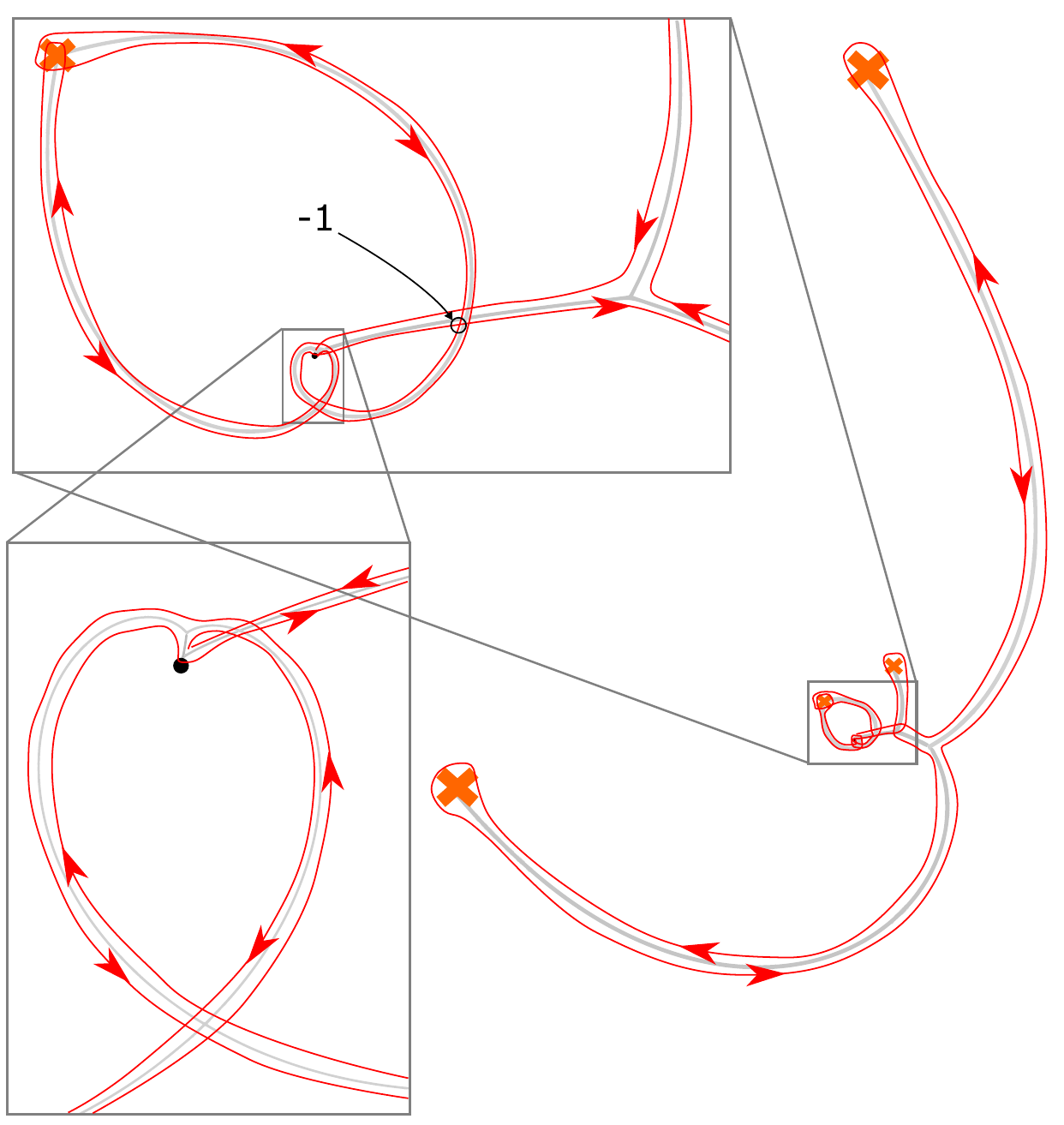}
    \caption{Concatenation of $a_5$ and $a_4^{[+1]}$ for the computation of $C_{45}$.}
    \label{fig:figureeight_f0_45}
\end{figure}

We collect details on the computation of $C_{ij}$ for $i<j$ in Figures \ref{fig:figureeight_f0_12}, \ref{fig:figureeight_f0_13}, \ref{fig:figureeight_f0_14}, \ref{fig:figureeight_f0_15}, \ref{fig:figureeight_f0_23}, \ref{fig:figureeight_f0_24}, \ref{fig:figureeight_f0_25}, \ref{fig:figureeight_f0_34}, \ref{fig:figureeight_f0_35} and \ref{fig:figureeight_f0_45}.

\newpage
\bibliographystyle{unsrt}
\bibliography{bibliography.bib}

\begin{thebibliography}{10}

\bibitem{Gopakumar:1998ii}
Rajesh Gopakumar and Cumrun Vafa.
\newblock {M theory and topological strings. 1.}
\newblock 9 1998.

\bibitem{Gopakumar:1998jq}
Rajesh Gopakumar and Cumrun Vafa.
\newblock {M theory and topological strings. 2.}
\newblock 12 1998.

\bibitem{Ooguri:1999bv}
Hirosi Ooguri and Cumrun Vafa.
\newblock {Knot invariants and topological strings}.
\newblock {\em Nucl. Phys. B}, 577:419--438, 2000.

\bibitem{Labastida:2000zp}
J.~M.~F. Labastida and Marcos Marino.
\newblock {Polynomial invariants for torus knots and topological strings}.
\newblock {\em Commun. Math. Phys.}, 217:423--449, 2001.

\bibitem{Labastida:2000yw}
J.~M.~F. Labastida, Marcos Marino, and Cumrun Vafa.
\newblock {Knots, links and branes at large N}.
\newblock {\em JHEP}, 11:007, 2000.

\bibitem{Kucharski:2017ogk}
Piotr Kucharski, Markus Reineke, Marko Stosic, and Piotr Su\l{}kowski.
\newblock {Knots-quivers correspondence}.
\newblock {\em Adv. Theor. Math. Phys.}, 23(7):1849--1902, 2019.

\bibitem{Ekholm:2018eee}
Tobias Ekholm, Piotr Kucharski, and Pietro Longhi.
\newblock {Physics and geometry of knots-quivers correspondence}.
\newblock {\em Commun. Math. Phys.}, 379(2):361--415, 2020.

\bibitem{Ekholm:2019lmb}
Tobias Ekholm, Piotr Kucharski, and Pietro Longhi.
\newblock {Multi-cover skeins, quivers, and 3d $\mathcal{N}=2$ dualities}.
\newblock {\em JHEP}, 02:018, 2020.

\bibitem{Aganagic:2013jpa}
Mina Aganagic, Tobias Ekholm, Lenhard Ng, and Cumrun Vafa.
\newblock {Topological Strings, D-Model, and Knot Contact Homology}.
\newblock {\em Adv. Theor. Math. Phys.}, 18(4):827--956, 2014.

\bibitem{Eager:2016yxd}
Richard Eager, Sam~Alexandre Selmani, and Johannes Walcher.
\newblock {Exponential Networks and Representations of Quivers}.
\newblock {\em JHEP}, 08:063, 2017.

\bibitem{Banerjee:2018syt}
Sibasish Banerjee, Pietro Longhi, and Mauricio Romo.
\newblock {Exploring 5d BPS Spectra with Exponential Networks}.
\newblock {\em Annales Henri Poincare}, 20(12):4055--4162, 2019.

\bibitem{Terashima:2011qi}
Yuji Terashima and Masahito Yamazaki.
\newblock {SL(2,R) Chern-Simons, Liouville, and Gauge Theory on Duality Walls}.
\newblock {\em JHEP}, 08:135, 2011.

\bibitem{Dimofte:2011ju}
Tudor Dimofte, Davide Gaiotto, and Sergei Gukov.
\newblock {Gauge Theories Labelled by Three-Manifolds}.
\newblock {\em Commun. Math. Phys.}, 325:367--419, 2014.

\bibitem{Dimofte:2011py}
Tudor Dimofte, Davide Gaiotto, and Sergei Gukov.
\newblock {3-Manifolds and 3d Indices}.
\newblock {\em Adv. Theor. Math. Phys.}, 17(5):975--1076, 2013.

\bibitem{Dimofte:2010tz}
Tudor Dimofte, Sergei Gukov, and Lotte Hollands.
\newblock {Vortex Counting and Lagrangian 3-manifolds}.
\newblock {\em Lett. Math. Phys.}, 98:225--287, 2011.

\bibitem{Gupta:2024ics}
Kunal Gupta and Pietro Longhi.
\newblock {Vortices on cylinders and warped exponential networks}.
\newblock {\em Lett. Math. Phys.}, 114(5):123, 2024.

\bibitem{Seiberg:1994rs}
N.~Seiberg and Edward Witten.
\newblock {Electric - magnetic duality, monopole condensation, and confinement
  in N=2 supersymmetric Yang-Mills theory}.
\newblock {\em Nucl. Phys. B}, 426:19--52, 1994.
\newblock [Erratum: Nucl.Phys.B 430, 485--486 (1994)].

\bibitem{Galakhov:2014xba}
Dmitry Galakhov, Pietro Longhi, and Gregory~W. Moore.
\newblock {Spectral Networks with Spin}.
\newblock {\em Commun. Math. Phys.}, 340(1):171--232, 2015.

\bibitem{Seiberg:1996bd}
Nathan Seiberg.
\newblock {Five-dimensional SUSY field theories, nontrivial fixed points and
  string dynamics}.
\newblock {\em Phys. Lett. B}, 388:753--760, 1996.

\bibitem{Morrison:1996xf}
David~R. Morrison and Nathan Seiberg.
\newblock {Extremal transitions and five-dimensional supersymmetric field
  theories}.
\newblock {\em Nucl. Phys. B}, 483:229--247, 1997.

\bibitem{Douglas:1996xp}
Michael~R. Douglas, Sheldon~H. Katz, and Cumrun Vafa.
\newblock {Small instantons, Del Pezzo surfaces and type I-prime theory}.
\newblock {\em Nucl. Phys. B}, 497:155--172, 1997.

\bibitem{Intriligator:1997pq}
Kenneth~A. Intriligator, David~R. Morrison, and Nathan Seiberg.
\newblock {Five-dimensional supersymmetric gauge theories and degenerations of
  Calabi-Yau spaces}.
\newblock {\em Nucl. Phys. B}, 497:56--100, 1997.

\bibitem{Intriligator:2013lca}
Kenneth Intriligator and Nathan Seiberg.
\newblock {Aspects of 3d N=2 Chern-Simons-Matter Theories}.
\newblock {\em JHEP}, 07:079, 2013.

\bibitem{Cecotti:1992rm}
Sergio Cecotti and Cumrun Vafa.
\newblock {On classification of N=2 supersymmetric theories}.
\newblock {\em Commun. Math. Phys.}, 158:569--644, 1993.

\bibitem{Cecotti:1991me}
Sergio Cecotti and Cumrun Vafa.
\newblock {Topological antitopological fusion}.
\newblock {\em Nucl. Phys. B}, 367:359--461, 1991.

\bibitem{Cecotti:2013mba}
Sergio Cecotti, Davide Gaiotto, and Cumrun Vafa.
\newblock {$tt^*$ geometry in 3 and 4 dimensions}.
\newblock {\em JHEP}, 05:055, 2014.

\bibitem{Klemm:1996bj}
Albrecht Klemm, Wolfgang Lerche, Peter Mayr, Cumrun Vafa, and Nicholas~P.
  Warner.
\newblock {Selfdual strings and N=2 supersymmetric field theory}.
\newblock {\em Nucl. Phys. B}, 477:746--766, 1996.

\bibitem{Gaiotto:2011tf}
Davide Gaiotto, Gregory~W. Moore, and Andrew Neitzke.
\newblock {Wall-Crossing in Coupled 2d-4d Systems}.
\newblock {\em JHEP}, 12:082, 2012.

\bibitem{Gaiotto:2012rg}
Davide Gaiotto, Gregory~W. Moore, and Andrew Neitzke.
\newblock {Spectral networks}.
\newblock {\em Annales Henri Poincare}, 14:1643--1731, 2013.

\bibitem{Garoufalidis:2020pax}
Stavros Garoufalidis and Rinat Kashaev.
\newblock {Resurgence of Faddeev's quantum dilogarithm}.
\newblock 8 2020.

\bibitem{Grassi:2022zuk}
Alba Grassi, Qianyu Hao, and Andrew Neitzke.
\newblock {Exponential Networks, WKB and Topological String}.
\newblock {\em SIGMA}, 19:064, 2023.

\bibitem{Alim:2022oll}
Murad Alim, Lotte Hollands, and Iv\'an Tulli.
\newblock {Quantum Curves, Resurgence and Exact WKB}.
\newblock {\em SIGMA}, 19:009, 2023.

\bibitem{Gaiotto:2009hg}
Davide Gaiotto, Gregory~W. Moore, and Andrew Neitzke.
\newblock {Wall-crossing, Hitchin systems, and the WKB approximation}.
\newblock {\em Adv. Math.}, 234:239--403, 2013.

\bibitem{Aganagic:2000gs}
Mina Aganagic and Cumrun Vafa.
\newblock {Mirror symmetry, D-branes and counting holomorphic discs}.
\newblock 12 2000.

\bibitem{Aganagic:2001nx}
Mina Aganagic, Albrecht Klemm, and Cumrun Vafa.
\newblock {Disk instantons, mirror symmetry and the duality web}.
\newblock {\em Z. Naturforsch. A}, 57:1--28, 2002.

\bibitem{Iqbal:2004ne}
Amer Iqbal and Amir-Kian Kashani-Poor.
\newblock {The Vertex on a strip}.
\newblock {\em Adv. Theor. Math. Phys.}, 10(3):317--343, 2006.

\bibitem{Panfil:2018faz}
Mi\l{}osz Panfil and Piotr Su\l{}kowski.
\newblock {Topological strings, strips and quivers}.
\newblock {\em JHEP}, 01:124, 2019.

\bibitem{Ekholm:2021irc}
Tobias Ekholm, Angus Gruen, Sergei Gukov, Piotr Kucharski, Sunghyuk Park, Marko
  Sto\v{s}i\'c, and Piotr Su\l{}kowski.
\newblock {Branches, quivers, and ideals for knot complements}.
\newblock {\em J. Geom. Phys.}, 177:104520, 2022.

\bibitem{Ekholm:2021gyu}
Tobias Ekholm, Piotr Kucharski, and Pietro Longhi.
\newblock {Knot homologies and generalized quiver partition functions}.
\newblock {\em Lett. Math. Phys.}, 113(6):117, 2023.

\bibitem{PhysRevD.96.121902}
Piotr Kucharski, Markus Reineke, Marko Sto\ifmmode \check{s}\else
  \v{s}\fi{}i\ifmmode~\acute{c}\else \'{c}\fi{}, and Piotr Su\l{}kowski.
\newblock {BPS states, knots, and quivers}.
\newblock {\em Phys. Rev. D}, 96:121902, Dec 2017.

\bibitem{2011arXiv1103.2736E}
Alexander~I. {Efimov}.
\newblock {Cohomological Hall algebra of a symmetric quiver}.
\newblock {\em arXiv e-prints}, page arXiv:1103.2736, March 2011.

\bibitem{Faddeev:1993rs}
L.~D. Faddeev and R.~M. Kashaev.
\newblock {Quantum Dilogarithm}.
\newblock {\em Mod. Phys. Lett. A}, 9:427--434, 1994.

\bibitem{Cecotti:1992qh}
Sergio Cecotti, Paul Fendley, Kenneth~A. Intriligator, and Cumrun Vafa.
\newblock {A New supersymmetric index}.
\newblock {\em Nucl. Phys. B}, 386:405--452, 1992.

\bibitem{Bullimore:2018yyb}
Mathew Bullimore and Andrea Ferrari.
\newblock {Twisted Hilbert Spaces of 3d Supersymmetric Gauge Theories}.
\newblock {\em JHEP}, 08:018, 2018.

\end{thebibliography}

\end{document}